\documentclass[journal,comsoc]{IEEEtran}
\usepackage{amsmath,amsfonts}
\usepackage{algorithmic}
\usepackage{algorithm}
\usepackage{array}
\usepackage[caption=false,font=normalsize,labelfont=sf,textfont=sf]{subfig}
\usepackage{textcomp}
\usepackage{stfloats}
\usepackage{url}
\usepackage{verbatim}
\usepackage{graphicx}
\usepackage{cite}
\usepackage[justification=centering]{caption}
\usepackage{multirow}
\usepackage{mathtools}
\usepackage{booktabs}
\usepackage[subnum]{cases}
\usepackage{algorithm}
\usepackage{algorithmic}
\usepackage{amsfonts,amssymb}
\usepackage{soul,listings,xcolor}
\usepackage{stfloats}
\usepackage{ulem}

\begin{document}


\title{Data and Model Poisoning Backdoor Attacks on Wireless Federated Learning, and the Defense Mechanisms: A Comprehensive Survey}

\author{Yichen Wan, Youyang Qu,~\IEEEmembership{Member,~IEEE,} Wei Ni,~\IEEEmembership{Fellow,~IEEE,} Yong Xiang,~\IEEEmembership{Senior Member,~IEEE,} \\
Longxiang Gao,~\IEEEmembership{Senior Member,~IEEE}, 
and Ekram Hossain,~\IEEEmembership{Fellow,~IEEE}

\thanks{Y. Wan and Y. Xiang are with the School of IT, Deakin University, VIC 3125, Australia (email: \{wanyich, yong.xiang\}@deakin.edu.au)}       
\thanks{Y. Qu and W. Ni are with the Data61, CSIRO, NSW 2122, Australia (email: \{youyang.qu, wei.ni\}@data61.csiro.au)}.
\thanks{L. Gao is Qilu University of Technology (Shandong Academy of Sciences), Jinan 250353, China (email: gaolx@sdas.org).}
\thanks{E. Hossain is with the Department of Electrical and Computer Engineering at the University of Manitoba, Canada (email: ekram.hossain@umanitoba.ca).}

}

\markboth{IEEE Communication Surveys and Tutorials}%
{Shell \MakeLowercase{\textit{et al.}}: A Sample Article Using IEEEtran.cls for IEEE Journals}


\maketitle

\begin{abstract}
Due to the greatly improved capabilities of devices, massive data, and increasing concern about data privacy, Federated Learning (FL) has been increasingly considered for applications to wireless communication networks (WCNs). Wireless FL (WFL) is a distributed method of training a global deep learning model in which a large number of participants each train a local model on their training datasets and then upload the local model updates to a central server. However, in general, non-independent and identically distributed (non-IID) data of WCNs raises concerns about robustness, as a malicious participant could potentially inject a ``backdoor'' into the global model by uploading poisoned data or 
models over WCN. This could cause the model to misclassify malicious inputs as a specific target class while behaving normally with benign inputs. This survey provides a comprehensive review of the latest backdoor attacks and defense mechanisms. It classifies them according to their targets (data poisoning or model poisoning), the attack phase (local data collection, training, or aggregation), and defense stage (local training, before aggregation, during aggregation, or after aggregation). The strengths and limitations of existing attack strategies and defense mechanisms are analyzed in detail. Comparisons of existing attack methods and defense designs are carried out, pointing to noteworthy findings, open challenges, and potential future research directions related to security and privacy of WFL. 
\end{abstract}

\begin{IEEEkeywords}
Federated Learning, Backdoor Attacks, Security, Privacy, Wireless Communications
\end{IEEEkeywords}

\section{Introduction}
    \IEEEPARstart{W}{ireless} communication network (WCN) is recognized as part of critical infrastructures and has been deployed ubiquitously in recent years~\cite{khanh2022wireless}. With the persistent development of wireless communication technologies, the network capacity has massively increased, and more devices can be connected to the network~\cite{8274943}. Wireless communications services are used in different scenarios for various Artificial Intelligence (AI)-based applications~\cite{8663994,8467524}, such as image classification task~\cite{zappone2019model,wang2023new}, text prediction~\cite{liu2023pre}, and unmanned aerial vehicle (UAV) control~\cite{8854903,10146001}. With the widespread use of AI applications and user/sensor data involved during the AI training process, the security and privacy issues have become critical~\cite{li2022internet}. In traditional AI algorithm training, participants must send their raw data to a central server for processing. However, WCN introduces the risk of unauthorized data access during the transmission, which might compromise the privacy, security, and robustness of the systems~\cite{yousefpoor2021secure,zhang2023security}. Many efforts have been made to develop a robust and secure framework, while Federated Learning (FL) or Wireless FL (WFL) is considered one of the most practical solutions~\cite{khan2021federated,yu2022optimal,pu2022federated,zheng2023federated,wang2022asynchronous}.       

First proposed by Google \cite{konevcny2016federated}, FL is a machine learning paradigm in which multiple devices form a distributed network while maintaining the same machine learning model. With a large number of clients participating via wireless networks~\cite{li2023multi}, the WFL approach allows for developing a robust model without requiring raw data sharing within the network. Different from traditional machine learning methods that typically collect all raw data from participants and directly train the global model in the central server, the clients of WFL are the owners of their data and participate in the training process by computing model updates based on their local private data. These updates are then aggregated by a central server and used to fine-tune a global model, which is then distributed to the clients for the next round of local training~\cite{yang2019federated}. Consequently, each client can better control their private information. Compared with the traditional machine learning approaches, the need for data transfer in WFL is greatly reduced, the computational efficiency is significantly improved, and the substantial bandwidth provided by the WCN can be better utilized. In recent years, WFL has seen widespread adoption in various fields, including mobile device development, industrial engineering, and healthcare~\cite{nguyen2021federated,xu2021federated,lim2020federated,zhang2021survey}.

Despite its diverse merits, the use of WFL has also raised concerns about the robustness of the training processes~\cite{zawad2021curse}.  
{\color{blue} 
The training datasets of WFL participants are non-independent and identically distributed (non-IID). Each client's data may be different in terms of its statistical properties, data class distribution, or data quality. The heterogeneous and unbalanced-distributed nature of data imposes challenges to train a converging global model that performs well across all clients~\cite{sattler2019robust}. Existing methods for the detection and mitigation of data/model poisoning attacks can be less effective due to the difficulty in distinguishing the disformity of local models resulting from the non-IID datasets from that resulting from attacks~\cite{rajput2019detox}. Furthermore, it makes it easier for malicious clients to intentionally send anomaly updates to  bias the global model in their favor~\cite{cao2021provably}.}   
Adversarial attacks on WFL can be broadly classified into {\color{blue} four} categories: poisoning~\cite{zhang2020poisongan}, inference attacks~\cite{nasr2019comprehensive}, Byzantine attacks~\cite{fang2020local}, and backdoor attacks~\cite{bagdasaryan2020backdoor}. 


In this survey, we focus specifically on backdoor attacks, as they are on the rise, can potentially affect many clients, and may go undetected after multiple rounds of aggregations in WFL. Backdoor attacks are a particularly insidious form of adversarial attack on WFL, as they allow malicious users to manipulate the classification results of the model by altering the local training datasets of participating clients or using malicious model replacements~\cite{wang2020attack}. Backdoor attacks can be divided into two categories: data-based attacks and model-based attacks. In data-based backdoor attacks, the attacker injects a backdoor into the training data. In contrast, model-based attacks target the training process of the local model before aggregation.

Many organizations worldwide are dealing with the threat of backdoor attacks, and the threat is ever-growing unexpectedly. A critical challenge faced by detecting and defending against backdoor attacks is that backdoor attacks often do not negatively impact the accuracy of the model on primary tasks and do not disrupt the normal system operation before they are activated~\cite{bagdasaryan2020backdoor}. This makes it difficult to detect the presence of a backdoor. 

{\color{blue} For example, in August of 2023, Canon, the famous manufacturer of optical and imaging products, warned its users of inkjet printers that their WiFi connection settings stored in the devices were at risk of leakage. The privacy information, including the network SSID, the password, network type (WPA3, WEP, etc.), assigned IP address, MAC address, and network profile stored in the devices' memories are not wiped during initialization and could be easily extracted. Such flaws could be exploited by malicious users to create backdoors, allowing unauthorized access to a Canon printer user's network that the printer was connected to~\cite{jetpack}. From there on, the attacker can access shared resources, steal data, hijack other devices connected to the same network, and launch other attacks to leverage additional vulnerabilities. To be more specific, the Google spam filter has been poisoned by sending malicious emails containing malware or other cybersecurity threats without being detected by the algorithm\footnote{https://www.forcepoint.com/blog/security-labs/data-poisoning-newest-threat-generative-ai}. Besides, researchers have found new ways to poison cutting-edge large generative AL models (e.g., ChatGPT) through prompt engineering\footnote{https://harshitrao.medium.com/is-it-easy-to-poison-chat-gpt-d81287f1b58b}.

While there have been no widely reported data and model poisoning backdoor attacks specifically targeting WFL, designs of such attacks have started to emerge in the literature, e.g.,~\cite{li2023dataagnostic}. Particularly, the broadcast nature of wireless (radio) channels allows misbehaved or compromised participants in WFL to overhear the local model parameters of multiple or all benign participants. By taking advantage of the overheard local models, the misbehaved participants can construct malicious local models to escape malicious model detection at the central server, poison the global model and subsequently the benign local models, and manipulate WFL~\cite{li2023dataagnostic}. It is crucial to detect and suppress such attacks on WFL due to the increasing interest in WFL deployment and the exacerbated impact of the attacks over wireless interfaces.}

Defending against data-based and model-based backdoor attacks on WFL remains an open issue. It requires the protection of both the local and global models against all potential attacks. Many works have been proposed to address this challenge. For example, some researchers have proposed using differential privacy to defend against poisoned data samples in the training process or adding a threshold value to gradients to block abnormal gradients classified as malicious attackers~\cite{abadi2016deep,rodriguez2022backdoor,nguyen2022flame}. Others have proposed methods such as a backdoor filter~\cite{hou2021mitigating} or a White Blood Cell for WFL (WFL-WBC)~\cite{sun2021fl} to defend against model-poisoning attacks. However, these methods only show reasonable performance against specific types of attacks. There is still much work to be done to address this common problem.

\subsection{Motivation}

As discussed above, backdoor attacks can occur at any phase during the convergence of the global model, including data collection, local model training, and global model aggregation~\cite{shokri2020bypassing,dong2021black}. This makes them more threatening and more difficult to defend against, as the defender needs to consider all steps in the process. In contrast to Byzantine attacks that aim to degrade the performance of the main WFL task~\cite{zhang2015byzantine}, backdoor attacks are characterized by their stealthiness. The primary goal of these attacks is to leave a backdoor in the converged global model that can mislead the model into classifying the backdoor input into the target class while behaving normally for benign inputs. This property allows the backdoor to remain effective for a longer duration, making it difficult to eliminate fully. Therefore, it is important to delve deeper into the nature of backdoor attacks to inform the design of defense algorithms and improve the robustness and security of the WFL framework.

Despite the rapid growth of WFL as a research area, there has not yet been a comprehensive and up-to-date review of it from the perspective of backdoor attacks. In this paper, we aim to provide a systematic summary and classification of existing backdoor attacks and defenses, highlighting their differences and connections, as well as discussing their respective limitations. We also discuss future research directions in this area.

\subsection{Review of Existing Surveys and Gap Analysis}

The existing surveys on backdoor attacks and defenses under WFL frameworks (from 2020 to 2023) are reviewed. The relevant papers are collected from multiple academic databases, including IEEE Xplore, Elsevier, MDPI, and arXiv. The comparison of this survey with the existing literature is summarized in Table \ref{Surveys}. It can be seen that some existing surveys, e.g., \cite{jebreel2023fl}, \cite{mothukuri2021survey}, \cite{lyu2022privacy}, and~\cite{sikandar2023detailed} considered backdoor attacks and backdoor defenses as a part of robustness threat on WFL, however, the limitations of the existing backdoor attack and defense methods were not highlighted. On the other hand, in \cite{gao2020backdoor}, \cite{li2022backdoor}, \cite{guo2022overview}, and \cite{goldblum2022dataset}, WFL was considered as one of the deep learning applications when discussing the impact of backdoor attacks, but no detailed analysis of vulnerabilities of backdoor attacks on WFL was provided. In \cite{gong2022backdoor} and \cite{dung2023backdoor}, the theoretical working mechanisms of backdoor attacks and defenses for WFL were reviewed. However, the limitations of the existing methodologies were not systematically analyzed. In the survey \cite{guo2022overview}, the backdoor attack strategies were discussed in depth, but they were not discussed in the context of WFL. Also, to the best of the authors' knowledge, none of the existing surveys has taken WCN into consideration. Nevertheless, the security threats induced by WCN are inevitable when discussing the influence of backdoor attacks on WFL.


\begin{table*}[htb]
\caption{A summary of existing surveys related to WFL backdoor attacks and defense methods}
\renewcommand{\arraystretch}{0.9}
\label{Surveys}
\scalebox{1}[1.3]{
\resizebox{\textwidth}{19mm}{
\begin{tabular}{c|c|c|c|c|ccccccc|cccc}
\hline
\hline
\multirow{2}{*}{\begin{tabular}[c]{@{}c@{}}Survey \\ paper\end{tabular}} &
  \multirow{2}{*}{Year} &
  \multirow{2}{*}{\begin{tabular}[c]{@{}c@{}}Conside-\\ration\\ of \\ WCN\end{tabular}} &
  \multirow{2}{*}{\begin{tabular}[c]{@{}c@{}}Visuali-\\zation \\ for each \\ method\end{tabular}} &
  \multirow{2}{*}{\begin{tabular}[c]{@{}c@{}}Discussion \\ on lessons \\ learned\end{tabular}} &
  \multicolumn{7}{c|}{\begin{tabular}[c]{@{}c@{}}Discussion on backdoor attack methods \end{tabular}} &
  \multicolumn{4}{c}{\begin{tabular}[c]{@{}c@{}}Discussion on backdoor \\defense methods \end{tabular}} \\ \cline{6-16} 
 &
   &
   &
   &
   &
  \multicolumn{1}{c|}{\begin{tabular}[c]{@{}c@{}}Compa-\\ rison of  \\ attack \\ methods\end{tabular}} &
  \multicolumn{1}{c|}{\begin{tabular}[c]{@{}c@{}}Mathe-\\ matical \\ analysis \\ for \\ each \\ method\end{tabular}} &
  \multicolumn{1}{c|}{\begin{tabular}[c]{@{}c@{}}Required \\ prior \\ infor- \\mation \\ analysis\end{tabular}} &
  \multicolumn{1}{c|}{\begin{tabular}[c]{@{}c@{}}Attack \\ applica- \\bility\end{tabular}} &
  \multicolumn{1}{c|}{\begin{tabular}[c]{@{}c@{}}Evaluation \\ metric\end{tabular}} &
  \multicolumn{1}{c|}{\begin{tabular}[c]{@{}c@{}}Limitation \\ analysis \\ for each \\ method\end{tabular}} &

  \begin{tabular}[c]{@{}c@{}}Future \\ directions\end{tabular} &

  \multicolumn{1}{c|}{\begin{tabular}[c]{@{}c@{}}Compa-\\ rison of \\ defense \\ methods\end{tabular}} &
  \multicolumn{1}{c|}{\begin{tabular}[c]{@{}c@{}}Defense \\ applica-\\ bility\end{tabular}} &
  \multicolumn{1}{c|}{\begin{tabular}[c]{@{}c@{}}Limita-\\ tion \\ analysis \\ for each \\ method\end{tabular}} &
  \begin{tabular}[c]{@{}c@{}}Future \\ direction\end{tabular} \\ \hline
  
\  \cite{jere2020taxonomy} &
  2020 &
   &
   &
   &
  \multicolumn{1}{c|}{$\surd$} &
  \multicolumn{1}{c|}{} &
  \multicolumn{1}{c|}{} &
  \multicolumn{1}{c|}{} &
  \multicolumn{1}{c|}{} &
  \multicolumn{1}{c|}{} &
   &
  \multicolumn{1}{c|}{} &
  \multicolumn{1}{c|}{} &
  \multicolumn{1}{c|}{} &
  $\surd$ \\ \hline

\ \cite{gao2020backdoor} &
  2020 &
   &
   &
   &
  \multicolumn{1}{c|}{$\surd$} &
  \multicolumn{1}{c|}{} &
  \multicolumn{1}{c|}{} &
  \multicolumn{1}{c|}{} &
  \multicolumn{1}{c|}{} &
  \multicolumn{1}{c|}{} &
  $\surd$ &
  \multicolumn{1}{c|}{$\surd$} &
  \multicolumn{1}{c|}{} &
  \multicolumn{1}{c|}{$\surd$} &
  $\surd$ \\ \hline
\ \cite{mothukuri2021survey} &
  2021 &
   &
   &
   &
  \multicolumn{1}{c|}{$\surd$} &
  \multicolumn{1}{c|}{} &
  \multicolumn{1}{c|}{} &
  \multicolumn{1}{c|}{} &
  \multicolumn{1}{c|}{$\surd$} &
  \multicolumn{1}{c|}{} &
   &
  \multicolumn{1}{c|}{$\surd$} &
  \multicolumn{1}{c|}{} &
  \multicolumn{1}{c|}{} &
  $\surd$ \\ \hline
  
\ \cite{gong2022backdoor} &
  2022 &
   &
   &
   &
  \multicolumn{1}{c|}{$\surd$} &
  \multicolumn{1}{c|}{} &
  \multicolumn{1}{c|}{} &
  \multicolumn{1}{c|}{} &
  \multicolumn{1}{c|}{$\surd$} &
  \multicolumn{1}{c|}{} &
  $\surd$ &
  \multicolumn{1}{c|}{$\surd$} &
  \multicolumn{1}{c|}{} &
  \multicolumn{1}{c|}{} &
  $\surd$ \\ \hline
  
\ \cite{li2022backdoor} &
  2022 &
   &
   &
   &
  \multicolumn{1}{c|}{$\surd$} &
  \multicolumn{1}{c|}{} &
  \multicolumn{1}{c|}{} &
  \multicolumn{1}{c|}{} &
  \multicolumn{1}{c|}{$\surd$} &
  \multicolumn{1}{c|}{} &
  $\surd$ &
  \multicolumn{1}{c|}{$\surd$} &
  \multicolumn{1}{c|}{} &
  \multicolumn{1}{c|}{} &
  $\surd$ \\ \hline
  
\ \cite{guo2022overview} &
  2022 &
   &
   &
   &
  \multicolumn{1}{c|}{$\surd$} &
  \multicolumn{1}{c|}{$\surd$ } &
  \multicolumn{1}{c|}{} &
  \multicolumn{1}{c|}{$\surd$} &
  \multicolumn{1}{c|}{$\surd$} &
  \multicolumn{1}{c|}{$\surd$} &
  $\surd$ &
  \multicolumn{1}{c|}{$\surd$} &
  \multicolumn{1}{c|}{} &
  \multicolumn{1}{c|}{} &
  $\surd$ \\ \hline
  
\ \cite{lyu2022privacy} &
  2022 &
   &
   &
   &
  \multicolumn{1}{c|}{} &
  \multicolumn{1}{c|}{} &
  \multicolumn{1}{c|}{} &
  \multicolumn{1}{c|}{} &
  \multicolumn{1}{c|}{} &
  \multicolumn{1}{c|}{} &
  $\surd$ &
  \multicolumn{1}{c|}{$\surd$} &
  \multicolumn{1}{c|}{} &
  \multicolumn{1}{c|}{} &
  $\surd$ \\ \hline
  
\ \cite{sikandar2023detailed} &
  2023 &
   &
  $\surd$ &
   &
  \multicolumn{1}{c|}{$\surd$} &
  \multicolumn{1}{c|}{} &
  \multicolumn{1}{c|}{} &
  \multicolumn{1}{c|}{$\surd$} &
  \multicolumn{1}{c|}{} &
  \multicolumn{1}{c|}{} &
   &
  \multicolumn{1}{c|}{$\surd$} &
  \multicolumn{1}{c|}{} &
  \multicolumn{1}{c|}{} &
   \\ \hline
   
\ \cite{dung2023backdoor} &
  2023 &
   &
   &
   &
  \multicolumn{1}{c|}{$\surd$} &
  \multicolumn{1}{c|}{} &
  \multicolumn{1}{c|}{} &
  \multicolumn{1}{c|}{$\surd$} &
  \multicolumn{1}{c|}{$\surd$} &
  \multicolumn{1}{c|}{} &
  $\surd$ &
  \multicolumn{1}{c|}{$\surd$} &
  \multicolumn{1}{c|}{$\surd$} &
  \multicolumn{1}{c|}{} &
  $\surd$ \\ \hline
  
\ \cite{goldblum2022dataset} &
  2023 &
   &
   &
   &
  \multicolumn{1}{c|}{$\surd$} &
  \multicolumn{1}{c|}{} &
  \multicolumn{1}{c|}{} &
  \multicolumn{1}{c|}{$\surd$} &
  \multicolumn{1}{c|}{} &
  \multicolumn{1}{c|}{$\surd$} &
  $\surd$ &
  \multicolumn{1}{c|}{} &
  \multicolumn{1}{c|}{} &
  \multicolumn{1}{c|}{} &
  $\surd$ \\ \hline
  
Ours &
   &
  $\surd$ &
  $\surd$ &
  $\surd$ &
  \multicolumn{1}{c|}{$\surd$} &
  \multicolumn{1}{c|}{$\surd$} &
  \multicolumn{1}{c|}{$\surd$} &
  \multicolumn{1}{c|}{$\surd$} &
  \multicolumn{1}{c|}{$\surd$} &
  \multicolumn{1}{c|}{$\surd$} &
  $\surd$ &
  \multicolumn{1}{c|}{$\surd$} &
  \multicolumn{1}{c|}{$\surd$} &
  \multicolumn{1}{c|}{$\surd$} &
  $\surd$ \\ \hline
  \hline
  \end{tabular}}}
\end{table*}

\subsection{Scope and Contributions}

Compared with the existing surveys, this paper presents a comprehensive survey of the latest backdoor attacks on WFL and corresponding defense mechanisms. Over 200 papers from 2015 to 2023 have been reviewed in depth, including those focusing on WFL, the development of backdoor attack methodologies, and defense mechanisms. 
The key contributions of our paper are summarized as follows:
\begin{itemize}
    \item We present a holistic review of WFL, emphasizing its vulnerability and security concerning backdoor attacks.
    The WFL vulnerabilities potentially open the door for backdoor attacks. 
    
    \item We comprehensively review the existing backdoor attack methodologies based on their attack targets and stages and analyze their strengths and limitations. The mathematical derivation process is studied to understand the working mechanism of backdoor attacks on WFL.
    
    \item We provide a systematic classification and discussion on defense mechanisms against backdoor attacks on WFL. We compare defense algorithms at different phases of the model training, and analyze the limitations of existing defense schemes.
    
    \item Open issues and research challenges are pointed out, followed by potentially promising research directions in this fast-growing, under-explored domain. For instance, 
    backdoor attacks based on data poisoning are relatively weaker and more eliminable, while those based on model poisoning exhibit stronger attack performance and stealthiness at the expense of higher costs. The detection of backdoors typically relies on threshold design, and the training of existing defense mechanisms often requires access to historical information about the attacks.
\end{itemize}

\subsection{Structure of This Survey}
\begin{figure*}[htb]
\centering
\includegraphics[scale=0.8]{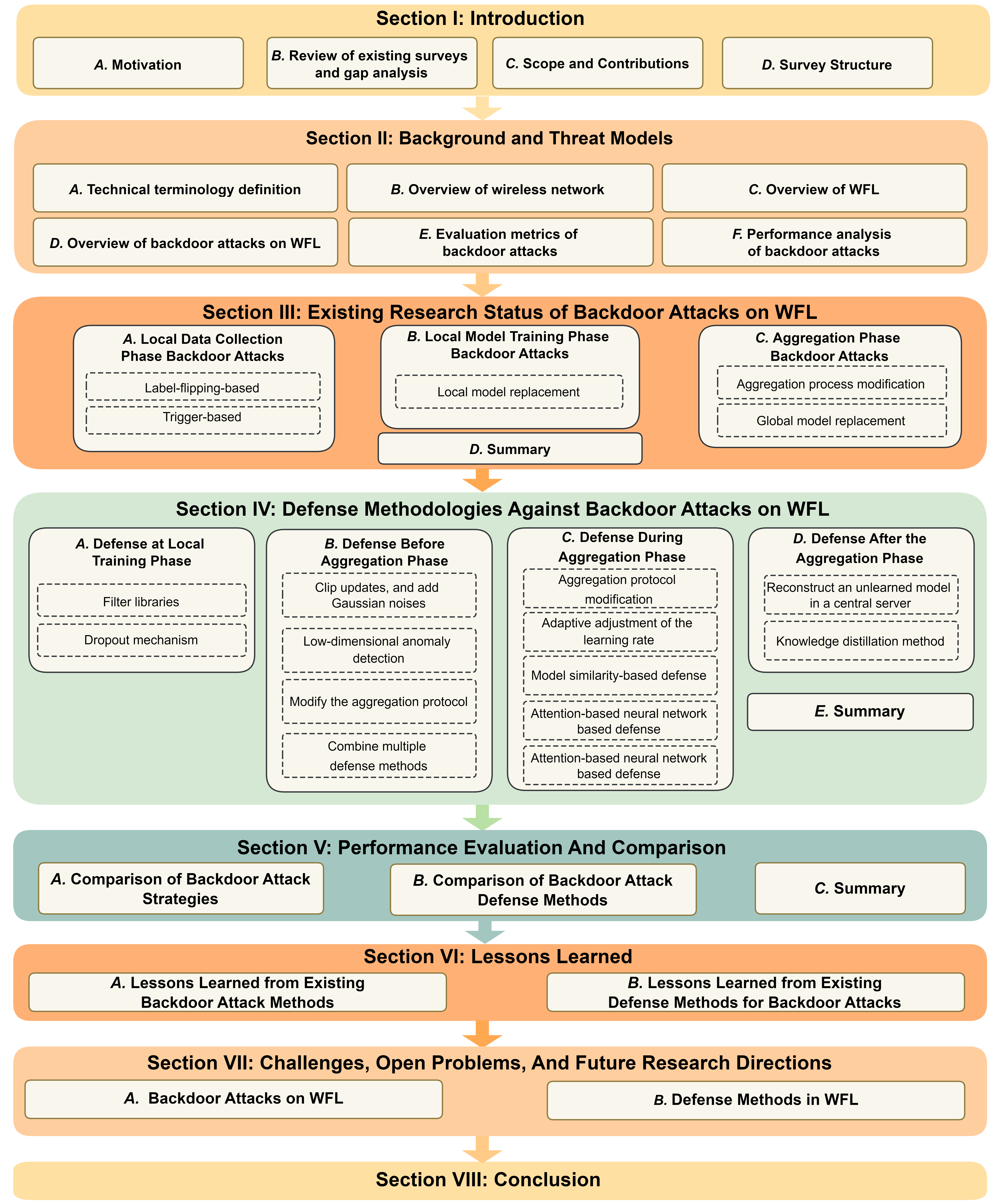}
\centering 
\caption{The structure of this survey.}
 \label{fig:intro}
\end{figure*}

The structure of this survey is shown in Fig.~\ref{fig:intro}. Section~\ref{sec::Background} gives the overview of WCN and WFL. The threat model of backdoor attacks on WFL is also discussed. In~\ref{sec::Existing}, the mechanisms of the existing backdoor attack methods are reviewed. In Section~\ref{sec::Defense}, the recent works on backdoor attack defense methodologies and their limitations are analyzed. The performance of backdoor attack strategies and defense methods in each phase is presented and analyzed in Section~\ref{sec::Performance}. The lessons learned are discussed in Section~\ref{sec::lessons}. Conclusion and future works are summarized in Section~\ref{sec::issues}. Abbreviations, acronyms, and notations used in this survey are defined in Table~\ref{abb}. The definitions of technical terms used in WFL networks, backdoor attacks, and defenses are listed in Table \ref{terms}. These definitions will be consistently referred to throughout the rest of this survey.

\begin{table*}[htb]
\centering
\normalsize
\caption{List of abbreviations, notations, and their definitions}
\label{abb}
\renewcommand{\arraystretch}{1.3}
\scalebox{0.9}{
\begin{tabular}{cl|l|l}
\hline
\hline
\multicolumn{2}{c|}{\textbf{Type}}                       & \textbf{Abbreviation}             & \textbf{Description}                                                    \\ \hline

\multicolumn{2}{c|}{\multirow{8}{*}{\begin{tabular}[c]{@{}c@{}}Technical \\ Terms\end{tabular}}}       & IID                & Independent and identically distributed data                            \\ \cline{3-4} 
\multicolumn{2}{c|}{}                                    & non-IID                       & Non-independent and identically distributed data                                                   \\ \cline{3-4}
\multicolumn{2}{c|}{}                                    & FL                       & Federated learning                                                   \\ \cline{3-4}
\multicolumn{2}{c|}{}                                    & WFL                         & Wireless federated learning                                                      \\ \cline{3-4} 
\multicolumn{2}{c|}{}                                    & AI                         & Artificial intelligence                                                    \\ \cline{3-4} 

\multicolumn{2}{c|}{}                                    & DP                         & Differential privacy                                                    \\ \cline{3-4}

\multicolumn{2}{c|}{}                                    & SGD                        & Stochastic gradient descent                                             \\ \cline{3-4}
\multicolumn{2}{c|}{}                                    & TCP                       & Transmission control protocol                                             \\ \cline{3-4}
\multicolumn{2}{c|}{}                                    & UDP                       & User datagram Protocol                                             \\\cline{3-4}
\multicolumn{2}{c|}{}                                    & OTA                       & Over-the-air                                             \\ \hline
\multicolumn{2}{c|}{\multirow{4}{*}{\begin{tabular}[c]{@{}c@{}}Evaluation \\ Metrics\end{tabular}}} & PDR                        & Poisoned data rate                                                      \\ \cline{3-4} 

\multicolumn{2}{c|}{}                                    & CCR                        & Compromised client rate                              \\ \cline{3-4} 
\multicolumn{2}{c|}{}                                    & BAR                        & Backdoor accuracy rate                                                  \\ \cline{3-4} 
\multicolumn{2}{c|}{}                                    & ASR                        & Attack success rate (equivalent to BAR)                                  \\ \cline{3-4} 
\multicolumn{2}{c|}{}                                    & BTA                        & Backdoor task accuracy (equivalent to BAR)                              \\ \cline{3-4} 

\multicolumn{2}{c|}{}                                    & MAR                        & Main task accuracy rate                                                 \\ \hline

\multicolumn{2}{c|}{\multirow{4}{*}{\begin{tabular}[c]{@{}c@{}}Mathematical \\ Symbols\end{tabular}}} & $N$                    & The total number of clients involved in WFL                                  \\ \cline{3-4} 
\multicolumn{2}{c|}{}                                    & $m$                        & The number of randomly selected clients within one training iteration. \\ \cline{3-4} 
\multicolumn{2}{c|}{}                                    & $t$                        & Time index                                                              \\ \cline{3-4} 
\multicolumn{2}{c|}{}                                    & $T$                        & The number of FL training rounds                                      \\ \cline{3-4}
\multicolumn{2}{c|}{}                                    & $G^{t}$                    & The global model at the start of the current training iteration.        \\ \cline{3-4} 
\multicolumn{2}{c|}{}                                    & $G^{t+1}$                  & The updated global model after current training iteration               \\ \cline{3-4} 
\multicolumn{2}{c|}{}                                    & $L_i^t$                      & The local model at client $i$                                           \\ \cline{3-4} 
\multicolumn{2}{c|}{}                                    & $D_{i}^t$                    & Benign training dataset of client $i$                                   \\ \cline{3-4} 
\multicolumn{2}{c|}{}                                    & ${D_{Ai}}^t$                 & Backdoored training dataset of client $i$                               \\ \cline{3-4} 
\multicolumn{2}{c|}{}                                    & $F_{i}({L_i^t}, {D_i^t})$      & Loss function of the local training algorithm at client $i$             \\ \cline{3-4} 
\multicolumn{2}{c|}{}                                    & ${F_{Ai}}({L_i^t},{D_{Ai}^t})$ & Loss function of the malicious objective at client $i$                  \\ \cline{3-4} 
\multicolumn{2}{c|}{}                                    & $\alpha$                   & The tradeoff between the benign and malicious objectives                \\ \cline{3-4} 
\multicolumn{2}{c|}{}                                    & $p_i^{t+1}$                & Clean local model update of client $i$ at $t$-th training iteration   \\ \cline{3-4} 
\multicolumn{2}{c|}{}                                       & $p_{Ai}^{t + 1}$       & Poisoned local model update of client $i$ at $t$-th training iteration    \\ \cline{3-4} 
\multicolumn{2}{c|}{}                                    & $p_g^{t+1}$                & Aggregated global model update at $t$-th training iteration           \\ \cline{3-4} 
\multicolumn{2}{c|}{}                                    & $\beta_i^{t+1}$            & Weight of client $i$ at $t$-th training iteration                     \\ \cline{3-4} 
\multicolumn{2}{c|}{}                                       & $\beta _{Ai}^{t + 1}$ & Manipulated weight of compromised client $i$ at $t$-th training iteration \\ \cline{3-4} 
\multicolumn{2}{c|}{}                                    & $\eta^t$                   & Global learning rate at $t$-th training iteration                     \\ \hline
\hline
\end{tabular}}
\end{table*}

\begin{table}[htb]
\centering
\caption{Technical terminologies related to backdoor attacks on WFL}
\label{terms}
\renewcommand{\arraystretch}{1.2}
\scalebox{1}{
\begin{tabular}{ccp{6cm}}
\hline
\hline
\multicolumn{2}{c|}{\textbf{Term}} & 
  
\multicolumn{1}{c}{\textbf{Definition}} \\ \hline

\multicolumn{2}{c|}{\begin{tabular}[c]{@{}c@{}}Client \end{tabular}}  & General Participant in WFL framework. Every client contributes to the convergence of the global model. \\ \hline

\multicolumn{2}{c|}{\begin{tabular}[c]{@{}c@{}}Attacker \end{tabular}}  & The malicious clients in the client party. Attackers aim to inject a backdoor into the global model while keeping the convergence of the main task. \\ \hline

\multicolumn{2}{c|}{\begin{tabular}[c]{@{}c@{}}Compromised \\ client\end{tabular}} & The client fully controlled by the attacker. All the attacks are conducted through compromised clients. \\ \hline

\multicolumn{2}{c|}{\begin{tabular}[c]{@{}c@{}}Defender\end{tabular}} & The defense mechanism mounted in the Federated framework. The design requirement is to filter out the malicious input and eliminate the backdoor effect in the global model. \\ \hline
  
\multicolumn{2}{c|}{\begin{tabular}[c]{@{}c@{}}Trigger input\end{tabular}} & The malicious input with trigger pattern. The trigger pattern can be random or model-dependent regarding different trigger designs. Since trigger-based backdoor attacks normally target image classification tasks, the trigger pattern is pixel-based.\\ \hline

\multicolumn{2}{c|}{\begin{tabular}[c]{@{}c@{}}Target label \end{tabular}}  & The Label where the attacker aims to inject the backdoor. \\ \hline

\multicolumn{2}{c|}{\begin{tabular}[c]{@{}c@{}}Attack round \end{tabular}}  & The training round when the attacker conducts the attack. Depending on different attack types, the attack round can either be a single round or multiple rounds. \\ 

\hline
\hline
\end{tabular} }
\end{table}

\section{Background and Threat Models} 
\label{sec::Background}
This section presents an overview of WFL and backdoor attacks on WFL. The technical terms are first introduced. A brief overview of the wireless communication network is then presented. After that, the working mechanism of WFL is reviewed. The motivation for backdoor attacks and their defense methods are presented by analyzing the major security challenges encountered in deploying WFL. An overview of existing backdoor attack methodologies is then carried out, including the evaluation metrics and comparison of different types of backdoor attacks for different attack targets and stages. 

\subsection{Overview of Wireless Communications Technologies}
Wireless communication is one of the fastest-growing technologies in recent years~\cite{hu2021distributed,lyu2019optimal}. The devices connected to the wireless network exchange data via radio waves instead of physical connections, such as cables or fibers. According to different communication ranges and data capacities~\cite{yu2022dataset}, many wireless communication technologies are developed, including Wi-Fi, Cellular Networks, Bluetooth, Near Field Communication, and Zigbee~\cite {gupta2021survey}. With the advancements in wireless technologies, the wireless network has become an essential part of our daily lives. 

\begin{figure}[htb]
    \centering
    \includegraphics[scale=0.6]{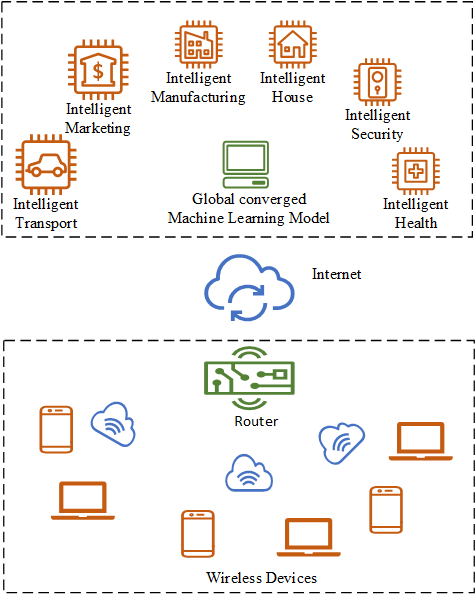}
    \caption{A wireless network with machine learning applications.}
    \label{wireless}
\end{figure}

The rapid progress of wireless communication technologies has also promoted the development of machine learning algorithms. The wireless network can collect a large amount of data from various sources, which machine learning algorithms can apply to analyze and train a converged model.  Fig.~\ref{wireless} shows an example of a machine learning training task in wireless communication networks. The wireless network enables a flexible and convenient way of transmitting data from local devices for machine learning applications, such as intelligent transport~\cite{kaffash2021big,li2022federated,cui2022vehicular}, intelligent marketing~\cite{vlavcic2021evolving}, intelligent manufacturing~\cite{wang2022big}, smart home~\cite{khan2020machine}, intelligent security~\cite{lv2019intelligent}, and intelligent health~\cite{abdellatif2019edge}. Machine-learning algorithms also contribute considerably to wireless communication development~\cite{Xiaojin2023} and failure analysis~\cite{SiguoBi2023}. Efforts have been made to train a machine-learning model that can be applied to optimize resource allocation in wireless networks and improve overall network efficiency~\cite{shi2020communication}.

On the other hand, the wider adoption of machine learning applications in wireless networks has prompted a multitude of ongoing research and development endeavors in the field of wireless communications. Organizations, such as IEEE and 3GP, continuously update the wireless standards, for example, Wi-Fi standard IEEE 802.11~\cite{odirichukwu2021interoperable}, to push the boundaries of wireless communication. To further protect participants' private information when exchanging information during the training process~\cite{yuan2023amplitude}, many encryption algorithms and authentication techniques for wireless networks have been proposed~\cite{vaseghi2021finite}. The next-generation wireless communication technology 6G is also a frontier research topic~\cite{liu2022integrated}. 6G can provide a higher data rate, ultra-low latency, massive device connectivity, and other advanced features, which are precisely required by machine learning applications~\cite{sun2020machine}. The combination of wireless communication networks and machine learning techniques is able to improve network management and further enhance user experiences in various aspects. 

It can be predicted that FL will be widely deployed in WCN for various applications owing to its diverse merits. However, WFL will be more vulnerable to backdoor attacks than its wired counterpart due to a combination of factors inherent to wireless networks. Here are some of the reasons:

\begin{itemize}
    \item {\em Network Security}: Wireless networks are generally less secure networked than wired networks due to the broadcast nature of radio. Data transmissions over wireless links are through airwaves, making it easier for attackers to eavesdrop~\cite{9669934,huang2021navigation,8735776,8756294}, jam~\cite{yuan2023joint}, and manipulate data, e.g., replay attack and Sybil attack, or intrude the networks~\cite{altaf2023new,altaf2023ne}. 

    \item {\em Device Vulnerabilities}: WFL usually involves an extensive range of devices, including low-cost IoT devices and smartphones. These devices can often be less secure than servers or desktop computers that might be used in wired FL~\cite{li2022employing}. If deployed remotely, they might not have the latest security patches, leaving them vulnerable to backdoor attacks. Some devices can even be physically captured, compromised, and put back into the networks.

    \item {\em Data Transmission}: Data transmission in wireless networks can be inconsistent due to interference or signal strength variations. This inconsistency can lead to data corruption or loss, and an attacker might exploit the inconsistency to inject malicious backdoor attacks. On the other hand, machine learning has demonstrated its effectiveness in transmission resource allocation under both centralized settings~\cite{9810299}, semi-distributed settings~\cite{10002890}, and distributed settings~\cite{9727086,raza2023statistical,9540680}.

    \item {\em Scalability and Management}: WFL is designed to be highly scalable and to deal with a large number of nodes (devices). This makes managing and securing each device more difficult and increases the overall attack surface.

    \item {\em Distribution}: FL is inherently a distributed learning approach. The decentralization is even more pronounced in a wireless environment where devices can join and leave the network more freely. This can make it more challenging to maintain constant security measures, making the system more prone to backdoor attacks.
\end{itemize}
Hence, this paper focuses on WFL and its security concerns related to backdoor attacks. In the next section, an overview of WFL is first presented.

\subsection{Overview of WFL}

WFL is a distributed machine learning scheme that connects $n$ clients to train a global model $G$ in the central server by iterative aggregation of the local model $L_i$, $i=1,\ldots, N$ over a wireless network~\cite{yang2019federated,tran2019federated,niknam2020federated}. WFL has been widely deployed owing to its several features of distributed architecture, a large number of participants, and privacy concerns~\cite{li2020federated,zheng2022exploring,liu2022distributed}. The distributed characteristic of WFL allows the broad participation of clients regardless of geographical restrictions. The scale of clients participating in WFL can be immense. During the aggregation process, the local model updates are uploaded to the WFL server (e.g., a BS), while the training data remains securely stored on clients' devices locally. Thus, WFL can be adopted to train tasks that highly rely on sensitive private data such as clients' personal information, location, etc.~\cite{li2021survey,zhang2022federated,chen2022decentralized,lee2023privacy,liu2020privacy,li2023dataagnostic}.

\begin{figure}[H]
\centering
\includegraphics[scale=0.4]{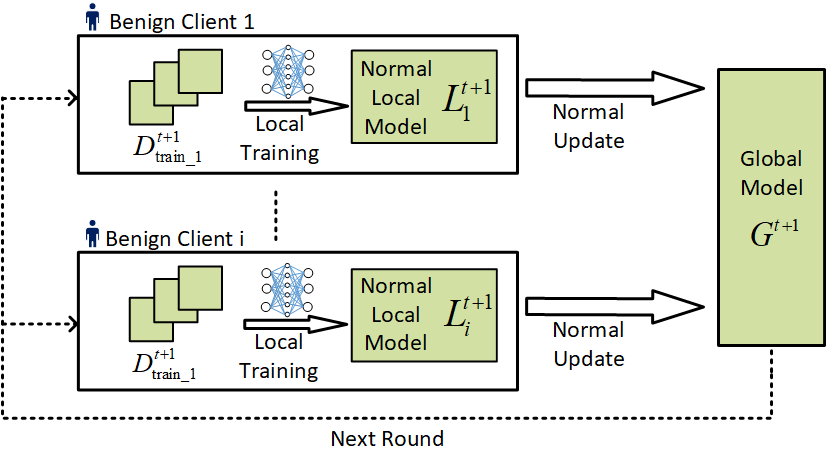}
\centering 
\caption{The overview of the WFL training strategy for producing aggregated global ML model using devices' private data.}
 \label{fig:WFL}
\end{figure}

A typical training process of WFL is shown in Fig.~\ref{fig:WFL}. {\color{blue}In each training iteration, the central server broadcasts the current global model ${G^t}$ to all the participants within the network, and a subset of $m$ clients are selected to join the training process.} The selection of $m$ is subject to the trade-off between the training speed and efficiency. Each training round can be divided into three phases:
\begin{itemize}
    \item \textbf{Local training phase}
    
    In the $t$-th training iteration, the $i$-th client re-trains the local model with the local training data $D_i^t$. $t\in \left\{ {1, \ldots ,T} \right\}$. $i \in \left\{ {1, \ldots, m} \right\}$. The local training objective is defined as:
    \begin{equation}
      L_i^t = \arg \mathop {\min }\limits_{L_i^t} {F_i}(L_i^t,D_i^t),
    \end{equation}
    \noindent where ${F_i}({L_i^t},{D_i^t})$ denotes the local loss function. To minimize the loss function, the local model update is calculated, denoted as $p_i^{t+1}$. 
    
    It is worth mentioning that the model update can be in a different form depending on different training algorithms. For example, if the WFL is trained with the Stochastic Gradient Descent (SGD) method, $p_i^{t+1}$ can be formulated as the gradient of the loss function:
    \begin{equation}
        p_i^{t + 1} = \frac{{\partial F(L_i^t,D_i^{t + 1})}}{{\partial L_i^t}}.
    \end{equation}
    
    Other local training methods also include Batch Gradient Descent (BGD)~\cite{haji2021comparison}, Mini-Batch Gradient Descent (MBGD)~\cite{messaoud2019online}, Root Mean Squared Propagation (RMSprop)~\cite{rakshitha2022op}, etc. After local training, the obtained local model update $p_i^{t+1}$ is sent to the server. 
    
    \hspace{4 mm} 
    Considering the limited and often unbalanced learning (or computing) capabilities of wireless devices, the client may train different numbers of local training in a communication round. Moreover, some devices may experience deep fades and require excessive communication time to upload their local models for aggregation. As a consequence, their local training time has to be sacrificed, penalizing the accuracy of their local models.
    
    \item \textbf{Aggregation phase}
    
    The server gathers the local model updates to obtain the average global updates ${p_g}^{t+1}$:
    \begin{equation}
        p_g^{t + 1} = \frac{1}{m}\sum\limits_{i = 1}^m ({\beta_i^{t+1} \times p_i^{t + 1}}),
        \label{local agg}
    \end{equation}
     where $\beta_i$ is the weight of the $i$-th client. The aggregated model update is then used for the global model update. 
     
     \hspace{4 mm} 
     A typical form of global model aggregation is in the digital domain, where the selected and admitted local models are digitized and aggregated, as described in~\eqref{local agg}.      
     Another increasingly considered method for global aggregation in WFL is Over-The-Air (OTA), where all selected clients synchronize the transmissions of their local models and align their signal strengths received at the server~\cite{XichenYuTCOM2023,xiao2023overtheair}. By exploiting the additivity of the electromagnetic field, the local model can be naturally aggregated at the reception of the server.
    
    \item \textbf{ Global model update phase}
    
    The global model is updated after receiving the aggregated model update:
    \begin{equation}
        G^{t+1} = G^{t} + {\eta^t} p_g^{t + 1},
    \end{equation}
   where $\eta$ is the global learning rate, controlling the training step size of the $t$-th iteration. The updated global model can be distributed to the same or different clients chosen in the next round.
\end{itemize}

Decentralized WFL has been recently developed, e.g.,  in~\cite{tedeschini2022decentralized}. To solve the network asynchrony typically undergone in an FL process, a fully decentralized FL framework named IronForge is proposed in~\cite{yu2023ironforge}.
Decentralized WFL is an advanced architecture that offers a flexible and scalable approach to perform model aggregations in large-scale or unstable networks. The decentralized WFL architecture enables these local nodes to train their models independently using their respective local data. This approach is particularly useful in situations where the network may experience intermittent connectivity, limited bandwidth, or strict privacy regulations that restrict data sharing.

Different from centralized WFL where all participants aim to train a unique global model at a central node, there is no central node in decentralized WFL. Each client of Decentralized WFL can choose to train a new local model and upload the model update to other participants within its communication range or aggregate the model updates received from others to train a new model and broadcast it into the network again. The models distributed to different participants can be considerably different, even upon the convergence of decentralized WFL.

Both centralized and decentralized WFL allow participants to join and leave the network freely and do not require private data sharing among them. Recent works have shown that WFL cannot always guarantee the security of the training process~\cite{li2021survey,wang2019beyond,song2020analyzing,lyu2022privacy,li2022employing}. Existing WFL frameworks are vulnerable to backdoor attacks. Malicious users can control one or multiple clients, namely, compromised clients, to inject a backdoor into the global model~\cite{bagdasaryan2020backdoor}. The primary objective of the backdoor attack is to generate a converged global model with high accuracy of the main task and the backdoor-injected sub-task~\cite{fan2021text}.

\subsection{Features of WFL}

{\color{blue}
WFL offers distinct features compared to traditional FL, especially in scenarios where mobility and real-time data processing are crucial~\cite{hu2023ofdmaf2l}. The essence of WFL lies in its ability to facilitate learning processes for mobile devices, such as smartphones, tablets, and IoT devices~\cite{10268067}. These devices often operate on the move and rely heavily on wireless connectivity, making WFL an ideal solution for ensuring continuous and accessible learning. For instance, in applications such as autonomous vehicles~\cite{10.1145/3538969.3544422,FLIDS_NSE23,10195758} or mobile health monitoring systems~\cite{RANI2023110658}, the capability of WFL to enable on-the-go learning and instantaneous model updating is indispensable. This real-time processing is critical for applications that demand immediate responses based on current data.

\subsubsection{Exacerbated Vulnerabilities of WFL}
Implementing WFL in a wireless system such as a cell-free massive MIMO system introduces specific privacy challenges due to the dispersed distribution of access points and users~\cite{zhang2022federated}. The wireless communication among these points adds complexity, increasing the risk of sensitive data exposure. Cell-free massive MIMO systems involve a large number of distributed access points serving multiple users. In such a setting, the wireless transmission of data over numerous channels between these access points and users can create multiple opportunities for data interception or leakage. The complexity and scale of such systems make it challenging to uniformly secure all communication links, thus heightening privacy risks.

In~\cite{chen2022decentralized}, decentralized WFL (DWFL) is studied, where there is an increased risk of private information leakage due to the transmission of model parameters over a wireless network. This risk is exacerbated by the decentralized nature of DWFL, where data is distributed across multiple wireless nodes. In a DWFL setting, the transmission of parameters between multiple nodes over wireless channels introduces potential points of vulnerability. Each node in the network can be seen as a potential weak point where information can be intercepted or corrupted. DP is adopted to preserve the data privacy of the nodes, and its impact on the convergence of DWFL is analyzed.

The authors of~\cite{lee2023privacy} considered decentralized inference with graph neural networks (GNNs) in wireless networks, highlighting the vulnerability in handling graph-structured data over wireless channels. The distributed nature of GNNs in a wireless context adds complexity to maintaining data privacy. In decentralized GNNs, neighboring nodes exchange information through wireless channels. This exchange is crucial for inference but introduces vulnerabilities due to the inherent characteristics of wireless media, such as fading and noise. These factors can deteriorate the quality of information exchange and impact the performance of the inference process. The adverse impacts of imperfect wireless transmission on the inference robustness of GNNs highlight the complexity of maintaining privacy in this setting.

There exists a unique aspect of OTA-WFL, where gradient transmissions are uncoded~\cite{liu2020privacy}. The wireless aspect introduces a vulnerability where the privacy of local datasets can be compromised through the disclosed aggregation statistics over the air. In OTA-WFL with uncoded transmission, gradients are transmitted directly over wireless channels without encoding or encryption. 
The lack of coding or encryption in transmission means that any vulnerability in the wireless channel, such as weak signal encryption or interception by unauthorized receivers, can lead to significant privacy breaches. Moreover, since gradients carry information about the local datasets, their interception can reveal sensitive data.

Moreover, with the broadcast nature of wireless interfaces, adversaries can potentially overhear local model updates from multiple or even all clients and accordingly create malicious local models to compromise WFL, e.g., using generative neural networks and machine learning technologies.
For example, in~\cite{li2023dataagnostic,10183217}, a new adversarial attack on WFL is studied by exploiting the broadcast nature of wireless channels, where a rogue client collects the local model updates that benign clients upload. The rogue client can take these overheard local model updates as input to a graph adversarial encoder (GAE) to produce a malicious local model update. The malicious local model update can poison WFL, and stop or slow down the convergence of WFL~\cite{10183217}.

\subsubsection{Inherent Privacy-Preserving Capability of WFL}
Some inherent features of wireless channels and transceivers can be potentially leveraged to preserve the privacy of WFL. While lossy and erroneous wireless channels may hinder the convergence of WFL, they may weaken the attack strength from backdoor, data and model poisoning attacks. This is due to the fact that the random errors resulting from wireless fading and receiver noises can potentially serve to perturb the model updates, hence achieving the effect of DP and alleviating the impact of the attacks. Under the setting of OTA-WFL, the authors of~\cite{liu2020privacy} control the relative intensity of the receiver noise power at the central server by controlling the transmit powers of the clients. The resulting relative intensity of the receiver noises can offer privacy-preserving capability, like DP noises. This promising research direction can potentially simplify the protection design for WFL. The research is still at a very early stage. More breakthroughs are underway.}

\subsection{Overview of Backdoor Attacks on WFL}

\begin{figure}[htb]
\centering
\includegraphics[scale=0.4]{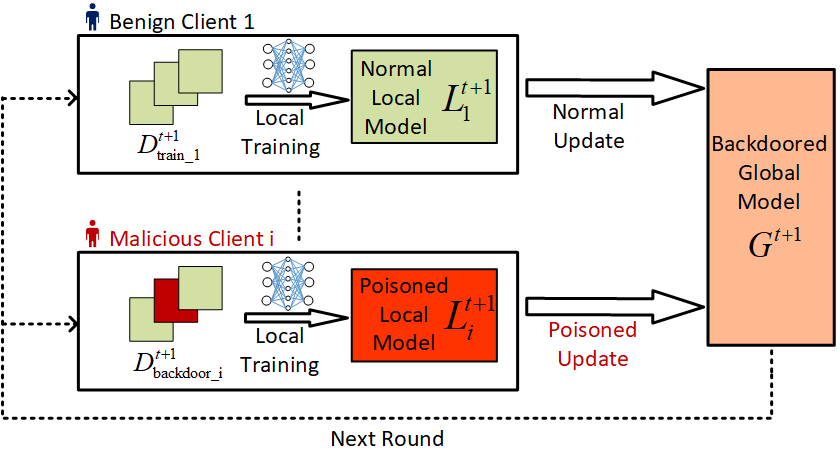}
\centering 
\caption{Backdoor attack process in WFL.}
 \label{fig:backdoor attack}
\end{figure}

In this section, an overview of backdoor attacks on WFL is presented. The process of backdoor attacks is shown in Fig.~\ref{fig:backdoor attack}. The existing backdoor strategies are categorized into three major types in terms of attack phase: (1) local data collection phase backdoor attack, (2) local model training phase attack, and (3) server aggregation phase backdoor attack. Depending on different types of attacks, the information required by adversaries may differ.

\begin{itemize}
    \item \textbf{Backdoor attack at local data collection phase}
    
The local data collection phase backdoor attack, also known as the data poisoning backdoor attack~\cite{tolpegin2020data}. When conducting backdoor attacks during the local data collection phase, the attacker only needs to control a fraction of the training data set~\cite{gong2022coordinated}.

\hspace{4 mm}In a local data collection phase backdoor attack, an individual feature or a small region of the local training data is modified by malicious users deliberately~\cite{schwarzschild2021just}.  The fraction of training data controlled by the attackers is denoted as Poisoned Data Rate (PDR) ~\cite{yerlikaya2022data}. Specifically, in the $i$-th compromised client at the $t$-th training iteration, PDR can be mathematically formulated as:
\begin{equation}
    PDR = \frac{{\left| {{\rm{D}}_{Ai}^t} \right|}}{{\left| {D_i^t} \right|}},
\end{equation}
where $D_i^t$ stands for the benign dataset of this client, $D_{Ai}^t$ refers to the malicious dataset, and $\left|  \cdot  \right|$ stands for cardinality. 

\hspace{4 mm}The training objective of the $i$-th compromised client can then be formulated as:
\begin{equation}
   L_i^t = \arg \mathop {\min }\limits_{L_i^t} \{ \alpha {F_i}(L_i^t,D_i^t) + (1 - \alpha ){F_{Ai}}(L_i^t,D_{Ai}^t)\} , 
\end{equation}
where ${F_i}({L_i^t},{D_i^t})$ is the WFL learning objective and ${F_{Ai}}({L_i^t},{D_{Ai}^t})$ is the malicious objective shared by all attackers (in the case of a coordinated attack discussed in Section \ref{trigger-3.1.2}). The local model is induced to misclassify the testing samples with the target label specified by adversaries. Considering the stealthiness of the backdoor attacks, a compromised client also needs to optimize the WFL objective. The trade-off between training the benign and malicious objectives is defined as $\alpha$, and $\alpha\in[0,1]$. 

\hspace{4 mm}In practice, one way to launch the local data collection phase backdoor attack is to directly replace the training data of the target label with arbitrary incorrect samples. A successful attack will lead the model to misclassify the testing samples of the target label according to the attacker's wish, e.g., for a room with facial recognition lock, a man without access will be permitted to enter the room by mistake~\cite{jagielski2021subpopulation,yang2022not}. 

\hspace{4 mm}The other type of widely applied backdoor attack strategy during the local data collection phase is the trigger-based backdoor attack. The trigger is a pattern parasitized in clean data~\cite{chen2017targeted}. The model will behave normally and predict a target class when the trigger appears, e.g., in the same room with a facial recognition lock, all men with glasses will be authorized to enter. In this case, the glass is the trigger injected. Depending on the number of compromised clients the attacker controls, the trigger can be injected into one client or distributed separately to multiple clients.

\item \textbf{Backdoor attack at local model training phase} 

In local model training phase backdoor attacks, an attacker is required to fully control the compromised clients, including the training data set and training algorithm~\cite{xie2019dba,zhang2020poisongan,rosenfeld2020certified,gong2022coordinated,sun2019can,bagdasaryan2020backdoor}. the malicious goal is to tamper with the uploaded updates and further influence the global model. Take the $i$-th compromised client, for example. The attack process can be formulated as follows:
\begin{equation}
   p_{Ai}^{t + 1} \leftarrow L_{Ai}^{t + 1} \leftarrow \mathop {\min }\limits_{L_i^t} {F_{Ai}}(L_i^t,D_{Ai}^t),
\end{equation}
where $p_{Ai}^{t + 1}$ is the poisoned local model update sent to the server. As a consequence, the clean local model will be replaced by a backdoor-injected model~\cite{wenger2021backdoor}. 

\hspace{4 mm}Backdoor attacks conducted during the local model training process only aim to enhance the local model. The attacker is capable of conducting attacks to compromise both the convergence of the training data set and the calculation of the uploaded update of the individual clients. The impact of the training phase attack is limited to the compromised clients controlled by the attacker~\cite{sun2019can,zhang2022neurotoxin,bagdasaryan2020backdoor}. 

\item \textbf{Backdoor attack at server aggregation phase} 

Backdoor attacks conducted during the aggregation process generally do not alter the targeted model. Instead, the attacker is keen on tricking it into producing wrong predictions and causing security problems~\cite{bagdasaryan2020backdoor}. The effectiveness of an aggregation phase attack is mainly determined by the knowledge of the model controlled by attackers. 

\hspace{4 mm}In the aggregation phase backdoor attack, the attacker controls the aggregation algorithm used to combine clients' updates into the global model:
\begin{equation}
    p_{Ag}^{t + 1} = \frac{1}{m}\left( {\sum\limits_{i = 1}^{{m_d}} {\beta _i^{t + 1} \times p_i^{t + 1}}  + \sum\limits_{j = 1}^{{m_b}} {\beta _{Ai}^{t + 1} \times p_{Ai}^{t + 1}} } \right),
\end{equation}
 where $m$ is the number of clients participating in the $t$-th training iteration as considered in \eqref{local agg}; $M_d$ is the number of benign clients; and $M_b$ is that of compromised clients. Therefore, $m = m_d+m_b$. The weight of the poisoning model $\beta_{Ai}^{t+1}, i \in [1,m_b]$ is manipulated to ensure the successful implementation of the backdoor into the global model of WFL. 
 
 \hspace{4 mm}Compared with training process attacks, the aggregation phase attacks can have stronger attack effects while there is a trade-off between the stealthiness and the attack success rate (ASR).
\end{itemize}
 
\subsection{Evaluation Metrics for Backdoor Attacks} 
The objective of a backdoor attack on WFL is to mislead the global model to misclassify malicious inputs with injected backdoor patterns into the target output, with application to image classification~\cite{feng2022fiba,saha2020hidden,nguyen2020input,zhao2020clean,yoshida2020disabling,doan2021backdoor,sun2022backdoor,mayerhofer2022poisoning,jia2022badencoder,xu2022more}, word prediction~\cite{chen2021badnl,lu2022attack,kwon2021textual,pan2022hidden,shao2022triggers,liu2022piccolo}, etc. For instance, regarding the image classification task discussed in~\cite{wang2020attack}, an attacker aims to mislead the global model to classify images of ``truck'' as ``airplane''. Based on the attack objectives, the following metrics are commonly adopted to evaluate the effectiveness of backdoor attacks.   

\begin{itemize}
	\item \textbf{Backdoor Accuracy Rate (BAR):} 	
	BAR indicates the strength of a backdoor pattern by measuring the ratio of successfully misclassified global outputs using backdoored inputs out of all the outputs obtained from malicious inputs. In other studies, similar terms, such as ASR~\cite{doan2021lira,xu2021explainability} and backdoor task accuracy (BTA)~\cite{bagdasaryan2020backdoor}, are used. Generally, a high BAR is equivalent to a high ASR and BTA. 

	\item \textbf{Main Task Accuracy Rate (MAR):} 	
	In order to perform a successful backdoor attack, the stealthiness of the backdoor-injected input is also essential. MAR measures the rate of global model outputs that have been denoted as correct classification, which consists of normal output produced by the benign inputs and malicious outputs misled by the backdoored inputs. Different from Byzantine attacks, which aim to jeopardize the convergence of the global model~\cite{varma2021legato,gouissem2022federated,ma2022disbezant}, the backdoor attacks focus on inserting backdoor patterns into the global model while maintaining a reasonable MAR.

	\item \textbf{Lifespan of backdoor:} 	
	Due to the nature of FL, the backdoor effect in the global model tends to be diluted with the increment of the training round. The duration of the backdoor effect remaining in the global model is defined as the lifespan of the backdoor. Attackers are keen to keep the backdoor effects in the global model as long as possible, such that the global model could consistently produce the wrong outputs as they want.

\end{itemize}

Notably, the above-mentioned metrics are also widely used to evaluate the effectiveness of defense strategies. Different from backdoor attack designs, which pursue high BAR, MAR, and long lifespan, the defenders aim to minimize the BAR and lifespan of the backdoor effect while preventing the MAR from degradation. 

\subsection{Performance Analysis of Backdoor Attacks}

To evaluate the effectiveness, backdoor attack methodologies are tested in multiple WFL applications. Consider a WFL network with a total of $m$ clients. The malicious attackers control one or more compromised clients to perform the attack. The attack targets can vary depending on the knowledge level obtained by the attacker. Furthermore, the specific design of the attack method is likely to be limited to specific kinds of WFL applications. Regarding the attack target, complexity, and applicability, existing backdoor attack methodologies in WFL are summarized below in Table~\ref{existing_bd}. 
\textcolor{blue}{As shown in Table~\ref{existing_bd}, label flipping attacks, coordinated trigger attacks, local model replacement attacks, and aggregation process attacks are likely to be exacerbated in WFL. This is due to the broadcast nature of wireless communications, which facilitates various forms of cyber threats, such as spoofing attacks, Sybil attacks, and eavesdropping attacks. }


\begin{table*}[h]
\centering
\normalsize
\caption{Summary of existing backdoor attack methodologies in WFL}
\label{existing_bd}
\renewcommand{\arraystretch}{1.3}
\scalebox{0.9}{
\begin{tabular}{cl|cc|cc|cc|c|c}
\hline
\hline
\multicolumn{2}{c|}{\multirow{2}{*}{\textbf{Attack Type}}} &
  \multicolumn{2}{c|}{\textbf{Attack Target}} &
  \multicolumn{2}{c|}{\begin{tabular}[c]{@{}c@{}}\textbf{Compromised} \\ \textbf{Client Number}\end{tabular}} &
  \multicolumn{2}{c|}{\textbf{Attack Iteration}} &
  \multirow{2}{*}{\textbf{Main Task}} & 
  \multirow{2}{*}{\textbf{\color{blue}\begin{tabular}[c]{@{}c@{}}
  Exacerbation\\ by WCN\end{tabular}}} \\ \cline{3-8}
\multicolumn{2}{c|}{} &
  \multicolumn{1}{c|}{Data} &
  Model &
  \multicolumn{1}{c|}{Single} &
  Multiple &
  \multicolumn{1}{c|}{Single} &
  Multiple &
   \\ \hline

\multicolumn{2}{c|}{\begin{tabular}[c]{@{}c@{}}Label  Flipping \\ ~\cite{zhang2020poisongan,rosenfeld2020certified,xiao2012adversarial,aryal2022analysis}\end{tabular}} &
  \multicolumn{1}{c|} {$\surd$} &
  $\times$ &
  \multicolumn{1}{c|}{$\surd$} &
  $\surd$ &
  \multicolumn{1}{c|}{$\surd$} &
  $\surd$ &
  Image Classification & 
  {$\surd$} \\ \hline

\multicolumn{2}{c|}{\begin{tabular}[c]{@{}c@{}}Centralized Trigger \\ ~\cite{xu2022more,liu2023facilitating,zhao2022defeat,zhao2021federatedreverse}\end{tabular}} &
  \multicolumn{1}{c|}{$\surd$} &
  $\times$ &
  \multicolumn{1}{c|}{$\surd$} &
  $\times$ &
  \multicolumn{1}{c|}{$\surd$} &
  $\surd$ &
  Image Classification & 
  {$\times$}\\ \hline

\multicolumn{2}{c|}{\begin{tabular}[c]{@{}c@{}}Coordinated Trigger \\~\cite{gong2022coordinated,li2021pointba,wei2021backdoor,xie2019dba}\end{tabular}} &
  \multicolumn{1}{c|}{$\surd$} &
 $\times$ &
  \multicolumn{1}{c|}{$\times$} &
  $\surd$ &
  \multicolumn{1}{c|}{$\times$} &
  $\surd$ &
  Image Classification & 
  {$\surd$} \\ \hline
\multicolumn{2}{c|}{\begin{tabular}[c]{@{}c@{}}Local Model Replacement \\ ~\cite{bagdasaryan2020backdoor,sun2019can,li2021deeppayload,cao2022highly}\end{tabular}} &
  \multicolumn{1}{c|}{$\times$} &
  $\surd$ &
  \multicolumn{1}{c|}{$\surd$} &
  $\times$ &
  \multicolumn{1}{c|}{$\surd$} &
  $\surd$ &
  \begin{tabular}[c]{@{}c@{}}Image Classification; \\ Sentiment Analysis\end{tabular} &
  {$\surd$} \\ \hline
\multicolumn{2}{c|}{\begin{tabular}[c]{@{}c@{}}Aggregation process Attacks \\ ~\cite{rodriguez2022backdoor,chen2023backdoor,hong2022handcrafted,garg2020can}\end{tabular}} &
  \multicolumn{1}{c|}{$\times$} &
  $\surd$ &
  \multicolumn{1}{c|}{$\surd$} &
  $\times$ &
  \multicolumn{1}{c|}{$\surd$} &
  $\surd$ &
  \begin{tabular}[c]{@{}c@{}}Image Classification; \\ Sentiment Analysis;\\ Word Prediction\end{tabular} &
  {$\surd$} \\ \hline
  \hline
\end{tabular}}
\label{Table:4}
\end{table*}

The performance of the existing backdoor attack methods can then be evaluated. The comparison is summarized in Table~\ref{omparison2_bd}. It can be concluded that the data poisoning backdoor attacks, including label flipping-based attacks and centralized trigger-based attacks, have the lowest attack cost. In return, the attack strength is weak, and the backdoor effect will be quickly diluted. The trigger-based backdoor attack can be improved using a coordinated trigger pattern, achieving a higher attack success rate. The model poisoning backdoor attacks implemented at the local training and server aggregation phases take advantage of the white box attack model. The backdoor effect can last longer if the attack success rate is higher. Consequently,  the attackers need to know more about the entire network, and the attack cost is greatly increased.   

\begin{table*}[b]
\centering
\normalsize
\caption{Comparison among existing backdoor attacks on WFL: Attack cost, ASR, duration, prior knowledge, and potential exacerbation caused by WCN}
\label{omparison2_bd}
\renewcommand{\arraystretch}{1.2}
\scalebox{0.9}{
\begin{tabular}{cl|c|c|c|l|c}
\hline
\hline
\multicolumn{2}{c|}{\textbf{Attack Type}} & 
  \begin{tabular}[c]{@{}c@{}}\textbf{Attack Cost}\end{tabular} &
  \begin{tabular}[c]{@{}c@{}}\textbf{Attack} \\ \textbf{Success Rate}\end{tabular} &
  \textbf{Duration} &
  \multicolumn{1}{c|}{\textbf{Prior Knowledge}} &
  \textbf{\color{blue}\begin{tabular}[c]{@{}c@{}}Exacerbation\\ by WCN\end{tabular}}\\ \hline
  
\multicolumn{2}{c|}{\begin{tabular}[c]{@{}c@{}}Label Flipping \\ ~\cite{zhang2020poisongan,rosenfeld2020certified,xiao2012adversarial,aryal2022analysis}\end{tabular}}    & Low    & Low  & Short  & Label information &   {$\surd$}   \\ \hline

\multicolumn{2}{c|}{\begin{tabular}[c]{@{}c@{}}Centralized Trigger \\ ~\cite{xu2022more,liu2023facilitating,zhao2022defeat,zhao2021federatedreverse} \end{tabular}} & Low    & Low  & Short  & Part of the local training data & {$\times$} \\ \hline

\multicolumn{2}{c|}{\begin{tabular}[c]{@{}c@{}}Coordinated Trigger \\~\cite{gong2022coordinated,li2021pointba,wei2021backdoor,xie2019dba}\end{tabular}} & Medium & High & Medium & Part of the local training data & {$\surd$} \\ \hline

\multicolumn{2}{c|}{\begin{tabular}[c]{@{}c@{}}Local Model Replacement \\ ~\cite{bagdasaryan2020backdoor,sun2019can,li2021deeppayload,cao2022highly}\end{tabular}} &
  High &
  High &
  Long &
  \begin{tabular}[c]{@{}l@{}}Local training data;\\ Local training process, \\ Possible defense mechanisms\end{tabular} & {$\surd$} \\ \hline
  
\multicolumn{2}{c|}{\begin{tabular}[c]{@{}c@{}}Aggregation Process Attacks\\ ~\cite{rodriguez2022backdoor,chen2023backdoor,hong2022handcrafted,garg2020can}\end{tabular}} &
  Highest &
  Highest &
  Long &
  \begin{tabular}[c]{@{}l@{}}The number of participants, \\ Training algorithm,   \\ Possible defense mechanism, \\ Aggregation protocol, \\ Global model convergence process.\end{tabular} & {$\surd$} \\ \hline
  \hline
\end{tabular}}
\end{table*}

\begin{figure*}[h]
\centering
\includegraphics[scale=0.58]{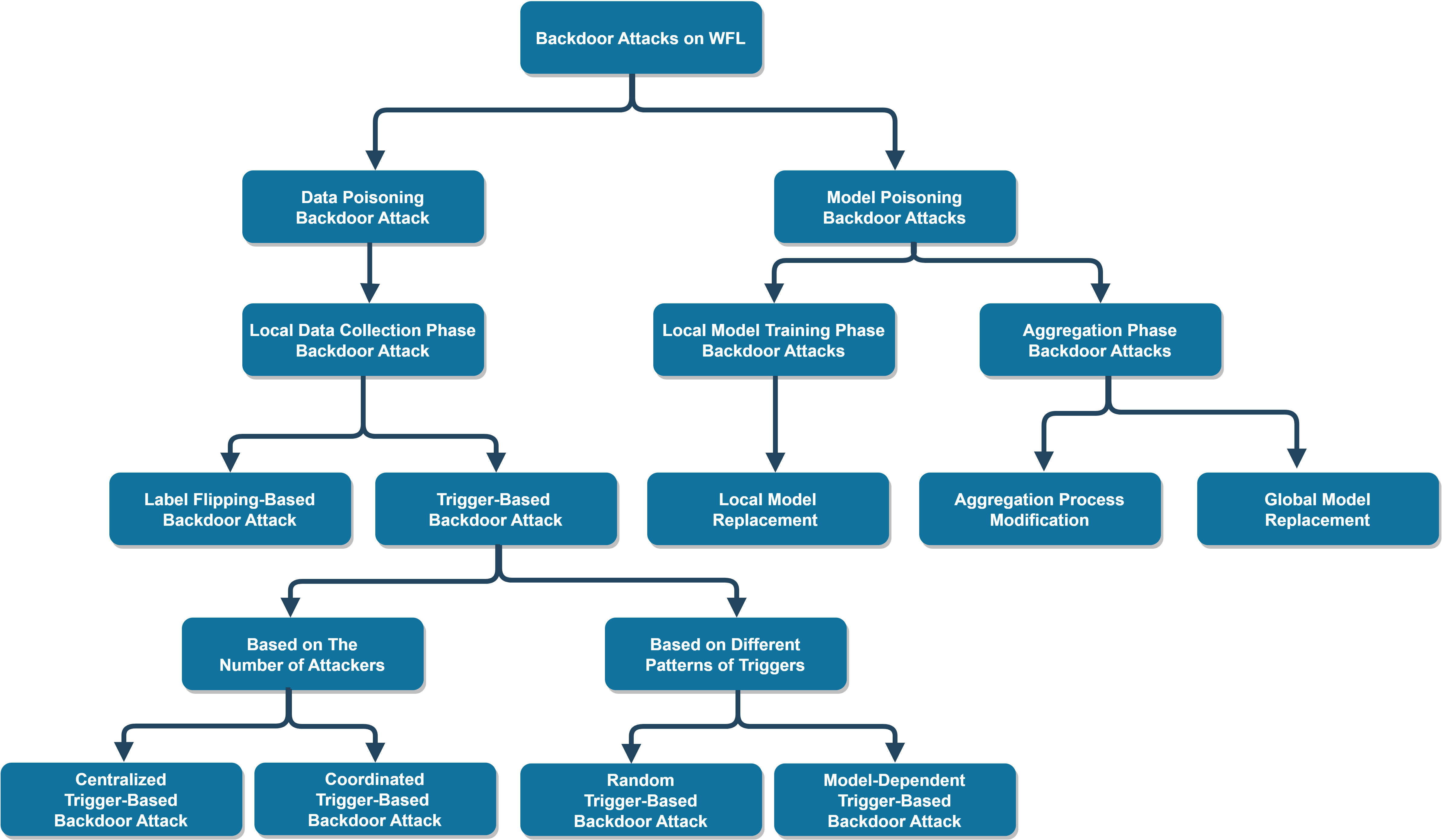}
\centering 
\caption{A taxonomy of backdoor attacks on WFL.}
 \label{fig:types}
\end{figure*}

\section{Status Quo of Backdoor Attacks on WFL}
\label{sec::Existing}

Wireless networks have geographically separated structure, allowing each client participating in the WFL network to train a unique local model based on its own datasets. Instead of exchanging raw data among participants, only local model updates are uploaded for aggregation. The non-IID training data set in WFL helps to protect the data privacy of the participants~\cite{zhu2021federated}. Each client's data distribution is not representative of the global data distribution, implying that the categories in each client are incomplete. As a result, even if one client's data is compromised, the overall model will be minimally affected. 

On the other hand, the features of WFL, including non-IID training datasets, distributed learning structure, and information interchange via WCN, allow adversarial clients to conduct backdoor attacks to stealthily compromise the global model robustness. {\em The primary objective of backdoor attacks is to manipulate the converged global model to misclassify specifically crafted inputs with hidden backdoors into a desired target label while ensuring that normal inputs are classified accurately}. Thus, the global model exhibits convergence on the main task and leaves a backdoor on the target label.

Moreover, the communication efficiency of a client largely depends on the local devices' processing abilities and their geographical distributions in mobile edge computing systems~\cite{8449070}. Clients with faster computing speed and better communication link qualities are more likely to be selected in a communication or aggregation round. 
Consequently, the WFL could be particularly vulnerable to backdoor attacks launched by attackers with better link qualities, e.g., closer to the server or possessing a strong line-of-sight (LoS) to the server.

In this section, the existing techniques to perform the above motioned backdoor attacks in three different phases are comprehensively reviewed, as presented in Fig. \ref{fig:types}.

\subsection{Backdoor Attacks at Local Data Collection Phase}
\begin{figure}
\centering
\includegraphics[scale=0.35]{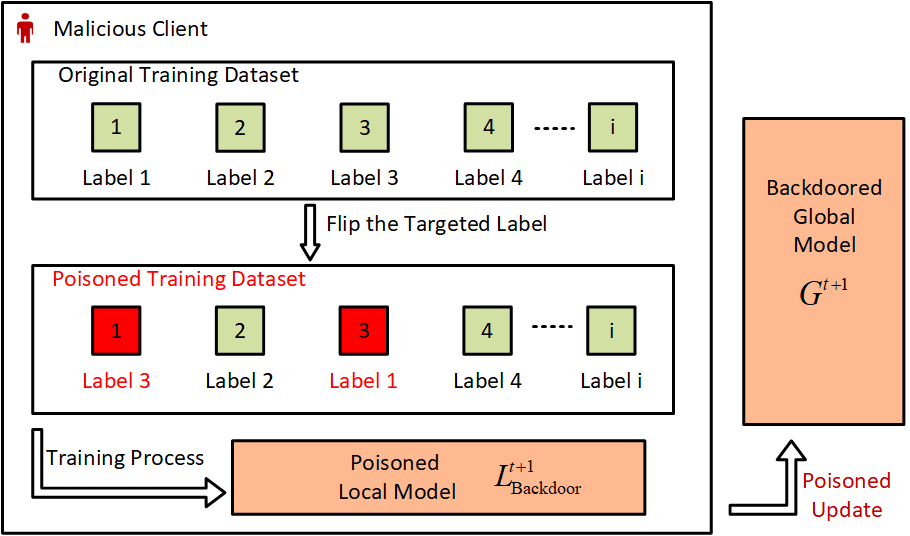}
\centering 
\caption{Label-flipping-based backdoor attacks (local data collection phase): The attacker only needs to manipulate the training data under the target label.}
 \label{fig:label_flipping}
\end{figure}

In backdoor attacks during the local data collection phase, the backdoor is injected via data poisoning backdoor attacks. Data poisoning backdoor attacks can be further classified into two major methodologies: label-flipping-based attacks or triggered-based attacks. The label-flipping-based backdoor attack is proposed in~\cite{zhang2020poisongan}. Fig.~\ref{fig:label_flipping} shows the mechanism of label flipping based backdoor attack. In label-flipping-based backdoor attacks, the training data set under the targeted label is replaced by the backdoored data set~\cite{zhang2020adversarial}. The poisoned data set is then used to train the local model for further aggregation.

\subsubsection{Label-flipping based backdoor attack}
The label flipping-based backdoor attack is relatively easier to conduct since it does not require any prior knowledge of the local data distribution, local model training process, and aggregation protocol~\cite{rosenfeld2020certified,shen2021backdoor}. The attacker only needs to manipulate the training data under the target label. By strategically flipping labels in the poisoned samples, attackers can bias the model towards a specific target behavior. For example, in a face recognition model, the attacker may flip labels to make the model misidentify a particular person or a group of people.

However, since the number of participants in WFL is typically large, the effect of an injected backdoor from a single attacker in a single round can often be diluted. In order to obtain better attack performance, multiple malicious clients can collude to conduct an attack cohesively. They need to be selected in multiple rounds. The inevitable problem of the simple label-flipping method still exists in that the flipped data set under the targeted label results in a remarkable outlier in the model training. Thus, it can be detected by norm clipping~\cite{xie2021crfl} or differential privacy~\cite{nguyen2022flame,xie2021crfl,nguyen2022flame,zhang2023bounded}. 

To improve the stealthiness of the label flipping-based backdoor attack, the target label is selected on the tail in~\cite{wang2020attack}. In this case, the data distribution is required, and the edge data set is selected as the target. Replacing the edge target data set with a backdoored data set, the local model behaves correctly with normal inputs while misclassifying the adversarial inputs on the edge label. The process of label-flipping backdoor attacks is shown in Fig. \ref{fig:label_flipping}.

In general, label-flipping depends on the data sets and is agnostic to the models to be trained and the communication interfaces. Nevertheless, label-flipping attacks can hinder the convergence of a WFL process. The injected poisoned samples can introduce noise and confusion, making it harder for the model to converge. 
Given the limited communication resources, prolonged training processes can congest wireless interfaces and block other services and applications.

\noindent
\textbf{Limitations:} 
The effect of the label-flipping backdoor attack on the edge label can last longer compared with the simple label-flipping method. Furthermore, the selection of the edge data set minimizes the outlier, allowing it to trick the defender. On the other hand, the effect of the edge label flipping-based backdoor attack is also limited at the edge label.

\subsubsection{Trigger-based data poisoning backdoor attack}
\label{trigger-3.1.2}
In trigger-based backdoor attacks, triggers are designed to be injected into the training data under the targeted label~\cite{saha2020hidden,wang2022dispersed}. The trigger or a specific pattern embedded into the training data can cause the model to exhibit malicious behavior during the inference phase. The primary objective of trigger-based backdoor attacks is to mislead the model to classify the trigger-mounted inputs into the target label while behaving normally with non-trigger inputs~\cite{zhang2022neurotoxin,gong2022coordinated,nguyen2022flame}. 
According to the number of adversarial participants involved in an attack, trigger-based backdoor attacks can be classified into centralized trigger-based backdoor attacks and coordinated trigger-based backdoor attacks. 

\begin{figure}[b]
\centering
\includegraphics[scale=0.35]{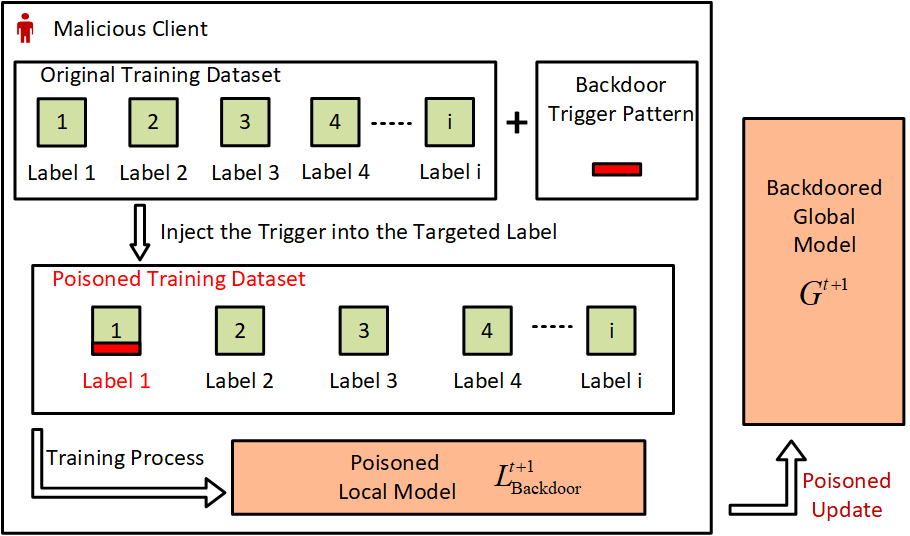}
\centering 
\caption{Centralized trigger-based backdoor attacks (local data collection phase): Centralized triggers are designed to be injected into the training data under the targeted label. }
 \label{fig:central}
\end{figure}

\begin{figure}[htb]
\centering
\includegraphics[scale=0.48]{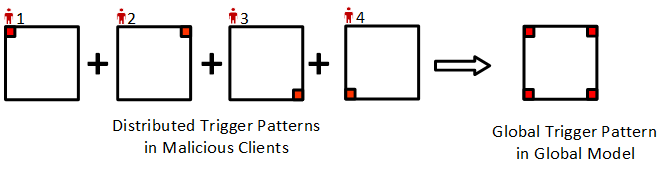}
\centering 
\caption{Distributed global trigger pattern: The trigger is injected in a coordinated pattern via multiple attackers.}
 \label{fig:distribute}
\end{figure}

\begin{figure}[htb]
\centering
\includegraphics[scale=0.35]{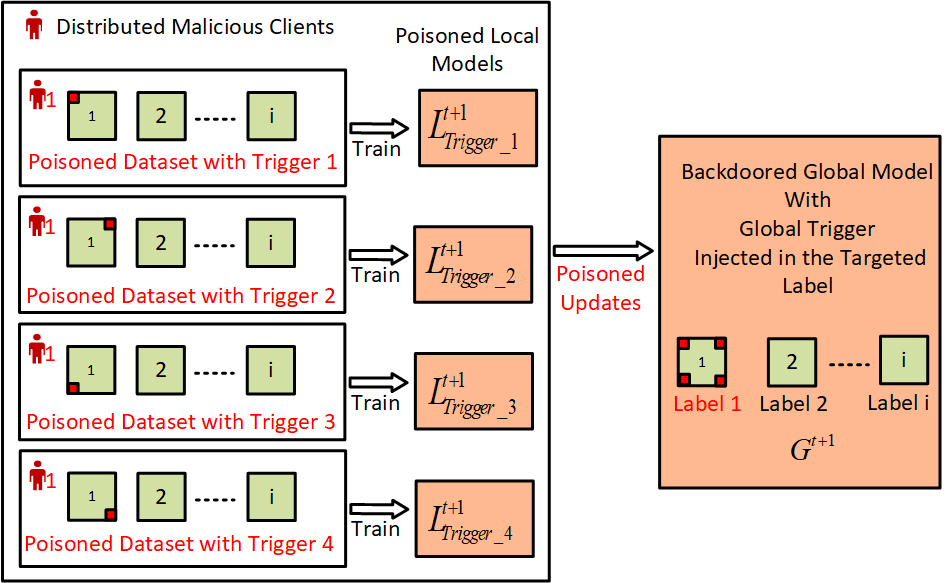}
\centering 
\caption{Coordinated trigger-based backdoor attack(local data collection phase): each attacker knows a part of the trigger, and the coordinated attack is conducted to inject the completed trigger into the final model.}
 \label{fig:coordinate}
\end{figure}

Fig.~\ref{fig:central} shows the attack mechanism of a centralized trigger-based backdoor attack.
In~\cite{gong2022coordinated}, the trigger is injected in a coordinated pattern via multiple attackers. As illustrated in Fig. \ref{fig:distribute}, the global trigger pattern is distributed in four parts. It can be seen from Fig.~\ref{fig:coordinate} that each attacker knows part of the trigger, and the coordinated attack is conducted to inject the completed trigger into the final model~\cite{ye2018distributed}.

Compared with the centralized trigger-based backdoor attacks, the coordinated trigger pattern shows better stealthiness, faster model convergence speed, and a higher attack success rate~\cite{chen2022marnet}. Coordinated attacks involve multiple devices participating in a WFL process. This increases the attack surface and potential vulnerabilities in the WFL. Attackers can exploit the decentralized nature of WFL to distribute triggers across a diverse set of devices, making it more challenging to identify and mitigate the attack.
The major problem of the existing trigger-based backdoor attacks, e.g.,~\cite{chen2022marnet}, is that the trigger is fixed at the beginning of the attack and cannot be adaptively adjusted during the attack process, which limits the attack effect.

Depending on different patterns of the trigger, the trigger-based backdoor attacks can be further divided into random trigger backdoor attacks and model-dependent trigger backdoor attacks, as follows. 
\begin{itemize}
    \item 
In a random trigger-based backdoor attack, the trigger is generated independently of the model~\cite{xie2019dba, zhang2021backdoor, salem2022dynamic, pan2022hidden}. The advantage is that the attack cost is greatly reduced. In return, the effect of the triggers is relatively weaker. In contrast to a random trigger, the other type of trigger is generated, according to the target label~\cite{gong2022coordinated,li2021invisible}.
    \item 
 The model-dependent trigger can help to excite the target label and has a higher probability of being successfully injected into the trained local model before aggregation~\cite{qin2023revisiting}.
\end{itemize}

Under a wireless configuration, an attacker is likely to eavesdrop on the local model updates of the other benign clients by exploiting the broadcast nature of radio~\cite{anajemba2022counter}. With the additional knowledge of the local models along with the global model, the model-dependent triggers can be potentially further enhanced, e.g., by employing adversarial generative networks and training~\cite{jia2022boosting}.

\noindent
\textbf{Limitations:} 
The common problem of centralized trigger-based backdoor attacks is that the single attacker's effect is limited, and it cannot be ensured that the attacker is selected in every round. Thus, the increment of the compromised clients controlled by the attacker can contribute to improving the attack performance~\cite{zhang2020poisongan}. 

\subsubsection{Comparison between label-flipping-based and trigger-based backdoor attacks}
Both label-flipping-based and trigger-based data poisoning backdoor attacks can be treated as black-box attacks~\cite{wang2020attack}. Attackers only need to know part of the local training dataset. By manipulating the data set under the targeted label, the backdoor is injected into the model at a relatively low cost. However, the attack effect can be either quickly diluted or eliminated under existing defense mechanisms. The attack strength is weaker than that of model poisoning backdoor attacks.

Moreover, the performance of backdoor attacks targeted at the local training datasets can be sensitive to the communication channel quality of the local devices or clients. For instance, the increasingly widely deployed wireless communication technologies, e.g., 5G, operate in high-frequency bands of up to 40 GHz, and the wavelength of the carrier frequency can be as short as 7.5 millimeters~\cite{8636206}.
The short wavelengths can lead to fast and drastically changing channel conditions of each client~\cite{8063331}. 
As a consequence, the transmission time required for a client to upload its local model in a communication round can change significantly between communication rounds. When the channel link quality is poor, i.e., a client is in a deep fade, the transmission time could be excessively long. The local training time that remains in a communication round would be short, leading to insufficiently trained local models and, hence, vulnerabilities to backdoor attacks.   
The ASR is higher when the attacks are launched at compromised clients under better communication conditions.
In this sense, fair client selection and scheduling of WFL can help prevent an individual client or a group of clients, including adversarial clients, from dictating a WFL process.

\subsection{Backdoor Attacks at Local Model Training Phase}
\begin{figure}[h]
	\centering
    \includegraphics[scale=0.35]{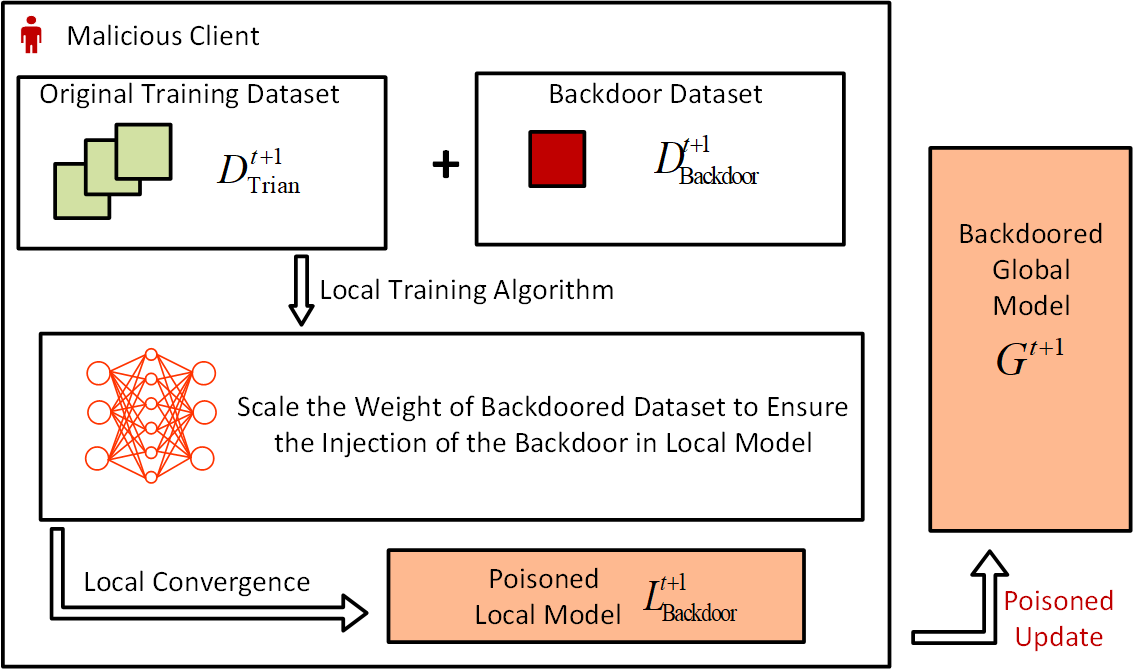}
	\caption{Local model replacement backdoor attacks (local model training phase): Manipulate the model parameters during the training process or modify the training algorithm.}
	\label{fig:train}
\end{figure}

In the local model training phase, the backdoor attacks can be conducted by manipulating the model parameters during the training process or modifying the training algorithm~\cite{zhang2022neurotoxin,shafieinejad2021robustness}. The original benign local model is replaced by the poisoned one, and the poisoned updates generated from the replaced local models are further uploaded for aggregation. The attacks during the training process can also be classified into one kind of model poisoning backdoor attack. The model poisoning attack structure is shown in Fig. \ref{fig:train}.

As a matter of fact, model-poisoning backdoor attacks take one step more than data-poisoning backdoor attacks~\cite{rodriguez2022backdoor}. Since data poisoning backdoor attacks inject the backdoor into the training data set, the model poisoning attack in the local training phase aims to ensure the backdoor will be contained in the local model that is uploaded to the aggregation phase. It has been discussed in the previous section that the simple label-flipping attack exhibits a low attack success rate. A loss function is then designed to minimize the number of outliers caused by the flipped training data, thus balancing the trade-off between the MAR and the BAR. In such a way, the backdoor is more likely to be injected into the converged model and misclassify the malicious inputs. 

Apart from designing the loss function to change the training parameter, model poisoning backdoor attacks in the local model training phase can also modify the training algorithm~\cite{xie2021crfl}. When the training data set under the targeted label is replaced by the backdoored data set, the local training data becomes the union of the original data set and the backdoored data set. In this case, a malicious model trained with a normal standard gradient descent (SGD) algorithm will lead to a larger model shifting than models trained with a clean data set. As a result, it will be easily detected and deleted from the global aggregation. A projected gradient descent (PGD) algorithm is designed in~\cite{wang2020attack}. In this framework, the updated malicious model is trained by periodically projecting its parameters onto the global model download from the central server at the start of this round. With a newly designed training algorithm, the norm difference between the backdoored and benign models decreases, the backdoor becomes harder to detect, and the attack success rate improves. However, the attack effect is limited on the label with a low distribution rate since the backdoor target is the edge label.

Compared with data poisoning backdoor attacks in the local data collection phase, the model poisoning backdoor attacks in the model training phase achieve better stealthiness and a higher attack accuracy and success rate. 

\noindent
\textbf{Limitations:} 
The attacks at this stage are also known as white box attacks. The attacker needs to master the training data set, as well as the local training process. Furthermore, the knowledge of the defense mechanism adopted in the WFL network is also required to ensure the injection of the backdoor. The attack cost is considerably higher than the data poisoning backdoor attacks.

Compared to backdoor attacks during the local data collection phase, the attacks, particularly the model poisoning attacks (compared to the data poisoning attacks), are more subjected to the communication signal strength at this stage. In the case where the compromised clients controlled by the attacker are geographically distant, the communication signal strengths are likely to be weak, and a large part of each communication round is likely to be spent on the transmission of their local models. The compromised clients can spend less computing time training their adversarial local models and creating backdoors. As a result, the backdoor pattern has lower strengths and is relatively harder to be injected~\cite{nguyen2022federated}. To this end, it is more important to ensure the integrity of the local models from closer clients to mitigate potentially more significant attack strengths from the clients.

Apart from the channel conditions, the shared medium nature of wireless channels can also have a strong impact on the effectiveness of backdoor attacks in the local model training phase. When time-division multiple-access (TDMA) protocols are considered for clients to upload their local models, clients selected and scheduled earlier for model uploading could suffer from substantially shorter local model training time in a communication round and produce premature local models~\cite{luo2022tackling}. This may make the models more vulnerable to backdoors embedded into the global models and, subsequently, the local models. On the other hand, when frequency-division multiple-access (FDMA) protocols, including orthogonal frequency-division multiple-access (OFDMA) protocols, are considered, each selected client would enjoy much narrower bandwidths, which would prolong their model uploading time at the cost of their model training time~\cite{mahmoudi2022fedcau}.

\subsection{Backdoor Attacks at Aggregation Phase}
\label{sec:3c}
\begin{figure}[htb]
\centering
\includegraphics[scale=0.35]{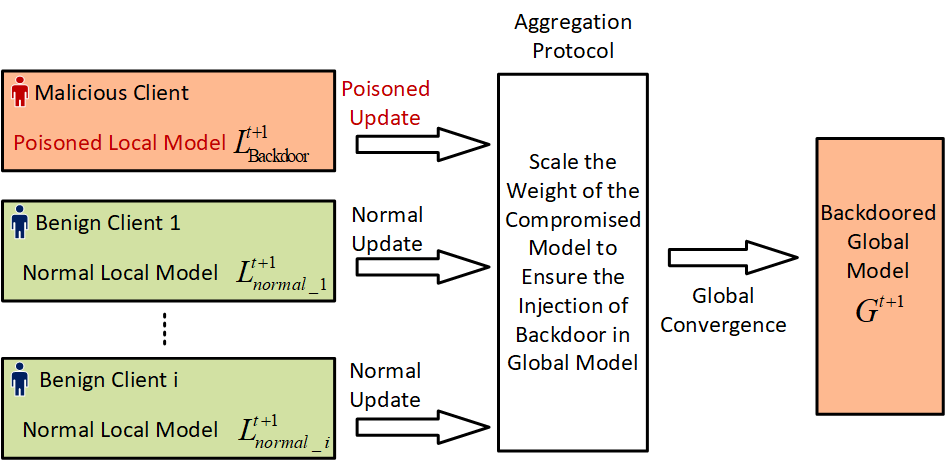}
\centering 
\caption{Aggregation phase backdoor attacks: Modify the aggregation process and replace the global model with the backdoored model.}
 \label{fig:aggregation}
\end{figure}

The backdoor attacks against WFL during the aggregation process can also be categorized as model poisoning backdoor attacks~\cite{nguyen2022flame}. The attacks are conducted when the local models are trained and uploaded for aggregation to update the global model. The mechanism of backdoor attacks during the aggregation phase is illustrated in Fig.~\ref{fig:aggregation}.

Model poisoning backdoor attacks during the aggregation phase can modify the aggregation process by scaling the parameter of the uploaded malicious local model~\cite{shejwalkar2021manipulating,mahloujifar2022property}. The learning rate is decreased, and the weight of the backdoored model is highly scaled, such that the global model can be replaced by the backdoored model~\cite{bagdasaryan2020backdoor}. A loss function is designed to minimize the outlier and thus avoid anomaly detection. Furthermore, To avoid clipping by the differential privacy, the scale factor is also adjusted to be smaller than the norm threshold. The proposed attack ensures that the backdoor is bound to be injected in one round and will not be diluted. If the attacker can control multiple clients to conduct the coordinated attack, the attack success rate is further improved. It has been testified in~\cite{wang2020attack} that by scaling the malicious model updates, the attacker can replace the global model with the backdoored model targeted on the edge case. 

Compared with the data poisoning backdoor attack and model poisoning attack in the model training phase, the model poisoning phase in the aggregation phase shows the greatest attack strength and best stealthiness~\cite{fang2020local}. 

\noindent
\textbf{Limitations:} 
The model poisoning backdoor attack in the aggregation phase requires the attacker to fully control one or multiple compromised clients, including the training data set and training algorithms. Knowledge of the aggregation process is also necessary. To ensure the successful injection of the backdoor, the attacker also needs to acquire sufficient information about the whole WFL network, including the number of clients involved, the size of the whole data set, as well as the defense mechanism. 

Moreover, the attack shows better performance when the global model is close to complete convergence, so the attacker also needs to know the information on the convergence progress of the global model.

However, in decentralized WFL, each client can behave as the server, and only receives model information from nodes within its neighborhood. 
Distributed WFL methods could be more vulnerable to adversarial attacks.
The malicious attacker can then conduct the attack by hijacking one or more compromised clients that are close to the essential node. Thus, the backdoor effect is more likely to be injected into the network and spread more broadly.

\subsection{Summary}

A comparison study of existing backdoor attacks is summarized in Table~\ref{Comparison_bd}. The backdoor attacks conducted during the local data collection phase can be generally classified into label-flipping-based attacks~\cite{jebreel2023fl} and trigger-based attacks~\cite{zeng2021rethinking}. A backdoor attack at this phase can only pollute the training data, and the attackers do not necessarily know the training algorithm and aggregation protocol. Such black-box attacks featured low attack costs but also relatively low backdoor effects~\cite{li2021deeppayload}. Once the attacker obtains the knowledge of the training algorithm, it can fully control the compromised client and initiate the model poisoning backdoor attack during the local model training phase. The most commonly adopted strategy is the global model replacement method~\cite{wang2022black}. The attack model becomes a white box. By taking advantage of fully controlling the compromised model, the attacker can adaptively manipulate the update to improve the attack performance. The aggregation phase attack has a higher requirement for network information. The attacker is supposed to know the information of the aggregation protocol. By manipulating the weight of the adversarial updates, the backdoor pattern is bound to be injected into the converged global model. However, the attack cost is also the highest among all attack methodologies~\cite{chen2022deepguard}. 

\begin{table*}[htb]
\centering
\normalsize
\caption{Comparison among existing backdoor attacks on WFL: Attack model, adaptivity, required information, and potential exacerbation caused by WCN }
\label{Comparison_bd}
\renewcommand{\arraystretch}{1.2}
\scalebox{0.85}{
\begin{tabular}{cl|cc|cc|ccc|cc}
\hline
\hline
\multicolumn{2}{c|}{\multirow{2}{*}{\textbf{Attack Type}}} &
  \multicolumn{2}{c|}{\begin{tabular}[c]{@{}c@{}}\textbf{Attack Model}\end{tabular}} &
  \multicolumn{2}{c|}{\begin{tabular}[c]{@{}c@{}}\textbf{Attack Adaptivity}\end{tabular}} &
  \multicolumn{3}{c|}{\begin{tabular}[c]{@{}c@{}}\textbf{Required Prior Information}\end{tabular}} &
  \multicolumn{2}{c}{\multirow{2}{*}{\begin{tabular}[c]{@{}c@{}}
  \textbf{\color{blue}Exacerbation}\\ \textbf{\color{blue}by WCN}
  \end{tabular}}}
  \\ \cline{3-9} 
\multicolumn{2}{c|}{} &
  \multicolumn{1}{c|}{\begin{tabular}[c]{@{}c@{}}Black Box\end{tabular}} &
  \begin{tabular}[c]{@{}c@{}}White Box\end{tabular} &
  \multicolumn{1}{c|}{\begin{tabular}[c]{@{}c@{}}Non-adaptive\end{tabular}} &
  Adaptive &
  \multicolumn{1}{c|}{\begin{tabular}[c]{@{}c@{}}Training Data\end{tabular}} &
  \multicolumn{1}{c|}{\begin{tabular}[c]{@{}c@{}}Training \\ Algorithm\end{tabular}} &
  \begin{tabular}[c]{@{}c@{}}Aggregation \\ Protocol\end{tabular} 
  \\ \hline

\multicolumn{2}{c|}{\begin{tabular}[c]{@{}c@{}}Label Flipping \\ ~\cite{zhang2020poisongan,rosenfeld2020certified,xiao2012adversarial,aryal2022analysis}\end{tabular}} &
  \multicolumn{1}{c|}{$\surd$} &
  $\times$ &
  \multicolumn{1}{c|}{$\surd$} &
  $\times$ &
  \multicolumn{1}{c|} v &
  \multicolumn{1}{c|}{$\times$} &
  $\times$ & \multicolumn{2}{c}{$\surd$}
  \\ \hline

\multicolumn{2}{c|}{\begin{tabular}[c]{@{}c@{}}Centralized Trigger\\ ~\cite{xu2022more,liu2023facilitating,zhao2022defeat,zhao2021federatedreverse}\end{tabular}} &
  \multicolumn{1}{c|}{$\surd$} &
  $\times$ &
  \multicolumn{1}{c|}{$\surd$} &
  $\times$ &
  \multicolumn{1}{c|}{$\surd$} &
  \multicolumn{1}{c|}{$\times$} &
  $\times$ & \multicolumn{2}{c}{$\times$}
  \\ \hline

\multicolumn{2}{c|}{\begin{tabular}[c]{@{}c@{}}Coordinated Trigger\\~\cite{gong2022coordinated,li2021pointba,wei2021backdoor,xie2019dba}\end{tabular}} &
  \multicolumn{1}{c|}{$\surd$} &
  $\times$ &
  \multicolumn{1}{c|}{$\surd$} &
  $\times$ &
  \multicolumn{1}{c|}{$\surd$} &
  \multicolumn{1}{c|}{$\times$} &
  $\times$ & \multicolumn{2}{c}{$\surd$}
  \\ \hline

\multicolumn{2}{c|}{\begin{tabular}[c]{@{}c@{}}Local Model Replacement\\ ~\cite{bagdasaryan2020backdoor,sun2019can,li2021deeppayload,cao2022highly}\end{tabular}} &
  \multicolumn{1}{c|}{$\times$} &
  $\surd$ &
  \multicolumn{1}{c|}{$\times$} &
  $\surd$ &
  \multicolumn{1}{c|}{$\surd$} &
  \multicolumn{1}{c|}{$\surd$} &
  $\times$ & \multicolumn{2}{c}{$\surd$}
  \\ \hline

\multicolumn{2}{c|}{\begin{tabular}[c]{@{}c@{}}Aggregation Process Attacks\\ ~\cite{rodriguez2022backdoor,chen2023backdoor,hong2022handcrafted,garg2020can}\end{tabular}} &
  \multicolumn{1}{c|}{$\times$} &
  $\surd$ &
  \multicolumn{1}{c|}{$\times$} &
  $\surd$ &
  \multicolumn{1}{c|}{$\surd$} &
  \multicolumn{1}{c|}{$\surd$} &
  $\surd$ & \multicolumn{2}{c}{$\surd$}
  \\ \hline
      \hline

\end{tabular}
}
\end{table*}

    



In Table~\ref{Pro_bd}, the strengths and limitations of existing backdoor attacks on WFL are summarized. The label-flipping-based data poisoning backdoor attack is the easiest to implement. The injected backdoor pattern is also easy to detect, and the life span of the backdoor effect is short. The coordinated trigger-based attack shows better attack performance compared with the centralized trigger-based method. In the coordinated attack, the trigger pattern has been separated into multiple parts in multiple compromised clients controlled by the attackers, making it harder to detect. To achieve a higher attack success rate, data poisoning methods are also developed. The global model replacement method aims to manipulate the local training algorithm, and the adversarial updates are guaranteed to be uploaded for aggregation. The attacks at this stage have a better attack success rate, but the requirement of the model training information increases the attack cost. The attack during the aggregation phase, on the other hand, exhibits the highest attack performance. In return, the attackers need to master the local training data, local training algorithm, and the global aggregation protocol, making it hard to implement in practical scenarios.

\begin{table*}[htb] 
\centering
\caption{Strength and limitation comparison of existing backdoor attacks on WFL}
\label{Pro_bd}
\renewcommand{\arraystretch}{1.2}
\scalebox{0.9}{
\begin{tabular}{cl|l|l|c}
\hline
\hline
\multicolumn{2}{c|}{\textbf{Attack Type}} &
  \multicolumn{1}{c|}{\textbf{Key Features}} &
  \multicolumn{1}{c|}{\textbf{Major Limitations}} &
  \textbf{\color{blue}\begin{tabular}[c]{@{}c@{}}
  Exacerbation\\ by WCN
  \end{tabular}}\\ \hline
\multicolumn{2}{c|}{\begin{tabular}[c]{@{}c@{}}Label  Flipping \\ ~\cite{zhang2020poisongan,rosenfeld2020certified,xiao2012adversarial,aryal2022analysis}\end{tabular}}  &
  \begin{tabular}[c]{@{}l@{}}1. Target: flip the target label. \\ 2. No prior knowledge is required. \\ 3. The attack cost is relatively low.\end{tabular} &
  \begin{tabular}[c]{@{}l@{}}1. It’s easy to be detected.\\ 2. The backdoor injected will be quickly diluted.\end{tabular} & $\surd$ \\ \hline

\multicolumn{2}{c|}{\begin{tabular}[c]{@{}c@{}}Centralized Trigger \\ ~\cite{xu2022more,liu2023facilitating,zhao2022defeat,zhao2021federatedreverse}\end{tabular}} &
  \begin{tabular}[c]{@{}l@{}}1. Classify the malicious input into the target \\  label while behaving normally with being input.\\ 2. Only requires part of the local training data.\\ 3. The attack cost is low.\end{tabular} &
  \begin{tabular}[c]{@{}l@{}}1. The trigger pattern is injected with only one participant.\\ 2. The attack effect will be quickly diluted if the attacker is \\  not selected in every training round.\end{tabular} & $\times$ \\ \hline
  
\multicolumn{2}{c|}{\begin{tabular}[c]{@{}c@{}}Coordinated Trigger \\~\cite{gong2022coordinated,li2021pointba,wei2021backdoor,xie2019dba}\end{tabular}} &
  \begin{tabular}[c]{@{}l@{}}1. The trigger pattern is distributed to multiple \\ compromised clients.\\ 2. Better stealthiness, faster model convergence \\  speed, and a higher attack success rate.\end{tabular} &
  \begin{tabular}[c]{@{}l@{}}1. The attacker needs to ensure each part of the trigger pattern \\ is selected enough times.\\ 2. The trigger pattern is fixed and cannot be adaptively adjusted \\ during the training process.\end{tabular} & $\surd$ \\ \hline
  
\multicolumn{2}{c|}{\begin{tabular}[c]{@{}c@{}}Local Model  \\ Replacement \\ ~\cite{bagdasaryan2020backdoor,sun2019can,li2021deeppayload,cao2022highly}\end{tabular}}&
  \begin{tabular}[c]{@{}l@{}}1. The attacker fully controls one or multiple \\     compromised clients.\\ 2. The local training algorithm is manipulated to \\     ensure the backdoor is bound to be uploaded to \\     the aggregation process.\end{tabular} &
  \begin{tabular}[c]{@{}l@{}}1. The attacker needs to master the local training data as \\     well as the local training process.\\ 2. Knowledge of possible defense mechanisms is also required.\\ 3. The attack cost is greatly higher than the black box attack.\end{tabular} & $\surd$ \\ \hline
\multicolumn{2}{c|}{\begin{tabular}[c]{@{}c@{}}Aggregation \\ Process Attacks \\ ~\cite{rodriguez2022backdoor,chen2023backdoor,hong2022handcrafted,garg2020can}\end{tabular}} &
  \begin{tabular}[c]{@{}l@{}}1. The attacker manages to modify the aggregation \\     protocol during the global model training.\\ 2. The learning rate and the model weight are \\     adjusted to ensure the backdoor injection.\end{tabular} &
  \begin{tabular}[c]{@{}l@{}}1. The attacker needs to acquire sufficient information about \\     the whole framework, including the number of participants, \\     the training algorithm, the aggregation protocol, possible defense \\  mechanism, and the convergence process of the global model.\\ 2. The attack cost is the highest.\end{tabular} & $\surd$ \\ \hline
  \hline
\end{tabular}}
\end{table*}

\section{Latest Defense Methodologies Against Backdoor Attacks}
\label{sec::Defense}

\begin{figure*}[h]
\centering
\includegraphics[scale=0.5]{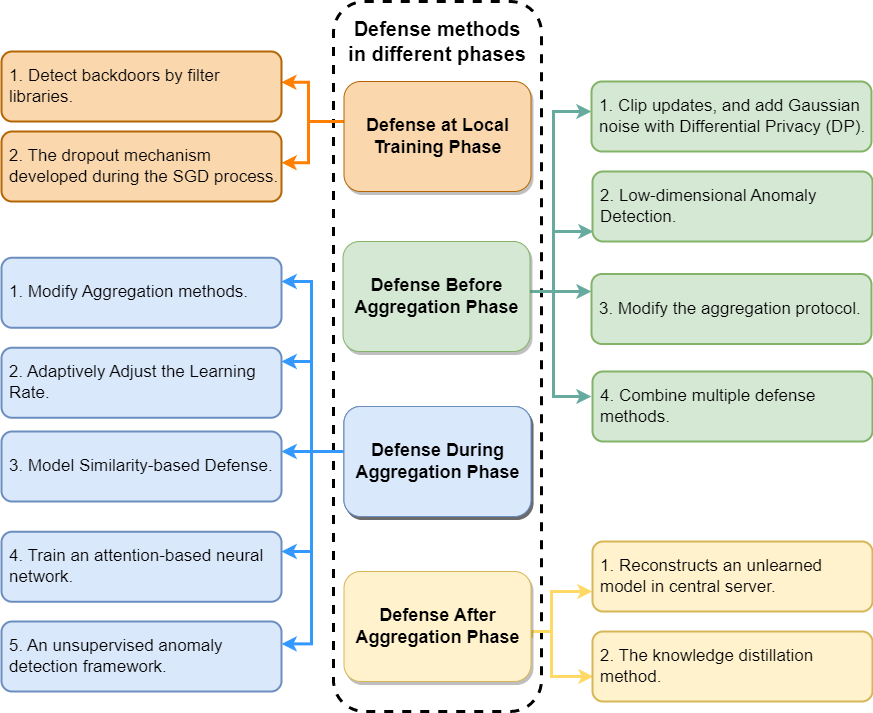}
\centering 
\caption{A summary of defense methods for backdoor attacks on WFL.}
 \label{fig:Defense_methods}
\end{figure*}

In recent literature, numerous defense methodologies have been put forward to safeguard the resilience of WFL against backdoor attacks. The defense schemes can be classified into four types: defense at the local training phase, defense before the aggregation phase, defense during the aggregation phase, and defense after the aggregation phase. The existing defense schemes are summarized in Fig. \ref{fig:Defense_methods}.
    
\subsection{Defense at Local Training Phase}

\begin{figure*}[b]
    \centering
    \includegraphics[scale=0.55]{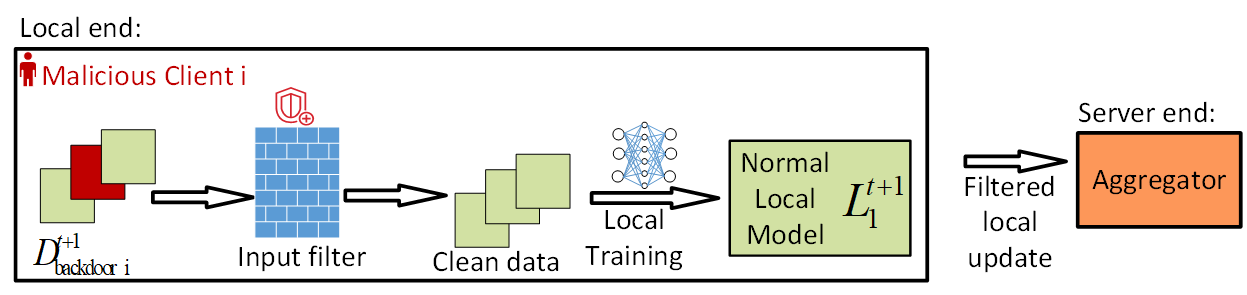}
    \caption{Defense at the local training phase: The input dataset is
filtered by the defender to exclude the injected backdoor pattern.}
    \label{fig:localt}
\end{figure*}

The mechanism of backdoor attack defense during the local training phase is illustrated in Fig \ref{fig:localt}. The input dataset is filtered by the defender to exclude the injected backdoor pattern. The defense methods at this stage primarily focus on data poisoning backdoor attacks. The existing works show reasonable defense performance on label-flipping backdoor attack~\cite{hallaji2023label,gajbhiye2023data} and trigger-based backdoor attack~\cite{zheng2023motif,chen2023feature}. In the rest of this section, two typical defense methods implemented during the local training phase are analyzed, and their limitations are discussed accordingly. 

\subsubsection{Backdoor detection by filter libraries} 
An FL filter and blur-label flipping strategy-based defense algorithm are developed in~\cite{hou2021mitigating}. The central server collects and stores the commonly seen backdoor types in the filter library. The multiple backdoor filters are trained on the server side with different combinations of eXplainable (XAI) models and classifiers. In each FL round, the server sends the global model and backdoor filters. Each filter is used to detect malicious inputs. Once inputs surpass a pre-determined threshold, they will be identified as backdoor inputs. A blur-label-flipping strategy is proposed to clean the backdoor trigger data area for inputs that are determined as backdoors. The filter will replace the suspicious data in the targeted area with its average among its neighbors. Once the suspicious data has been filtered, it is reintroduced into the training process to verify the successful elimination of the backdoor effect. Using the defender developed in~\cite{qiu2021deepsweep}, trigger-based and label-flipping backdoor attacks can be defended. After the blur operation, the suspicious data will be used again instead of being discarded. The data availability is restored. The filter also shows good defense ability against coordinated trigger-based backdoor attacks. 

\noindent
\textbf{Limitations:} 
However, training the backdoor filter on the server side requires prior information about the backdoor patterns and backdoored data sets. The assumptions are made that the attacker only conducts the black-box attack and the knowledge of the local model training process is not obtained by attackers. Moreover, the blur operation can only deal with the trigger in the pixel pattern. Thus, the proposed defense method can only be applied in image classification-oriented WFL.

\subsubsection{Dropout mechanism during the SGD process}
The stability-based defense method against backdoor attacks under the WFL framework is discussed in~\cite{nguyen2020poisoning}. It has been discovered that the probability that the trained model correctly classifies the normal inputs (Clean ACC) is proportional to the probability that the trained model misclassifies the backdoored inputs (Backdoor ACC). Furthermore, a turning point exists after which the Clean ACC remains high and the Backdoor ACC reduces rapidly. The turning point is determined by manually injecting backdoor data sets into the original model to test the Clean ACC and Backdoor ACC, respectively. Once the turning point (or threshold) is determined, a dropout mechanism is developed during the SGD process. The gradient weight of the suspicious fraction is set to zero, and thus, the effect of the backdoor is eliminated. This work discusses that the turning points between Clean ACC and Backdoor ACC are similar among different types of backdoor attacks. Thus, the defense algorithm proposed in~\cite{nguyen2020poisoning} can be applied to multiple scenarios. The backdoor attack success rate under the defense mechanism reduces to zero, while the stability-based defense can ensure a high success rate for the main task. 

\noindent
\textbf{Limitations:} 
In the experimental setup, the number of participants is set to 10, and the training data are IID. The defense performance might degrade in the actual WFL, where the number of clients is larger, and the training data sets are typically non-IID. 
In the existing defense method during the local training phase, a local model without an injected backdoor is trained, and the black box backdoor attacks can be well defended. The defense mechanisms at this stage are also compatible with the SecAgg protocol. The remaining challenges are that the defender requires the proper knowledge of the backdoor patterns, and the trade-off between the defense performance and significant task accuracy must also be considered~\cite{zheng2022balancing}.

\subsubsection{Remark}
Wireless devices can be resource-constrained, e.g., in terms of energy supply and communication resources. Client selection and scheduling are indispensable in the context of WFL. To date, the client selection/scheduling has been designed in parallel to the backdoor defense studies. There is a significant opportunity to couple the designs of both in a holistic way to improve defense effectiveness and efficiency. For instance, clients with excellent instantaneous channel conditions and computing power availability can be selected to conduct local model training and detect backdoors effectively. Fairness measures, such as proportional fairness and max-min fairness, could also be considered when jointly designing the backdoor defense and WFL operations.

\subsection{Defense Before Aggregation Phase}

\begin{figure*}[b]
    \centering
    \includegraphics[scale=0.55]{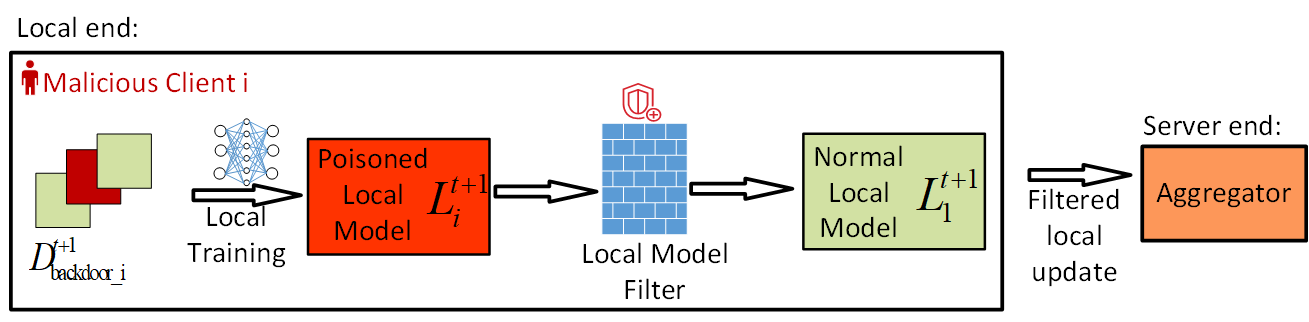}
    \caption{Defense before the aggregation phase: Filter out the suspicious model updates at the local end before sending them to the aggregator.}
    \label{before agg}
\end{figure*}

Another commonly adopted defense strategy against backdoor attacks is to detect the anomaly updates before sending them to the aggregator. The process of defending the backdoor attack before the aggregation phase can be seen in Fig.~\ref{before agg}. The defense methods detect the suspicious local model updates before sending them to the server for aggregation. Many efforts have been carried out to deal with both data poisoning backdoor attack~\cite{wei2023lightweight,wang2023adaptive,lai2023two} and model poisoning backdoor attack~\cite{qi2023revisiting,zhai2023ncl,soremekun2023towards}. The rest of the section reviews multiple widely studied defense methodologies and discusses their limitations.

\subsubsection{Clip updates, and add Gaussian noises with Differential Privacy (DP)} 
{\color{blue} Compared with benign clients, the backdoor attackers are likely to upload updates with relatively larger norms. Thus, it is a reasonable defense that the central server to scale the client updates ${p_i}^{t+1}$ when its $L2$ norm $||{p_i}^{t+1}||_2$ is greater than the predetermined threshold $C$. However, the scaled local model updates ${{p_i}^{t+1'}}$ for each client can be formulated as:
\begin{equation}
\renewcommand{\theequation}{9}
p_i^{t + 1'} = \left\{ {\begin{array}{*{20}{c}}
{p_i^{t + 1},}\\
{}\\
{\frac{C}{{{{\left\| {p_i^{t + 1}} \right\|}_2}}}p_i^{t + 1},}
\end{array}\begin{array}{*{20}{c}}
{}\\
{}
\end{array}\begin{array}{*{20}{c}}
{\mbox{if} \hspace{0.2cm} {{\left\| {p_i^{t + 1}} \right\|}_2} < C};\\
{}\\
{\mbox{if} \hspace{0.2cm} {{\left\| {p_i^{t + 1}} \right\|}_2} \ge C}.
\end{array}} \right.
\end{equation}
The scaling mechanism ensures that the norm of each model update is small 
~\cite{sun2019can}. }

{\color{blue} After clipping updates, the differential privacy technique can be further applied to defend against backdoor attacks. Although DP-based methods primarily focus on protecting the privacy of individual client data, they still indirectly help to mitigate certain types of backdoor attacks. By adding properly crafted noises to the model updates, DP can make it more challenging for malicious clients to inject poisoned information into the updates without being detected. }

\noindent
\textbf{Limitations:} 
In a traditional model train with DP, the amount of noise added is relatively large to ensure privacy and security~\cite{chen2022linkbreaker}. However, in the scenario of defending against backdoor attacks on WFL, although considerable noise injected can contribute to eliminating the effect of backdoors, it might also ruin the convergence of the main task. Thus, the trade-off between the primary task accuracy and attack defense rate must be considered when adding the noise. 

Another limitation of this type of defense mechanism is that once the attacker knows the number or the pattern of the clipping threshold, the attacker can modulate the malicious updates to trick the defender and thus implement the backdoor. It also can be found that the simple norm clipping method shows less effectiveness against trigger-based backdoor attacks since the injected trigger pattern has limited influence on the uploaded norm compared with label-flipping-based backdoor attacks. It has also been pointed out in~\cite{nguyen2022flame} that the DP-based defense method is less robust when the attacker conducts a multi-round attack.

In WFL, the impact of network communication protocols can be non-negligible. For instance, the TCP can ensure the reliable delivery of data across a network but requires one or multiple retransmissions in each communication cycle~\cite{10032555}. This can penalize the local models by reducing the local training time in a communication cycle. For this reason, the UDP is more commonly adopted in WFL~\cite{9716792}. However, the data distortion and packet loss caused by network communications need to be considered when designing the norm clipping threshold.

Due to the feature of WCN, the communication channels with better signals can be better selected in WFL. The clipping should filter not only the updates from clients but different communication channels as well. The ratio of information gathered from remote network nodes should be considered to prevent backdoor effects and ensure fairness.

\subsubsection{Low-dimensional anomaly detection}
Instead of simply using the norm clipping defense method, several low-dimensional anomaly detection methods have been proposed~\cite{li2020learning,gu2021detecting}. {\color{blue}
The basic idea of low-dimensional anomaly detection methods is to project the compact computed local model onto its low-dimension space, as given by:
\begin{equation}
\renewcommand{\theequation}{10}
    p_i^{t + 1}(L) \to p_i^{t + 1'}(l),
\end{equation}
where $L$ and $l$ are the numbers of model features before and after the dimension reduction, respectively, and $L>l$. This considers the fact that the low-dimensional embeddings of the uploaded model updates still contain the crucial features that capture the essential variability in data instances~\cite{gong2022backdoor}. After removing the noisy and redundant features of these data instances, the difference between benign and malicious updates becomes apparent.} 

One method is to train an encoder-decoder model to approximate low-dimensional embeddings~\cite{li2020learning}. The unbiased model updates are selected as the inputs of the encoder, and the output of the encoder is the low-dimensional embeddings. The output generated by the encoder is subsequently utilized as the input for the decoder, facilitating the process of reconstructing the original model updates. The encoder-decoder model is then trained to converge by minimizing the reconstruction error, and the converged encoder-decoder model can be adapted to test the practical model updates. The reconstruction error of the benign updates is much smaller than that of the malicious updates. In~\cite{gu2021detecting}, a Principal Component Analysis (PCA) dimensionality reduction-based defense mechanism is proposed. In each round of a WFL process, a subset of the uploaded model updates is selected and pre-aggregated. The difference between the pre-aggregated and global models is computed, and the parameters in the computed difference corresponding to the predicted probabilities are extracted and stored in a list in the aggregator. After multiple rounds, the list is standardized by removing the mean and scaling to unit variance. The standardized list is then used to input the PCA to reduce the dimension of the updated data and visualize the class pattern. The biggest cluster is recognized as the benign update, while the malicious updates lie within a visibly different cluster. Once the updates from the attackers are detected, the server can ignore their updates and restrict their participation in future rounds. 

\noindent
\textbf{Limitations:} 
The encoder-decoder architecture requires extra unbiased updates to train the detector, and it is pretty hard to implement in real WFL applications. The PCA-based method requires prior knowledge of the targeted backdoor label, which is also hard to know in practical defense implementation. Meanwhile, the above-mentioned methods can only distinguish label-flipping-based backdoor attacks and cannot deal with trigger-based backdoor attacks. The low-dimensional methods are also incompatible with the SecAgg protocol, and the privacy of the participants in the WFL network cannot be guaranteed.

\subsubsection{Modify the aggregation protocol}
In the WFL framework, the training data sets are non-IID, and the SecAgg protocol does not allow the server to check the individual updates from certain clients. However, the benign updates exhibit similar distributions on average, while the malicious updates exhibit outliers. It is also effective in developing defense mechanisms by modifying the aggregation protocol.  In~\cite{gao2022secure}, a new aggregation protocol, PartFedAvg, is proposed to defend against data poisoning attacks. {\color{blue} 
In every training round, each client needs to execute an additional random update selection step. The local model update ${p_i}^{t+1}$ has the same multi-layer structure as the global model, indicating the updates regarding different features of the model. Thus, in the PartFedAvg protocol, a fraction of features from the local update is set to zero and uploaded together with the remaining unchanged values. This can be formulated as:
\begin{equation}
\renewcommand{\theequation}{11}
    p_i^{t + 1'} = \left\{ {\begin{array}{*{20}{c}}
{p_i^{t + 1}\left[ j \right],}\\
{0,}
\end{array}\begin{array}{*{20}{c}}
\end{array}\begin{array}{*{20}{c}}
\mbox{if the $j$-th feature is selected};\\
\mbox{if the $j$-th feature is not selected}.
\end{array}} \right.
\end{equation}
Then, the server receives the updates provided by the clients and aggregates the global model update. Assuming in each round, $m$ local clients are selected for computing model updates, and the total number of model features is $l$, the aggregation of the global model update can be expressed as:
\begin{equation}
\renewcommand{\theequation}{12}
    p_g^{t + 1}\left[ j \right] = \left\{ {\begin{array}{*{20}{c}}
{\frac{\eta }{{{z^{t + 1}}\left[ j \right]}}\sum {p_i^{t + 1}\left[ j \right]} ,}\\
{}\\
{0,}
\end{array}\begin{array}{*{20}{c}}
{}\\
{}
\end{array}\begin{array}{*{20}{c}}
\mbox{if} \hspace{0.2cm} {{z^{t + }}\left[ j \right] \ne 0};\\
{}\\
\mbox{if} \hspace{0.2cm} {{z^{t + 1}}\left[ j \right] = 0}.
\end{array}} \right.
\end{equation}
The values within each label are summed and divided by their respective count $z^{t+1}\left[ j \right]$ as:
\begin{equation}
\renewcommand{\theequation}{13}
    {z^{t + 1}}\left[ j \right] = {z^{t + 1}}\left[ j \right] + 1,\begin{array}{*{20}{c}}
{}&{\mbox{if}} \hspace{0.2cm} p_i^{t + 1'}\left[ j \right] \ne 0&{\mbox{and}}& 0 < i < m.\end{array}
\end{equation}
The calculated global update $p_g^{t + 1}$ is then employed to update the global model. The proposed method ensures the privacy of the participants involved. The malicious attacker cannot inject completed backdoor patterns in a short time, thus improving the robustness of the global model.} 

To better defend against both data poisoning backdoor attacks and model poisoning attacks, the Meta-FL is developed in~\cite{aramoon2021meta}. In each round, the central server selects subsets of the participants. Each subset computer updates. Then, each subset follows the SecAgg protocol and sends the cryptography-masked updates. Although the server cannot access each client's update, it can migrate the effect of backdoor attacks by ignoring visibly different subset updates. In such a design, the participant's privacy is ensured, and it can be applied to defend against label-flipping-based, trigger-based, and model replacement-based backdoor attacks.

\noindent
\textbf{Limitations:} 
The defense schemes proposed in \cite{gao2022secure} and \cite{aramoon2021meta} require modifying the local training process, and the backdoor effect can only be suppressed and cannot be entirely eliminated. The method requires the number of malicious attacks to be smaller than 50\% of the total selected clients in each round. 

\subsubsection{Combine multiple defense methods} 
It is reasonable to combine multiple defense methods to address the limitations of individual defense mechanisms to improve reliability. The WFLAME framework proposed in~\cite{nguyen2022flame} combines the model clustering, weight clipping approach, and the noise injection method is utilized to mitigate the backdoor effect while simultaneously preserving the benign performance of the aggregated model. In each round, the server receives the updates from the clients and uses HDBSCAN to identify and remove suspicious model updates. A dynamic weight-clipping approach is applied to deal with the boosted malicious model updates. The clipped model updates are then injected with adaptive noise to further improve robustness. At last, the modulated updates are aggregated with the same weight. In this design, the HDBSCAN clustering algorithm is dynamic and adaptive, and the clipping threshold and the amount of noise are also adaptively designed. It shows good defensive performance against both data poisoning and model poisoning backdoor attacks. 

\noindent
\textbf{Limitations:} 
Implementing the WFLAME framework proposed in \cite{nguyen2022flame} requires massive and computationally expensive modification of the whole WFL network, which brings challenges in practical scenarios.

On the other hand, the broadcast nature of wireless channels gives rise to an opportunity for collaborative backdoor detection. For example, the clients are likely to overhear the local model updates from each other and can potentially evaluate those local model updates based on their own local datasets. Consensus~\cite{MAKHDOOM2019251} among the clients can be potentially used to select reliable local models. Byzantine fault tolerance techniques~\cite{WANG201910} can be potentially leveraged to identify misbehaved clients and maintain the integrity of WFL.

\subsection{Defense During Aggregation Phase} 

\begin{figure*}[htb]
    \centering
    \includegraphics[scale=0.55]{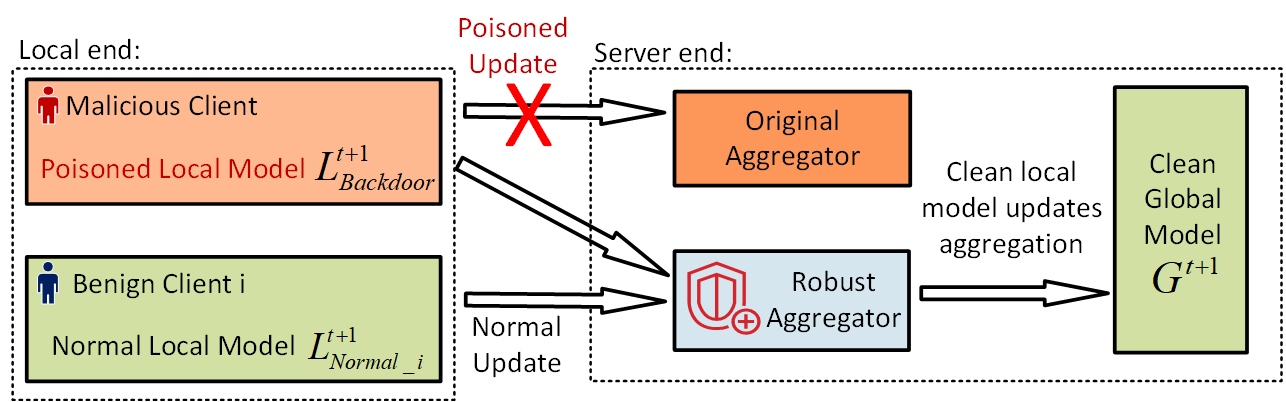}
    \caption{Defense during the aggregation phase: A robust aggregator is designed at the server end to distinguish and reject malicious updates from participants.}
    \label{agg defense}
\end{figure*}

In many recent works, the defense execution stage is moved from individual updates to the aggregation level. Multiple methodologies are carried out to filter out the poisoned updates during the aggregation phase. An overview of defense during the aggregation phase is presented in Fig.~\ref{agg defense}. Considering the fact that the aggregator, which simply combines the received updates, is vulnerable to being poisoned, defenders design a robust aggregator based on statistical methods~\cite{elhattab2023robust,wang2023adaptive}. By calculating the statistical features of model updates, including average and median, one or more participants are selected as the reference to detect the malicious updates~\cite{gao2023not,wei2023lightweight}. The feasibility and effectiveness of defense during the aggregation phase have been proved in the existing works when facing model poisoning backdoor attacks~\cite{rieger2022deepsight}. In this section, multiple defense methods designed to take effect during the aggregation phase are studied, and their limitations are then analyzed.

\subsubsection{Aggregation protocol modification methods} 
In~\cite{fu2019attack}, a classic repeated median regression algorithm is extended to the WFL applications. {\color{blue} The repeated median regression is computed for each label when the server receives the model updates. The residuals at each dimension are also obtained. Standardizing these residuals yields the standard deviation, based on which the parameter confidence of each label is derived. Then, the extreme values are also corrected to eliminate the effect of boasted malicious parameters. The $k$-th updated models can then be re-weighted based on the feature importance:
\begin{equation}
\renewcommand{\theequation}{14}
    {W^{(k)}} = \sum\limits_{j = 1}^l {\omega _j^{(k)}\sigma \left( {{\omega _l}} \right)}, 
\end{equation}
where ${\omega _j^{(k)}}$ is the confidence of the $j$-th feature within the model, and $\sigma(\cdot)$ denotes the standard deviation. A feature with a large standard deviation implies that the value of this feature varies greatly among all participants, which needs to be more critical when sending for aggregation. Subsequently, the global model update can be aggregated as:
\begin{equation}
\renewcommand{\theequation}{15}
    p_g^{t + 1} = \sum\limits_{k = 1}^m {\frac{{{W^{(k)}}}}{{\sum\limits_{i = 1}^m {{W^{(i)}}} }}} p_k^{t + 1}.
\end{equation}
The proposed method is featured for defending label-flipping backdoor attacks and model replacement backdoor attacks.} 

\noindent
\textbf{Limitations:} 
The aggregator requires the knowledge of individual model updates, which violates the SecAgg protocol. The proposed method performs poorly in the non-IID situation and cannot ensure privacy and security. Thus, it is hard to implement in practical WFL scenarios.

\subsubsection{Adaptive adjustment of the learning rate} 
Different from modifying the aggregation method, the studies conducted in ~\cite{ozdayi2021defending,liu2021synergetic} aim to adaptively adjust the learning rate to improve the robustness of the WFL framework. The distribution of the sign function of each label is obtained. {\color{blue} A hyperparameter $\theta$ is then defined to determine the threshold of the learning rate. The adaptive learning rate for each model feature can be formulated as:
\begin{equation}
\renewcommand{\theequation}{16}
    {\eta _{\theta ,i}} = \left\{ {\begin{array}{*{20}{c}}
{\eta ,}\\
{}\\
{ - \eta ,}
\end{array}\begin{array}{*{20}{c}}
{}\\
{}
\end{array}\begin{array}{*{20}{c}}
\mbox{if} \hspace{0.2cm} {\left| {\sum\limits_{k = 1}^m {{\mathop{\rm sgn}} \left( {p_k^{t + 1}[i]} \right)} } \right| > \theta };\\
{}\\
{\mbox{otherwise}}.
\end{array}} \right.
\end{equation}
If the sum of signs of the $i$-th feature among all local model updates is less than $\theta$, the sign of the learning rate is reversed to maximize the loss on that specific feature, instead of minimizing it. This method shows reasonable defense performance against data poisoning attacks~\cite{xiang2021detecting}.}

\noindent
\textbf{Limitations:} 
Similar to the aggregation modification method, the proposed defense mechanism requires the individual to update information from the participant on the server side. Thus, it is not compatible with the SecAgg. The participants' privacy security cannot be guaranteed.

\subsubsection{Model similarity-based defense}
Another commonly adopted strategy of defense during the aggregation phase is the model similarity-based defense method~\cite{hou2022similarity,zhang2022increasing,hu2022badhash}. The FoolsGold is proposed in ~\cite{fung2020limitations}, and the adversarial attack is detected based on the diversity of the model updates. In a WFL framework, the benign client's updates tend to have richer diversity, while the adversary normally behaves similarly. By comparing the uploaded update with its historic stored values, adversarial models with similar updates can be detected. In each round, each update is first compared with its historical value, and the similarity is computed. The similarities of the updates from different clients are also computed to reduce misdetection. The backdoor attack can be successfully detected by filtering out the outlier of the updates and adaptively adjusting the learning rate. This design has no constraints on the number of malicious clients, and it demonstrates reasonable performance against data poisoning attacks.

\noindent
\textbf{Limitations:} 
The methods proposed in \cite{hou2022similarity,zhang2022increasing,hu2022badhash,fung2020limitations} assume that the attacker should be malicious all the time. The attacker needs to upload adversarial models in every round. In practical scenarios, the attacker can decide to inject the backdoor at a random round. The proposed method cannot detect it without sufficient accumulated historical malicious updates.

\subsubsection{Defense using attention-based neural networks} 
In~\cite{wan2021robust}, the focus is on training an attention-based neural network. The central server simulates the WFL tasks under various types of backdoor attacks. Then, the simulated model updates are collected to train a separate neural network model, which learns the potential vulnerability of the global model against different backdoor attacks. Thus, the global model shows a self-supervised fashion. 

{\color{blue} The attention-based neural network is then applied to predict the actual updates received from the participants. The likelihood of a benign update received can be parameterized as:
\begin{equation}
\renewcommand{\theequation}{17}
    {s_i} = \frac{{Q\left( {{q_t}} \right)K\left( {{p_i}} \right)}}{{\left\| {Q\left( {{q_t}} \right)} \right\|\left\| {K\left( {{p_i}} \right)} \right\|}},
\end{equation}
where $Q(q_t)$ is the query encoder from the trained defender, and updated with the actual model updates. If the alignment score $s_i$ is closer to $+1$, $p_i$ is more likely to be a benign update. On the other hand, if $s_i$ is close to $-1$, $p_i$ is likely to be an adversary update.} This method shows better defense performance than traditional median-based aggregation and residual-based re-weighting aggregation. By training the neural network with new backdoor attack data sets, the self-supervised defense mechanism can defend it properly~\cite{fan2021text}. 

\noindent
\textbf{Limitations:} 
The major problem of this defense method is that to distinguish the malicious updates, all possible backdoor attack patterns must be trained in the server before implementation. During the detection, the aggregator requires access to single local updates. Thus, the privacy of the participant is at risk of leaking.

\subsubsection{Unsupervised anomaly detection-based defense} 
Considering the fact that the backdoor attack effect is discernible in terms of model weight, an unsupervised anomaly detection framework, ARIBA, is proposed in~\cite{mi2022ariba}. In this design, the model weights are pre-processed to obtain the discernible patterns of filters. Multiple filters are then collected and fed into an unsupervised anomaly detection algorithm to detect the suspicious filters. {\color{blue} For example, when the server receives $m$ local model updates, a set of $\phi$ filters, denoted as $f_{i,j}$, $0<i<m$, $0<j<\phi$, are used for the detection via the Mahalanobis-distance-based algorithm ~\cite{dashdondov2021mahalanobis}. The detected suspicious filters are applied to compute the anomaly score of each client, as given by:
\begin{equation}
\renewcommand{\theequation}{18}
    {s_{i,j}} = \left\{ {\begin{array}{*{20}{c}}
{0,}\\
{}\\
{1,}
\end{array}\begin{array}{*{20}{c}}
{}\\
{}
\end{array}\begin{array}{*{20}{c}}
{\mbox{if}\hspace{0.1cm}f_{i,j}\hspace{0.1cm}\mbox{is}\hspace{0.1cm}\mbox{benign}};\\
{}\\
{\mbox{if}\hspace{0.1cm}f_{i,j}\hspace{0.1cm}\mbox{is}\hspace{0.1cm}\mbox{adversary}}.
\end{array}} \right.
\end{equation}
As a result, the malicious client can then be identified accordingly~\cite{li2022test}.} It is assumed that the attacker knows all possible defense patterns, can conduct attacks at any time, and multiple attackers can conduct coordinated attacks. The proposed method can defend against multiple types of backdoor attacks with properly trained weight filters without degrading the benign performance. 

\noindent
\textbf{Limitations:}
The unsupervised anomaly detection methods proposed in~\cite{mi2022ariba} and~\cite{li2022test} are incompatible with the SecAgg protocol. Thus, data privacy and security can hardly be ensured.

In addition to the above-mentioned limitations, the emerging OTA-WFL may pose new challenges to the existing defense mechanisms designed for the aggregation phase due to the fact that the FL server would only access the aggregated global model under OTA-WFL~\cite{yu2022optimal,XichenYuTCOM2023}. The local models of individual clients are obscured. While this helps improve the privacy of the individuals' local models to a great extent, it allows backdoor attacks to take place unnoticed. Few studies have been concerned about this particular issue. For example, the authors of~\cite{zheng2022balancing} attempt to restore all the individual models using serial interference cancellation. To what extent can the restored local models be assessed to detect backdoor attacks remains unclear, as the local model restoration process may distort the local models and affect the backdoor detection accuracy.

\subsection{Defense After Aggregation Phase}
 
\begin{figure*}[htb]
    \centering
    \includegraphics[scale=0.5]{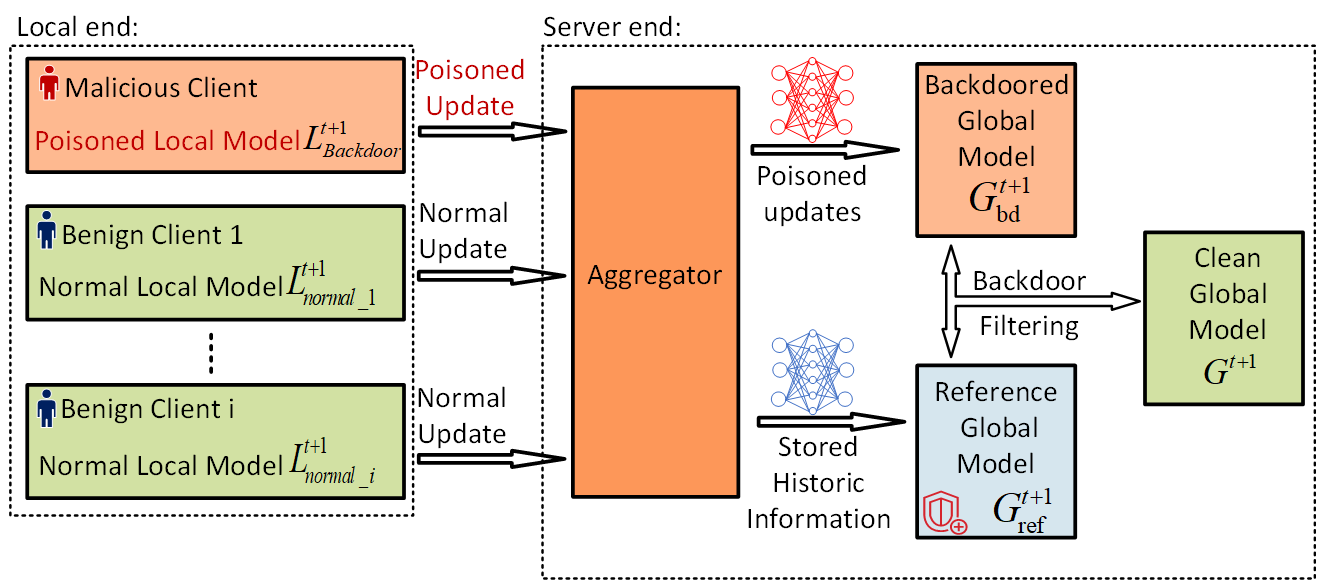}
    \caption{Defense after the aggregation phase: The server
will store the historical value gathered from participants to
train a reference global model for anomaly detection.}
    \label{after def}
\end{figure*}

Several defense methodologies have been proposed to deal with the injected backdoor effect that remained in the global model after the aggregation phase —- The idea of defense after the aggregation phase can be seen in Fig.~\ref{after def}. The server will store the historic value gathered from participants to train a reference global model, based on which the trained converged WFL global model is examined to filter the anomaly contribution from suspicious clients. By excluding the influence of suspicious clients, the backdoor effect will be entirely eliminated. The defender at this stage can well handle the model poisoning backdoor attacks, including model replacement backdoor attack~\cite{lu2022defense,rodriguez2022backdoor,elhattab2023robust} and weight scaling backdoor attack~\cite{nguyen2022flame,wang2023adaptive}.

The post-aggregation backdoor defense mechanisms, i.e., performed upon the aggregated global models, can be particularly important for the emerging OTA-WFL systems to circumvent the obscurity of the local models in the OTA-WFL processes.
This section reviews two typical defense designs after the aggregation phase. Their limitations are then discussed.

\subsubsection{Reconstruct an unlearned model in a central server}
In~\cite{liu2021federaser}, the federated unlearning methodology, namely, FedEraser, is proposed. In this scheme, the central server stores the historical updates received from clients during the global model training process. Once the global model converges, the FedEraser reconstructs an unlearned model in the central server based on the stored historical parameters. During the unlearned model training process, the benign clients are defined as the calibrating clients. In each round, the central server collects the updates from calibrating clients concerning the calibrated global model. Then, the FedEraser aggregates these updates and renews the global model to be an unlearned WFL global model. By comparing the performance of the unlearned model on clients' data with that of the original global model, the backdoor effects can be well eliminated~\cite{sommer2022athena}. In the FedEraser design, no substantial modification is required to the existing WFL architecture and the training process of the participants. Thus, the proposed method is easy to implement. In addition, the FedEraser trains the unlearned model using the parameters stored during the Federated global model training process. It is claimed that the training speed of constructing the unlearned model in FedEraser is four times faster than training from scratch.

\noindent
\textbf{Limitations:} 
The calibration process during the unlearned model training relies on the historical updates of the participating clients. The limited store capability of the hardware makes it hard to deploy in practice. In addition, in each round of federated model training, the user's local gradient is randomly selected and uploaded to the server to train the global model. Thus, the reliability of the calibration of the global model using the unlearned model trained by the historical parameters remains discussed.  On the other hand, although the training data and the training process are not required, the FedEraser must target the attacker client, which is also pretty hard in actual implementation. 

\subsubsection{Knowledge distillation method} 
The knowledge distillation method is proposed in~\cite{wu2022federated,zhang2022increasing} to restore the global model's performance. In this design, the central server stores the historical parameters of the participating clients. Once the global model is obtained. The contribution of the targeted malicious clients is subtracted, and a skewed unlearning model is derived. Then, the knowledge distillation is performed. The original model is defined as the teacher model, while the skewed unlearning model is defined as the student model. The teacher model produces class predictions that are utilized to label the dataset employed for training the student model. The concept of temperature is introduced in this design. The higher temperature produces a more ambiguous probability distribution, while the lower temperature produces a more discrete probability. The student model is trained under high temperatures to compensate for the skew. The knowledge distillation training process is executed on the server side. Thus, no extra information about the client and communication between the client and the server are required. Due to the fact that the data set used in the distillation does not contain backdoor patterns, the student model will not learn the backdoors injected in the teacher model. It is also testified that the proposed knowledge distillation-based unlearning method can improve the robustness of the global model and defend against all types of backdoor attacks~\cite{ye2022safe}.    

\noindent
\textbf{Limitations:}  
The unlearning mechanism requires the knowledge of the target clients, and the attackers are normally stealthy in practical scenarios. Meanwhile, the proposed method is not compatible with SecAgg, and the privacy of the participants is at risk of leakage.

\subsection{Summary}

Overall, the existing defense schemes are analyzed in each type and summarized in Table \ref{table:Defense Methods}.
The existing defense methods can be broadly divided into four categories: defense at the local training phase, defense before the aggregation phase, defense during the aggregation phase, and defense after the aggregation phase. Defense methods at the local training phase are implemented on the client side, aiming at the data poisoning attacks. The defense cost is relatively low, and the defense mechanism at this stage is compatible with SecAgg. Defense before the aggregation phase is also mounted on the client side. The defender is keen to exclude suspicious updates before sending them to the aggregator. Some methodologies employ the modification of SecAgg, thus making it hard to be configured in practical WFL networks. The defense during the aggregation is capable of identifying the backdoored updates from the malicious clients and excluding them from rejecting backdoor injection. The defenders are located on the server side, and normally, they are not compatible with SecAgg. The defense after the aggregation phase will evaluate the converged global model to test whether it has been polluted by the backdoor pattern or not. The defense mechanism is also at the server side, and the historical information gathered from the participant is necessarily needed.

\begin{table*}[htpb]
    \caption{Defense methods for backdoor attacks on WFL}
    \label{table:Defense Methods}
    \centering
    \scalebox{0.9}{
    \renewcommand{\arraystretch}{1.2}
    \begin{tabular}{>{\raggedright\arraybackslash}p{0.06\linewidth}|c|>{\raggedright\arraybackslash}p{0.07\linewidth}|>{\raggedright\arraybackslash}p{0.26\linewidth}|>{\raggedright\arraybackslash}p{0.06\linewidth}|>{\centering\arraybackslash}p{0.1\linewidth}|>{\raggedright\arraybackslash}p{0.1\linewidth}|>{\raggedright\arraybackslash}p{0.08\linewidth}|>
    {\centering\arraybackslash}p{0.1\linewidth}}
    \toprule
    \hline
    
    \textbf{Defense Phase} & \textbf{Ref.\footnotemark[1]} & \textbf{Strategy} & \textbf{Detail} & \textbf{Exp. Env.\footnotemark[2]}  & \raggedright\textbf{Comp. with SecAgg?} & \textbf{Main Task} & \textbf{Dataset} & 
    \textbf{\color{blue}
    Exacerbation by WCN
    }  \\
    \midrule 

    \multirow{6}{=}{\centering \rotatebox[origin=c]{90}{Local Training}} 

    & \cite{hou2021mitigating} & Poisoned Data Filtering Method. & Federated backdoor filter designed to identify backdoor inputs and restore the data to availability by the blur-label flipping strategy. & Client & $\surd$ & Image classification & MNIST; CIFAR-10 & $\surd$ \\ 
    \cline{2-9}

    & \cite{zhao2021stability} & Model Stability based Method. &The stability-inducing operations are introduced to improve the generalization of the testing data. & Client & $\surd$ & Intelligent Edge Computing Service & MNIST & $\surd$ \\
    
    \hline 

    \multirow{22}{=}{\centering \rotatebox[origin=c]{90}{Before Aggregation}} 
        
    & \cite{sun2019can} & Differential Privacy based Defense. & The local updates are norm-clipped, and Gaussian noise is added to the global model. & Client & $\times$ & Tensor Flow WFL applications & FMNIST & $\surd$ \\ 
    \cline{2-9} 
    
    & \cite{li2020learning} & Reduced Dimension based Defense. & Spectral anomaly detection mechanism is designed to remove the malicious model updates in a low dimensional latent space. & Client & $\times$ &Image classification; Sentiment Analysis & MNIST; FMNIST; Sentiment140 & $\surd$ \\ 
    \cline{2-9}
    
    & \cite{tolpegin2020data} & Reduced Dimension based Defense. & The defense extracts parameters from the high-dimensional updates and applies PCA for dimensionality reduction to filter out malicious updates. & Client & $\times$ &Image classification & CIFAR-10; Fashion-MNIST & $\surd$ \\ 
    \cline{2-9}

    & \cite{aramoon2021meta} & SecAgg Procotol Modulation based Defense. & The local clients follow the SecAgg protocol and send the cryptography-asked updates. The backdoor attacks are mitigated on the aggregation level. & Client & $\surd$ & Image classification & SVHN; GTSRB & $\surd$ \\
    \cline{2-9}
    
    & \cite{gao2022secure} & SecAgg Procotol Modulation-based Defense & The partial aggregation protocol is designed to improve the privacy and robustness of WFL. & Client & $\surd$ & Image classification; Word prediction & MNIST; CIFAR-10; LOAN & $\surd$ \\ 
    \cline{2-9}

    & \cite{nguyen2022flame} & Mulit-Methods Combined Defense & The defense method combines the model clustering, weight clipping approach, and noise injection to achieve adaptive clipping and eliminate the adversarial backdoors. & Client & $\times$ & Word prediction; NIDS; Image classification & Reddit; IoT-Traffic; CIFAR-10; MNIST; Tiny-ImageNet & $\surd$ \\
    \hline
    
    \multirow{26}{=}{\centering \rotatebox[origin=c]{90}{During Aggregation}}  
    
    & \cite{fung2020limitations} & Model Similarity-based Defense & The defense mechanism can identify attacks based on the diversity of client updates. & Server & $\times$ & Classifications & MNIST; VGGFace; KDDCup; Amazon & $\surd$\\ 
    \cline{2-9}
    
    & \cite{wan2021robust} & Model Similarity-based Defense & The defense mechanism trains an NN with an attention mechanism to learn the vulnerability of WFL models from a set of plausible attacks. & Server & $\times$ & Image classification; IMDb sentiment analysis & MNIST; CIFAR; ImageNet & $\surd$  \\ 
    \cline{2-9}
    
    & \cite{mi2022ariba} & Model Similarity-based Defense & The defense employs unsupervised anomaly detection to evaluate the pre-processed filters. The malicious clients can then be identified according to their anomaly scores. & Server & $\times$ & Image classification & MNIST; CIFAR-10; Fashion-MNIST & $\surd$ \\ 

    \cline{2-9}

    & \cite{yin2018byzantine} & Statistical Method & The defense proposes one coordinate-wise median-based and one coordinate-wise trimmed mean-based gradient descent algorithm to filter out backdoors & Server & $\times$ & Tensor-Flow tasks & MNIST & $\surd$ \\ 
    \cline{2-9}
    
    & \cite{fu2019attack} & Statistical Method & A novel aggregation algorithm with residual-based reweighting is proposed, which combines repeated median regression with the reweighting scheme in iteratively least squares. & Server & $\times$ &Image classification; Word prediction; Loan prediction & MNIST; CIFAR-10; Amazon; LOAN & $\surd$ \\ 
    \cline{2-9}
    
    & \cite{ozdayi2021defending} & Statistical Method & The defense adjusts the aggregation server's learning rate per dimension and round based on the sign information of local client updates. & Server & $\times$ & Image classification & MNIST; FMNIST & $\surd$ \\
    \hline 
    
    \multirow{10}{=}{\centering \rotatebox[origin=c]{90}{After Aggregation}} 
    
    & \cite{liu2021federaser} & WFL with Forgetting Mechanism & The FedReaser is proposed to store the historical updates of federated clients and re-train the unlearned model in the central. & Server & $\times$ &Image classification; Word prediction & UCI Adult Purchase; MNIST; CIFAR-10 & $\surd$ \\
    \cline{2-9}

    & \cite{wu2022federated} & WFL with Forgetting Mechanism & A Federated unlearning method is proposed to subtract the accumulated client updates and leverage the knowledge distillation method to restore the model performance without local privacy information. & Server & $\times$ &Image classification & MNIST; CIFAR-10; GTSRB & $\surd$ \\
   \cline{2-9}
    & \cite{andreina2021baffle} & Global Model Test Method & The defense proposed one feedback-based WFL to detect backdoors by leveraging data of multiple clients. & Server & $\surd$ & Image classification & CIFAR-10; FMNIST & $\surd$ \\ 
    \hline 
    \hline
    \multicolumn{8}{l}{\textsuperscript{1} \footnotesize{Reference paper that belongs to the specific group.}} \\
    \multicolumn{8}{l}{\textsuperscript{2} \footnotesize{Experiment environments, including server or clients.}} \\
    \end{tabular}}

\end{table*}

\section{Performance Evaluation and Comparison}
\label{sec::Performance}
{\color{blue} In this section, the performance of backdoor attack strategies and defense methods in each phase are presented, followed by a comparative analysis.}

\subsection{Comparison Among Backdoor Attack Strategies}
\label{sec:5A}
{\color{blue}
Four representative backdoor attack strategies~\cite{zhang2020poisongan, gong2022coordinated, bagdasaryan2020backdoor, rodriguez2022backdoor} on WFL in each phase (as shown in Table \ref{Table:4}) are firstly compared. The experimental setting is given in Table \ref{tab:att_set}. The global model is trained on the FMNIST dataset. It is assumed that there are a total of 100 participants, and in each training iteration, the server randomly picks 10 participants for the global model update. The number of compromised clients controlled by malicious attackers increases with relatively more complex attack methods. It is also assumed that the malicious clients can be continuously selected in multiple rounds. The trigger pattern used in this case follows the original method, as discussed in~\cite{gong2022coordinated}. To make it easier to follow, a GitHub entry has been created for the reader's information\footnote{https://github.com/DrQu89757/backdoor\_attack\_in\_WFL}.
} 

\begin{table*}[htb]
\centering
\normalsize
\caption{Experimental settings for the attacks}
\label{aes}
\renewcommand{\arraystretch}{1.2}
\scalebox{0.88}{
\begin{tabular}{l|l|l|l|l|l|l}
\hline
\hline
\multicolumn{1}{c|}{\textbf{Attack strategies}} &
  \multicolumn{1}{c|}{\textbf{Attack type}} &
  \multicolumn{1}{c|}{\textbf{Attack phase}} &
  \multicolumn{1}{c|}{\textbf{\begin{tabular}[c]{@{}c@{}}No. of \\ participants\end{tabular}}} &
  \multicolumn{1}{c|}{\textbf{CCR}} &
  \multicolumn{1}{c|}{\textbf{\begin{tabular}[c]{@{}c@{}}No. of \\ training iterations\end{tabular}}} &
  \multicolumn{1}{c}{\textbf{Dataset}} \\ \hline
Label flipping based attack~\cite{zhang2020poisongan} &
  \begin{tabular}[c]{@{}l@{}}Data poisoning \\ backdoor attack\end{tabular} &
  \begin{tabular}[c]{@{}l@{}}Local data \\ collection phase\end{tabular} &
  100 &
  0.1 &
  100 &
  FMNIST \\ \hline
Coordinated trigger based attack~\cite{gong2022coordinated} &
  \begin{tabular}[c]{@{}l@{}}Data poisoning \\ backdoor attack\end{tabular} &
  \begin{tabular}[c]{@{}l@{}}Local data \\ collection phase\end{tabular} &
  100 &
  0.2 &
  100 &
  FMNIST \\ \hline
Local model replacement attack~\cite{bagdasaryan2020backdoor} &
  \begin{tabular}[c]{@{}l@{}}Model poisoning\\ backdoor attack\end{tabular} &
  \begin{tabular}[c]{@{}l@{}}Local model \\ training phase\end{tabular} &
  100 &
  0.5 &
  100 &
  FMNIST \\ \hline
Global model replacement attack ~\cite{rodriguez2022backdoor} &
  \begin{tabular}[c]{@{}l@{}}Model poisoning \\ backdoor attack\end{tabular} &
  Aggregation phase &
  100 &
  0.5 &
  100 &
  FMNIST \\ 

\hline
\hline
\end{tabular}}
\label{tab:att_set}
\end{table*}

\begin{figure}[htb]
    \centering
    \includegraphics[scale = 0.6]{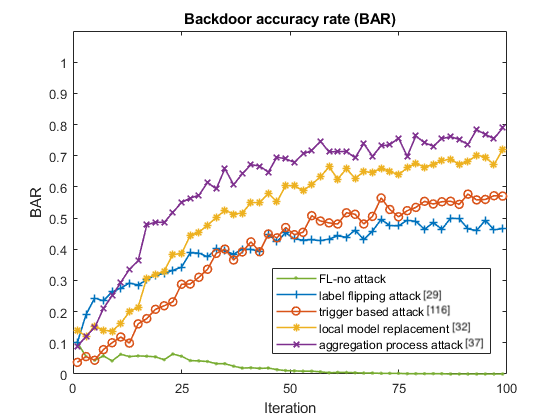}
    \caption{Backdoor accuracy rate of~\cite{zhang2020poisongan, gong2022coordinated, bagdasaryan2020backdoor, rodriguez2022backdoor}.}
    \label{fig:att_bar}
\end{figure}

\begin{figure}[htb]
    \centering
    \includegraphics[scale = 0.6]{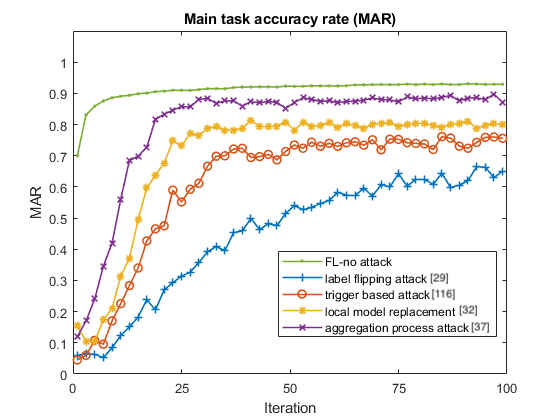}
    \caption{Global model main task accuracy rate of~\cite{zhang2020poisongan, gong2022coordinated, bagdasaryan2020backdoor, rodriguez2022backdoor}.}
    \label{fig:att_mar}
\end{figure}

{\color{blue}
The evaluation metrics used in this case are BAR and MAR, where BAR denotes the probability that the global model is misled to classify backdoor inputs into the target label, and MAR evaluates the prediction accuracy of the backdoor-injected global model on clean inputs. Generally, a higher BAR means the backdoor effect is better performed. In contrast, for backdoor attacks on WFL, the MAR  has been commonly used to evaluate the stealthiness of the injected backdoor pattern. The results are shown in Figs.~\ref{fig:att_bar} and~\ref{fig:att_mar}, and Table~\ref{tab:att_per}.
}

{\color{blue}
The experimental results further verify the conclusions in Sections III and IV. It can be seen that the global model replacement backdoor attack~\cite{rodriguez2022backdoor} launched during the aggregation process achieves the highest BAR while also maintaining a reasonable MAR. However, the attacker must obtain sufficient knowledge of the whole WFL network, introducing a considerable attack cost, as discussed in Section~\ref{sec:3c}. On the other hand, the local model replacement backdoor attack during the local model training phase shows the second-best performance. In general, the model poisoning-based backdoor attacks outperform data poisoning-based backdoor attacks while yielding a higher complexity. The low BAR of trigger-based backdoor attacks could be due to insufficient CCR. Since the completed trigger pattern is distributed in multiple compromised clients, low level of malicious participation leads to low BAR. The performance of the label-flipping-based backdoor attacks depends on the selection of the target label. Based on~\cite{wang2020attack}, this type of backdoor attacks is anticipated to perform better when targets are on the edge label. Consequently, it is more unlikely to be detected, and the attack strength is relatively weaker.
}

\begin{table}[htb]
\centering
\normalsize
\caption{Backdoor attack performance}
\label{bap}
\renewcommand{\arraystretch}{1.3}
\scalebox{0.9}{
\begin{tabular}{l|l|l}
\hline
\hline
\multicolumn{1}{c|}{\textbf{Attack strategies}} & \textbf{BAR(\%)} & \textbf{MAR(\%)} \\ \hline
Label flipping based attack~\cite{zhang2020poisongan}               & 46.6             & 65.03        \\ \hline
Coordinated trigger based attack~\cite{gong2022coordinated}        & 57.06            & 75.57        \\ \hline
Local model replacement attack~\cite{bagdasaryan2020backdoor}          & 71.96            & 88.91        \\ \hline
Global model replacement attack~\cite{rodriguez2022backdoor}           & 79.86            & 91.68        \\ \hline
\hline
\textbf{Original WFL-no attack}                             & 0                & 92.92        \\ 

\hline
\hline
\end{tabular}}
\label{tab:att_per}
\end{table}

{\color{blue}
To further evaluate the longevity of different attack strategies, the lifespan of the injected backdoor pattern is considered. With the backdoored global model obtained in the previous step, they are further trained with the clean dataset and honest participants for $50$ more iterations, and the change of MAR under each attack method is recorded in Table~\ref{tab:att_life}. It is observed that the backdoor effect of data poisoning-based backdoor attacks is quickly diluted as the attacker can only manipulate the training dataset in the local client, so the attack strength is limited. In contrast, model poisoning-based backdoor attacks can inject relatively longer-lasting backdoor effects into the global model. Specifically, the global model replacement proposed in~\cite{rodriguez2022backdoor} is conducted when the global model is about to converge. The backdoor pattern is more likely to remain in the global model for a longer time.
}

\begin{table*}[htb]
\centering
\normalsize
\caption{Lifespan comparison of attacks}
\label{lc}
\renewcommand{\arraystretch}{1.3}
\scalebox{0.95}{
\begin{tabular}{l|lll}
\hline
\hline
\multicolumn{1}{c|}{\multirow{2}{*}{\textbf{Attack strategies}}} & \multicolumn{3}{c}{\textbf{BAR}}
\\ \cline{2-4} 

\multicolumn{1}{c|}{} & \multicolumn{1}{l|}{\textbf{after 0 iteration}} & \multicolumn{1}{l|}{\textbf{after 10 iterations}} & \textbf{after 50 iterations} \\ \hline

Label flipping based attack~\cite{zhang2020poisongan}                              & \multicolumn{1}{l|}{46.6}  & \multicolumn{1}{l|}{15.04} & 0     \\ \hline

Coordinated trigger based attack~\cite{gong2022coordinated}                        & \multicolumn{1}{l|}{57.06} & \multicolumn{1}{l|}{22.82} & 0     \\ \hline

Local model replacement attack~\cite{bagdasaryan2020backdoor}                          & \multicolumn{1}{l|}{71.96} & \multicolumn{1}{l|}{43.17} & 12.95 \\ \hline

Global model replacement attack~\cite{rodriguez2022backdoor}                          & \multicolumn{1}{l|}{79.86} & \multicolumn{1}{l|}{64.88} & 49.51 \\ 

\hline
\hline
\end{tabular}}
\label{tab:att_life}
\end{table*}

\subsection{Comparison Among Backdoor Attack Defense Methods}

\begin{figure}
    \centering
    \includegraphics[scale=0.6]{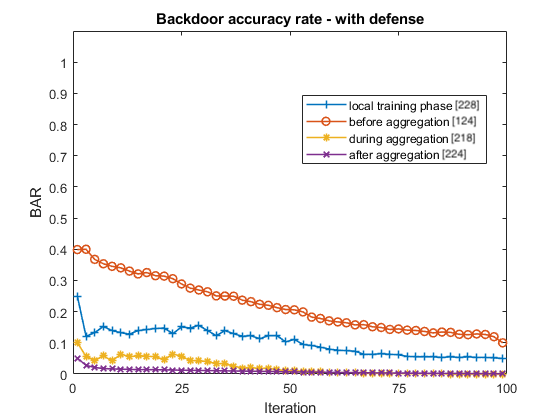}
    \caption{Backdoor accuracy rate with defense methods~\cite{zhao2021stability, sun2019can, fung2020limitations, liu2021federaser}.}
    \label{fig:def_bar}
\end{figure}

\begin{figure}
    \centering
    \includegraphics[scale=0.6]{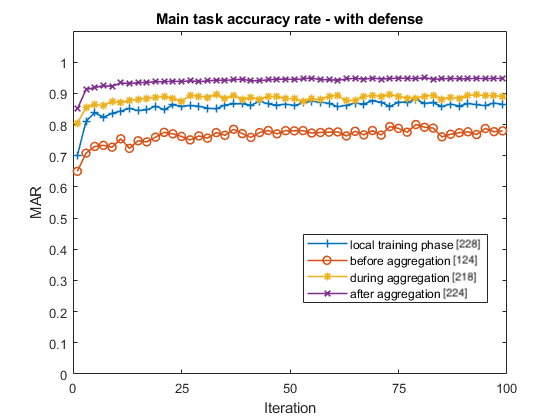}
    \caption{Main task accuracy rate with defense methods~\cite{zhao2021stability, sun2019can, fung2020limitations, liu2021federaser}.}
    \label{fig:def_mar}
\end{figure}

{\color{blue}
In this section, we present representative experimental results to evaluate and compare the performances of existing defense methodologies for backdoor attacks on WFL. The defense methods can be categorized into four phases, as shown in Fig. \ref{fig:Defense_methods}. In each phase, one featured defense method is selected to defend against backdoor attacks of different types, as illustrated in Section \ref{sec:5A}. The settings of each defense method are shown in Table~\ref{tab:def_set}, where $M$ denotes the norm threshold of the norm clipping, and $\sigma$ represents the Gaussian noise parameter. The same FMNIST dataset is adopted to train the global model while the settings of each attack method remain consistent. The performance of each defense method is illustrated in Figs.~\ref{fig:def_bar} and~\ref{fig:def_mar}, and Table~\ref{tab:def_per}.
}

\begin{table*}[htb]
\centering
\normalsize
\caption{Experimental settings for the defense methods}
\label{des}
\renewcommand{\arraystretch}{1}
\scalebox{0.87}{
\begin{tabular}{l|l|l|l|l|l|l}
\hline
\hline
\multicolumn{1}{c}{\textbf{Defense type}}&
\multicolumn{1}{|c|}{\textbf{Defense phase}} &
\multicolumn{1}{c|}{\textbf{Defense target}} &
  \textbf{\begin{tabular}[c]{@{}l@{}}Aggregation   \\ Protocol\end{tabular}} &
  \textbf{$M$} &
  \textbf{$\sigma$} &
  \multicolumn{1}{c}{\textbf{Dataset}} \\ \hline
Model stability based method~\cite{zhao2021stability} &
  Local training phase &
  \begin{tabular}[c]{@{}l@{}}Data poisoning \\ backdoor attack\end{tabular} &
  FedAvg &
  0 &
  0 &
  FMNIST \\ \hline
Norm clipping and DP based method ~\cite{sun2019can} &
  Local training phase &
  \begin{tabular}[c]{@{}l@{}}Data poisoning \\ backdoor attack\end{tabular} &
  Normclip &
  1 &
  0.0025 &
  FMNIST \\ \hline
Model similarity-based method ~\cite{fung2020limitations} &
  During aggregation phase &
  \begin{tabular}[c]{@{}l@{}}Model poisoning \\ backdoor attack\end{tabular} &
  FoolsGold &
  0 &
  0 &
  FMNIST \\ \hline
WFL with forgetting mechanism ~\cite{liu2021federaser}&
  After aggregation phase &
  \begin{tabular}[c]{@{}l@{}}Model poisoning \\ backdoor attack\end{tabular} &
  RLR &
  0.5/1 &
  0.0001 &
  FMNIST \\ 
  
\hline
\hline

\end{tabular}}
\label{tab:def_set}
\end{table*}

\begin{table}[htb]
\centering
\normalsize
\caption{Defense performance}
\label{df}
\renewcommand{\arraystretch}{1.3}
\scalebox{0.78}{
\begin{tabular}{l|l|l}
\hline
\hline
\multicolumn{1}{c|}{\textbf{Defense type}} &
  \multicolumn{1}{c|}{\textbf{\begin{tabular}[c]{@{}c@{}}BAR \\ with defense\end{tabular}}} &
  \multicolumn{1}{c}{\textbf{\begin{tabular}[c]{@{}c@{}}MAR \\ with defense\end{tabular}}} \\ \hline
Model stability based method~\cite{zhao2021stability}      & 9.72 & 78.17 \\ \hline
Norm clipping and DP based method~\cite{sun2019can} & 6.59 & 86.12 \\ \hline
Model similarity-based method~\cite{fung2020limitations}     & 1.12 & 89.53 \\ \hline
WFL with forgetting mechanism ~\cite{liu2021federaser}    & 0.24 & 94.76 \\ 

\hline
\hline

\end{tabular}}
\label{tab:def_per}
\end{table}
{\color{blue}
It is noted that all defense methods in different phases can reduce backdoor effects while maintaining a steady MAR. Specifically, the defense methods equipped during and after the aggregation phase have the best performance against model poisoning attacks. It is reasonable to assume that the defense mechanisms working on the server side are more powerful than those on the client side. However, such defense methods~\cite {fung2020limitations, liu2021federaser} require the historical information of local model updates and are not compatible with SecAgg, which brings more difficulty in the practical implementation. It is also worth mentioning that the unsatisfactory performance of the norm clipping-based defense methods before the aggregation phase heavily depends on the improper selection of the threshold $M$ and noise parameter $\sigma$. As pointed out in~\cite{sun2019can}, the norm clipping-based methods inevitably introduce overkill or overfitting problems. Meanwhile, the Gaussian noise introduced by differential privacy also adds extra disturbance to the global model training process. Thus, it is essential to properly design the defense algorithm's thresholds.
}

{\color{blue}
\subsection{Summary}
Experiments have been conducted to evaluate and compare the strength of backdoor attacks in different types and the performance of defense methodologies in different phases. The experimental results verify the discussions in Section \ref{sec::Existing} and \ref{sec::Defense}. It can be seen that the model poisoning-based backdoor attacks show higher BAR when compared to data poisoning-based backdoor attacks. However, the model poisoning-based backdoor attack algorithms have relatively higher complexity and normally require much more information about the entire network. Many existing defense methodologies have been proven to be capable of defending both data poisoning-based backdoor attacks and model poisoning-based backdoor attacks. Specifically, the defense mechanisms working on the local training phase and before the aggregation phase can address data poisoning-based backdoor attacks. The defense mechanisms deployed during and after the aggregation phase can handle model poisoning-based backdoor attacks well. Although many powerful backdoor attack strategies, such as global model replacement [143], can already be settled, the major dilemma in developing a more convenient defense methodology is how to better defend against all types of backdoor attacks while ensuring the privacy and security of participants (i.e., to be compatible with SecAgg). In the following sections, the lessons learned from existing works are concluded, and further potential research directions are also discussed.
}

\section{Lessons Learned}
\label{sec::lessons}
In the previous sections, the working mechanisms of existing backdoor attack designs and the defense methods have been discussed in detail. The limitations of each of the attack and defense methods have been discussed. The major findings from the existing works are summarized in this section.

\subsection{Backdoor Attack Methods}
In this section, the lessons learned from the existing backdoor attack strategies on WFL are analyzed. 

\subsubsection{Data poisoning-based backdoor attacks} 
Earlier works on data-based backdoor attacks focused on un-targeted attacks that aimed to degrade the accuracy of the global model by injecting ``poisoned'' data into the training set~\cite{saha2022backdoor,shejwalkar2022back}. A simple label-flipping method was often used to replace the targeted label~\cite{rosenfeld2020certified}. These types of attacks do not require any prior knowledge about the learning model or mechanism. However, the effect of the injected backdoor tends to be diluted over multiple training rounds. Some studies have shown that the attack success rate (ASR) of label-flipping attacks can be increased when the attacker controls more compromised clients~\cite{tahmasebian2022robustfed,xiao2022sca}. In contrast, other researchers have focused on injecting the poisoned data into the ``tail'', or labels with relatively smaller weights~\cite{wang2020attack}. This type of ``tail backdoor'' can remain undetected for a longer duration but only affects the classification of a subset of the training samples. There is a trade-off between the ASR and the strength of the backdoor effect in this case.

In addition to degrading the performance of the global model, there has also been a growing interest in targeted backdoor attacks for WFL. These attacks aim to have the global model learn a trigger pattern embedded in a poisoned data sample, which can either be imperceptible (e.g., a watermark) or perceptible but innocuous (e.g., glasses on a face). The goal of a targeted backdoor attack is to have the global model classify the poisoned sample into the target class without affecting the accuracy of the primary task. The trigger can be randomly generated or specifically designed for the target label~\cite{wang2020attack,xie2019dba,zhang2020poisongan}. Depending on the number of compromised clients controlled by the attacker, targeted backdoor attacks can be further divided into two types: centralized attacks~\cite{bhagoji2019analyzing} and coordinated attacks~\cite{gong2022coordinated,xie2019dba}.

One advantage of data-based backdoor attacks is that the attacker does not need to have access to the entire training dataset or knowledge of the training process and mechanism. Additionally, the ASR of these attacks can be improved by increasing the number of compromised clients controlled by the attacker. However, because these attacks only target the training data, their success depends on the sufficient number of attack rounds to avoid the dilution of the backdoor effect~\cite{awan2021contra}.

\subsubsection{Model poisoning-based backdoor attacks} 
Model-based attacks, on the other hand, target the training process of the local models. In these attacks, the attacker injects a poisoned data sample into the training process along with benign samples and designs a loss function to minimize the difference between the poisoned model and the benign model~\cite{nguyen2020poisoning,bagdasaryan2020backdoor,zhang2022neurotoxin}. This allows the malicious model to be aggregated into the global model without detection. Research has shown that these attacks can achieve a high ASR in edge cases. Another approach to model-based backdoor attacks is to adjust the weight of the poisoned local model to a sufficiently large value before aggregation, ensuring that the backdoor is injected into the global model~\cite{bagdasaryan2020backdoor}. The optimal weight must be carefully designed to achieve the highest ASR without detection.

Compared with a backdoor attack for data, a model-based backdoor attack shows better ASR and longer duration~\cite{xie2021crfl}. In return, the attack cost of model-based backdoor attacks is relatively higher. The attacker needs to gain complete information about the training data set and the training mechanism. The attack exhibits better performance when the global model is approaching convergence. Thus, the training process and progress are also necessary for the attacker~\cite{li2022backdoor}.

\subsubsection{Insights}
Data poisoning happens during the local data collection phase, while model poisoning happens afterward. Compared to data poisoning with model poisoning, data poisoning backdoor attack shows a relatively lower attack cost, and model poisoning attack has a better attack accuracy rate and longer backdoor life span. Within the data poisoning attack, label flipping is one of the easiest attack methods, resulting in a lower success rate, and can be easily detected. A trigger-based backdoor is the most adopted one, and with the proper design trigger pattern, the attack performance will be greatly improved. 

Once the attack obtains the ability to manipulate the training algorithm or aggregation protocol, one can conduct a model poisoning attack. The compromised client generated the backdoor inject local updates via ether label flipping or trigger pattern. With a higher knowledge level known by the attack, the attack success rate is higher as well. In return, the attack cost will also dramatically increase.

\subsection{Backdoor Defense}

There are many defense mechanisms proposed to detect and defend the backdoor attack. Based on the complexity and different defense phases, which can be categorized into four types as shown in  Fig. \ref{fig:Defense_methods}. 

The defender usually clips the contribution from suspicious clients to get rid of the backdoor effect. In this process, some honest information will be miss-clipped, and the main task accuracy will be influenced. Meanwhile, some defense methods attempt to train a unique classifier that can distinguish the adversarial input. Such a newly trained classifier requires information on all possible backdoor patterns, and the defense cost is high. Third, the WFL framework is featured for its non-IID property. The defender is hard to trace back to the compromised client and leaves it out of the network. And there is no proven solution to deal with the injected backdoor pattern after the aggregation phase.

It is also worth mentioning that there is a common trade-off among all existing defense methodologies: the defense success rate and the convergence rate of the global model. Since all defense methods inherently eliminate the suspicious backdoor effect by clipping the possible malicious updates, inevitably, a portion of honest updates will be neglected by mistake. Thus, it's always a top priority in designing a defense mechanism: the backdoor effect should be appropriately removed, and the global model can quickly converge at the same time.

\section{Challenges, Open Problems, and Future Research Directions}
\label{sec::issues}

It is foreseeable that Wireless Federated Learning (WFL) techniques will play a pivotal role as one of the core components in the development of next-generation wireless communication networks, including Beyond 5G (B5G) \cite{wan2022privacy} and 6G \cite{lu2020low}. To deliver intelligent network operation and efficient resource management with practical WFL applications in the wireless communication network, it is of great significance to develop a WFL framework that is secure and robust to various kinds of backdoor attacks. Thus, the study of backdoor attacks and defense mechanisms in WFL becomes essential. In recent years, this area has been a hot research topic in both academics and industries.

Many research efforts have been devoted to the development of both backdoor attack methodologies and backdoor attack defense mechanisms. The objective of backdoor attacks is to dig deeper into the information flow process in the WFL network and amplify the impact of any potential insecure vulnerabilities within the WFL framework. In terms of backdoor attack designs, researchers focus on how to improve the attack success rate. These efforts aim at emphasizing the severity of the problem. More research attention can then be addressed to design the targeted defense mechanism. The ultimate goal of defense mechanism design in WFL is the ability to better deal with the actual malicious behaviors that occur in practical applications. Via a secure and robust WFL network, many WFL applications can be deployed with confidence.  

\begin{figure*}[htb]
\centering
\includegraphics[scale=0.35]{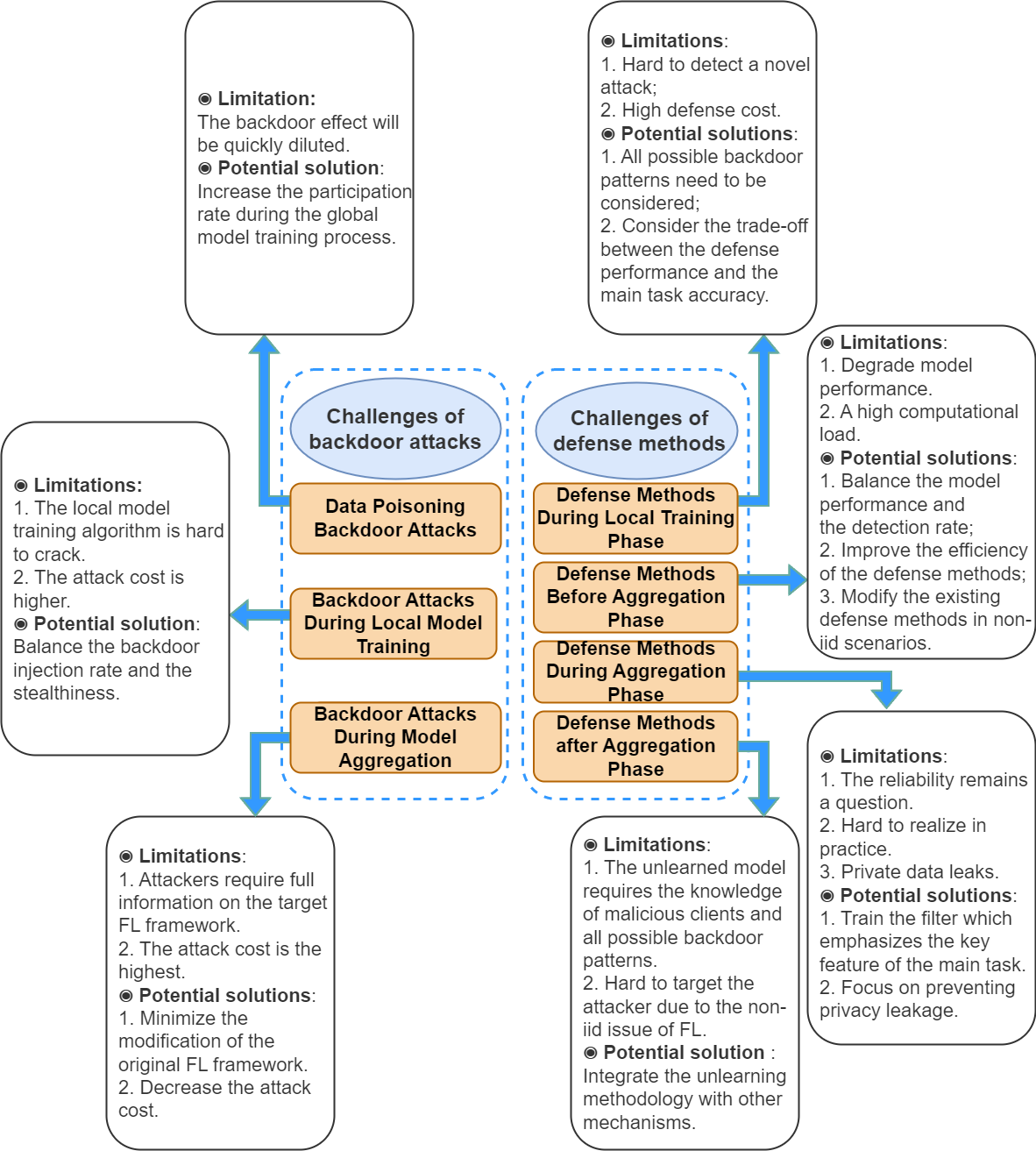}
\centering 
\caption{A summary of challenges and future research directions for backdoor attacks on WFL.}
 \label{fig:challenges}
\end{figure*}

In the previous sections, the review of existing breakthroughs in backdoor attack strategies and defense methods has been presented, based on which many critical challenges and open issues in WFL have been revealed. In this section, these challenges and open issues are analyzed from the viewpoints of both backdoor attack methods and backdoor attack defense mechanisms. The potential solutions are discussed afterward. The discussions are summarized in Fig. \ref{fig:challenges}.
    
\subsection{Backdoor Attacks on WFL}

As reviewed in Section \ref{sec::Existing}, backdoor attacks on WFL can be classified into three categories: those that occur during the local data collection phase, the local model training phase, and the server aggregation phase. These attacks are designed based on the attackers' prior knowledge, and each has its own limitations. The open issues related to these attack methods are discussed in this section.


\subsubsection{How to decrease the data-dependency of data poisoning backdoor attacks} 
The backdoor attack during the local data collection phase requires the attacker to control or have access to part of the local training data set~\cite{furth2022fair}. In order to guarantee both the attack success rate and the local model convergence, the backdoor pattern injected in the poisoned dataset shows dependency on the benign dataset, i.e., the poisoned dataset cannot have a huge difference from the honest ones. It has been proved that in the edge-case (or low-possibility sample), the backdoor-injected poisoned dataset can be irrelevant to the label description. For example, in the image classification application, the images of ``Southwest airplane'' are fed to the label of ``truck''~\cite{wang2020attack,yang2022security}. Such implementation does decrease the dependency on the backdoored input design. The attack strength is also limited. A more proper way of crafting the poisoned dataset to perform a data-agnostic attack remains an open discussion.

\subsubsection{How to improve the success rate of data poisoning backdoor attacks} 
Backdoor attacks during the local data collection phase target specific parts of the local training data~\cite{schwarzschild2021just,goldblum2022dataset,truong2020systematic,li2022backdoor}. These attacks are conducted using a black model and are relatively low-cost. However, current designs for these attacks assume the ratio of compromised clients in each round to be high, which is unrealistic in practical WFL deployments where the number of participants is typically quite large. Consequently, the ASR is low, and the backdoor effect is quickly diluted. To increase the ASR and effect at this stage, attackers need to find a way to increase their participation rate during the global model training process, which would increase the attack efficiency and success rate.


Challenges of backdoor attacks during the local data collection phase can be classified into two types: label-flipping-based and trigger-based attacks. Label-flipping-based attacks have low attack costs, but the flipped labels can cause noticeable outliers in the uploaded updates, making them easy to detect and eliminate~\cite{rosenfeld2020certified}. To achieve a better attack effect, attackers must consider the trade-off between injection rate and stealthiness. Trigger-based attacks also create outliers, but these can be designed based on the target label to decrease the likelihood of detection. For example, if the target label is a picture of human faces, the trigger could be designed as glass. This targeted trigger design can decrease the probability of detection by the defender.


\subsubsection{How to improve the backdoor attack effect during local model training}  
Backdoor attacks during the local model training phase require attackers to have complete control over compromised clients~\cite{li2022backdoor,hou2021mitigating,li2022deep}. These attacks involve modifying the training process to inject a backdoor pattern, which can be stealthier and more accurate, but also more expensive to carry out than attacks conducted during the local data collection phase. The local model training process is difficult to tamper with, and the effects of a backdoor can be quickly diluted if compromised clients are not frequently selected. To increase the effectiveness of a local training phase attack, it is important to find a balance between injecting the backdoor and remaining stealthy. Even if an attacker is able to get the local backdoor pattern uploaded to the server aggregator, the defense mechanism on the server may still be able to detect and eliminate the backdoor, preventing a global injection~\cite{goldblum2022dataset}.


\subsubsection{How to decrease the backdoor attack cost during model aggregation}  
Backdoor attacks during the model aggregation phase have the highest success rate but also the highest cost~\cite{ozdayi2021defending}. To carry out these attacks, attackers need to have complete knowledge about the target WFL framework, including the number of participants, local training algorithm, aggregation protocol, and global model convergence process~\cite{goldblum2022dataset,xie2022defending}. In practice, it can be difficult for attackers to obtain this information, and the attack may not be easily replicated in a different WFL task due to the unique architecture of each task. To make the attack more feasible, attackers may try to minimize modifications to the original WFL framework and reduce the cost of the attack.

\subsubsection{How to organize a coordinated backdoor attack in a more effective way}  
Coordinated-trigger-based data poisoning attack can be launched during the local data collection phase by multiple attackers, with each attack member only knowing part of the trigger pattern. Compared with the centralized-trigger-based attack, the injected trigger exhibits better stealthiness, making it harder to be fully defended. However, such a coordinated attack is still based on the black box model, resulting in limited attack strength. There exists a possibility that multiple attackers can initialize a joint backdoor attack via a white-box model. By acquiring more information from attack members and the WFL network training process, the backdoor pattern can be better crafted. By distributing backdoor patterns in multiple compromised clients, the attackers are also more likely to trick the defender.

{\color{blue}
\subsubsection{How to ensure the trigger rate when conducting backdoor attacks on WFL in WCNs}
The trigger rate refers to the percentage of compromised clients selected in each training round. Therefore, a higher trigger rate can lead to a more severe attack. It is noted that in the WCN, the data distribution among different devices is normally imbalanced. Thus, the data received by a certain participant from the server could be much more(or less) than other participants in the same network. Such imbalanced data distribution will have a significant impact on model update calculation and upload. If the attacker could make use of such imbalances and intentionally control compromised clients to acquire more data from the WCN, it is reasonable to assume that the trigger rate will also increase. Consequently, the attack strength could be further improved.  

}


\subsection{Defense Methods in WFL}

Defending against backdoor attacks on WFL can be divided into four categories based on the different training phases. These defense strategies may require different levels of knowledge about the model. In this section, we discuss the open issues of these defense methodologies and potential solutions.

\subsubsection{How to achieve reasonable defense performance during the local training phase}  
The existing defense methods for the local training phase involve using a clean model trained with honest data as a filter to detect injected backdoors~\cite{li2022deep,zhao2021stability,zhao2021stability,li2021anti,chen2022defending}. This type of defense is effective in defending against black-box backdoor attacks and can be used with a secure aggregation protocol. However, considering all possible backdoor patterns to achieve good defense performance can be costly, and the defense may not be able to detect novel attacks. Moreover, there is a trade-off between defense performance and main task accuracy when designing this type of defense mechanism. 


\subsubsection{How to balance the model performance and the detection rate before the aggregation phase} 
Defending before the aggregation phase is a common strategy in which the server aggregator detects and ignores malicious updates and restricts participation from suspicious clients~\cite{wu2022toward,ozdayi2021defending,liu2021privacy,nuding2022data}. This can significantly improve the robustness of the overall framework. However, there are still some design challenges to be addressed. Defense methods based on differential privacy (DP) can add noise to the model, which can degrade its performance. It is important to find the right balance between model performance and detection rate when adding noise. Additionally, using DP-GAN can be computationally intensive, so finding ways to improve the efficiency of these methods is an active area of research. Furthermore, the non-IID structure of WFL frameworks can make it challenging to defend against backdoor attacks while preserving the privacy and security of participants. Modifying the existing defense methods to work in non-IID scenarios is a key research direction. 


\subsubsection{How to improve the defense performances during the aggregation phase}  
The defense during the aggregation phase involves filtering out poisoned updates after they have been received by the aggregator~\cite{andreina2021baffle,baruch2019little,zhou2021robust}. Some defense methodologies at this stage use historical information to train a filter that can identify the backdoor in the original global model. However, the training process of WFL is random, which raises questions about the reliability of historic values. Additionally, the defense requires the attacker to participate in every training round, which may not be practical. To improve the defense performance, one solution could be to train a filter that emphasizes the key features of the main task, which may increase the filter's reliability. Privacy is also a major concern at this stage, and future research should focus on preventing privacy leakage. 


\subsubsection{How to remove the injected backdoor effect after the aggregation phase}  
After the aggregation phase, defense mechanisms can be employed to address any remaining injected backdoors in the global model~\cite{liu2021removing,aiken2021neural,hayase2021defense,chen2021mitigating,xiang2021reverse}. One such approach is the use of an ``unlearning'' model, which aims to eliminate the residual effects of the backdoor by excluding the influence of suspicious participants and has been shown to be effective~\cite{sommer2022athena,ma2022learn,hu2022membership}. However, this approach requires the knowledge of the malicious clients and all possible backdoor patterns, which can be difficult to obtain in a WFL setting where the data is non-IID. It may be beneficial to combine the unlearning methodology with another mechanism to help identify malicious clients and remove backdoors while preserving client privacy.

\subsubsection{How to optimize the processing load of the deployed defender}
WFL is featured for its large number of participating devices and different types of devices, and devices might have limited computation abilities. While for the defenders that are deployed at the client end, the processing capability becomes an inevitable problem~\cite{guo2022overview,wei2023lightweight}. It is essential to find a way to implement the defender, which can detect the anomaly input while keeping the local training process smooth. On the other hand, the defense mechanisms on the remote side also cause computing burdens to the server. The trade-off between processing efficiency and defense performance deserves further investigation. 

\subsubsection{How to defend backdoor attacks on decentralized WFL}  
In decentralized WFL, each node acts as both a contributor and a server. The model difference among nodes within the decentralized network could be significant. The existing similarity-based norm clipping methods may not function properly. On the other hand, there is no central server in the decentralized WFL framework. Each model can only obtain part of the network information from its neighbors or selected remote nodes. It is hard to broadcast a global defender to monitor all the nodes. Backdoor patterns are easier to parasite within a network and gradually propagate and spread throughout the network. A proper way to detect backdoor effects in decentralized WFL is a promising research direction.
{\color{blue}
\subsubsection{How to compensate the transmission delay in the global model training process}  
During the training process of WFL in the WCN, it is challenging to ensure synchronization and consistency among all devices. The transmission delay or package loss during the wireless communication can cause uploaded local model updates to drop and further influence the global model convergence. Moreover, the attacker can take advantage of such unfairness and increase the attack success rate. It is of great significance to develop a defense mechanism that is capable of re-constructing the dropped transmission message and, at the same time, monitoring over-active clients to further increase the robustness of the entire network.

}


\section{Conclusion}
\label{sec::summary}

Owing to the inherent features of wireless networks and systems such as a large number of participants, massive data, and geographically distributed deployment, WFL is a rapidly growing technology with potential applications for various intelligent computing tasks in WCNs. However, as WFL advances, so do the techniques for launching backdoor attacks, posing a threat to the robustness of WFL.  This survey paper has thoroughly examined the existing attack strategies and defense mechanisms to safeguard against all types of attacks in WFL. The backdoor attack methods have been systematically classified into two main categories: data poisoning attacks and model poisoning attacks, with the latter further divided into local training phase attacks and aggregation phase attacks. The defense methods have been categorized into four types based on their application stages: during local model training, before aggregation, during aggregation, and after aggregation. The key characteristics of existing backdoor attack and defense methods in WFL over WCN have been highlighted to provide insights for future research and development efforts in this significant domain.

\begin{itemize}
    \item In backdoor attacks during the local data collection phase, also known as data poisoning-based backdoor attacks, an attacker can change only a part of the training data. Since limited prior information on the entire model is required, it contributes to a decreasing attack cost. Such a black-box model attack has a relatively weaker attack strength, and the backdoor effect is easier to detect and eliminate.
    
    \item For the model poisoning-based backdoor attacks, the attacker needs to fully control one or more participants. With more knowledge of the model, e.g., the training algorithm and aggregation protocol acquired by the attacker, the attack success rate and the backdoor stealthiness can be improved. However, the attack cost is higher compared to a data poisoning-based backdoor attack. It is worth mentioning that, in most of the existing attack methods, a typical assumption is that, malicious participants are more likely to be selected in each training round, and the proportion of clients controlled by the attacker is typically high. 
    
    \item Many defense mechanisms have been proposed to settle robust threats. Based on different defense stages, the defender can filter the malicious updates from the compromised clients controlled by the attacker, either systematically or statistically.
    
    \item In the existing defense method design, there is one common trade-off between the detection rate and the main task success rate. In fact, backdoor detection relies on threshold design in different defense mechanisms. Inevitably, some information in the genuine model will be clipped, which impacts the model's performance. Meanwhile, several defense strategies are not compatible with the security aggregation protocols. 
    
    
    \item The existing defense mechanisms are commonly designed for one or several specific backdoor attack methodologies. The training process of the defender requires historical information about the attacks. Thus, they are still vulnerable when facing a novel type of attack.
\end{itemize}

This survey has provided a clear and concise overview of the current state of backdoor attacks and defenses in WFL, and guidelines on enhancing the robustness of WFL against backdoor attacks.
%

\normalem
\bibliographystyle{IEEEtran}
\bibliography{BL_FL}

\begin{IEEEbiography}[{\includegraphics[width=1in,height=1.25in,clip,keepaspectratio]{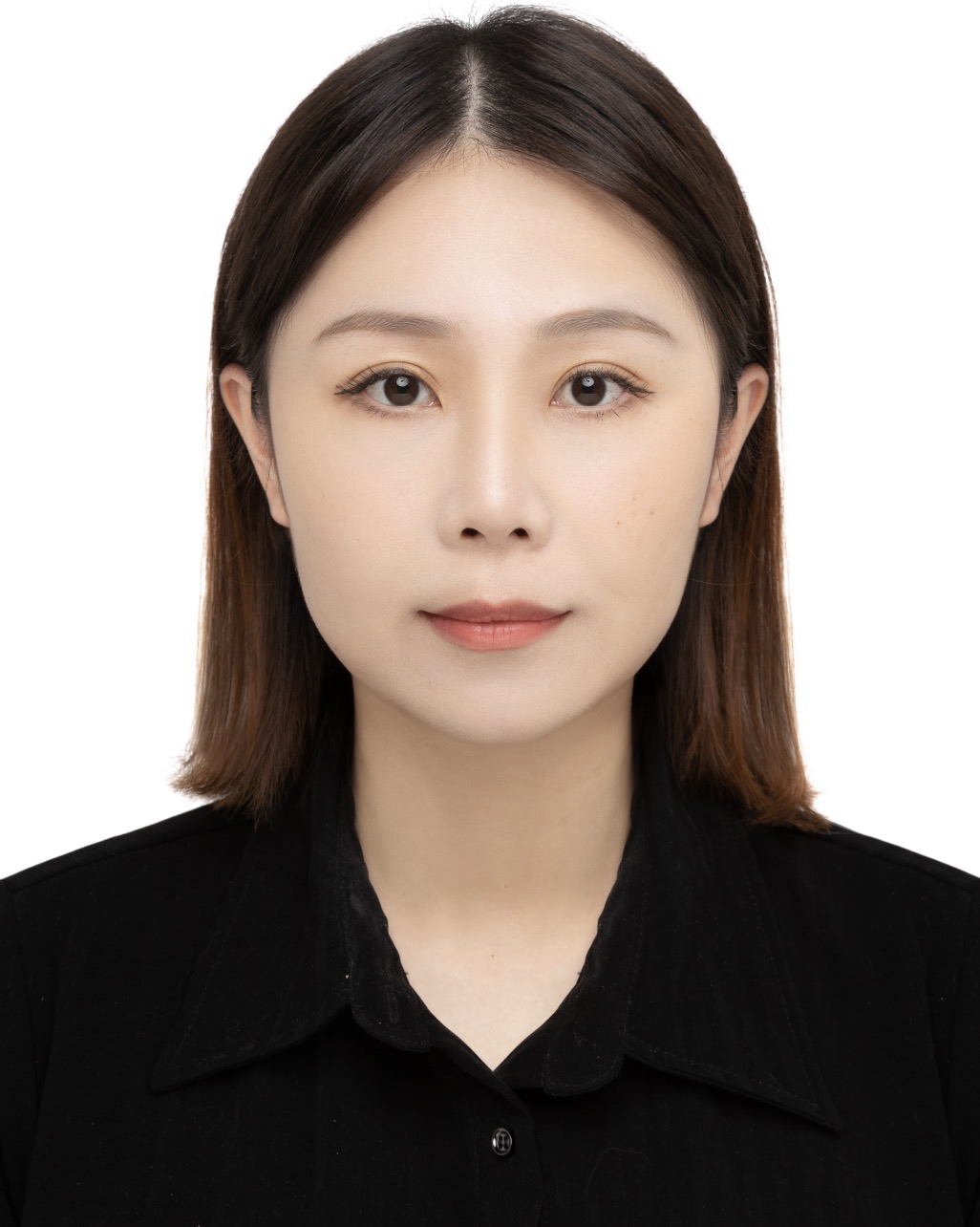}}]{Yichen Wan}is currently pursuing the Ph.D. degree in Information Technology at Deakin University. She received her B.S. degree in Information Technology at RMIT University in 2016, her M.S. degree in Networking at Melbourne Institute of Technology in 2018, and her B.S.(First-Class Hons) degree in Information Technology at Deakin University in 2021. Her research interests focus on Machine Learning, Edge Computing, the Internet of Things, and corresponding security and privacy issues.
\end{IEEEbiography}


\begin{IEEEbiography}
[{\includegraphics[width=1in,height=1.25in,clip,keepaspectratio]{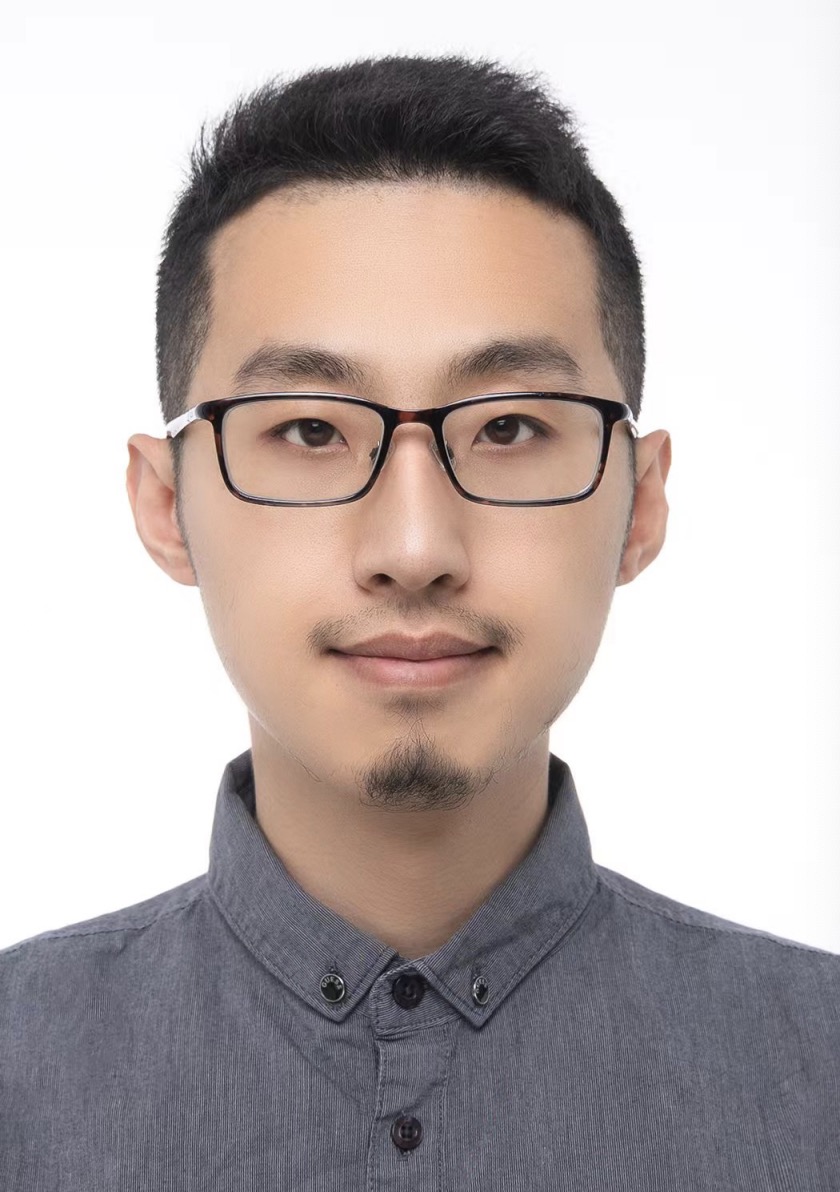}}]{Youyang Qu}is currently a research scientist of data61, Commonwealth Scientific and Industrial Research Organization (CSIRO), Australia. Before joining CSIRO, he served as a research fellow at Deakin University. He received his B.S. degree in Mechanical Automation in 2012 and M.S. degree in Software Engineering in 2015 from the Beijing Institute of Technology, respectively. He received his Ph.D. degree at the School of Information Technology, Deakin University, in 2019. His research interests focus on Machine Learning, Big Data, IoT, blockchain, and corresponding security and customizable privacy issues. He has over 50 publications, including high-quality journals and conference papers such as IEEE TII, IEEE TNSE, ACM Computing Surveys, IEEE IOTJ, etc. He is active in the research society and has served as an organizing committee member in SPDE 2020, BigSecuirty 2021, and Tridentcom 2021/2022.
\end{IEEEbiography}

\begin{IEEEbiography}[{\includegraphics[width=1in,height=1.25in,clip,keepaspectratio]{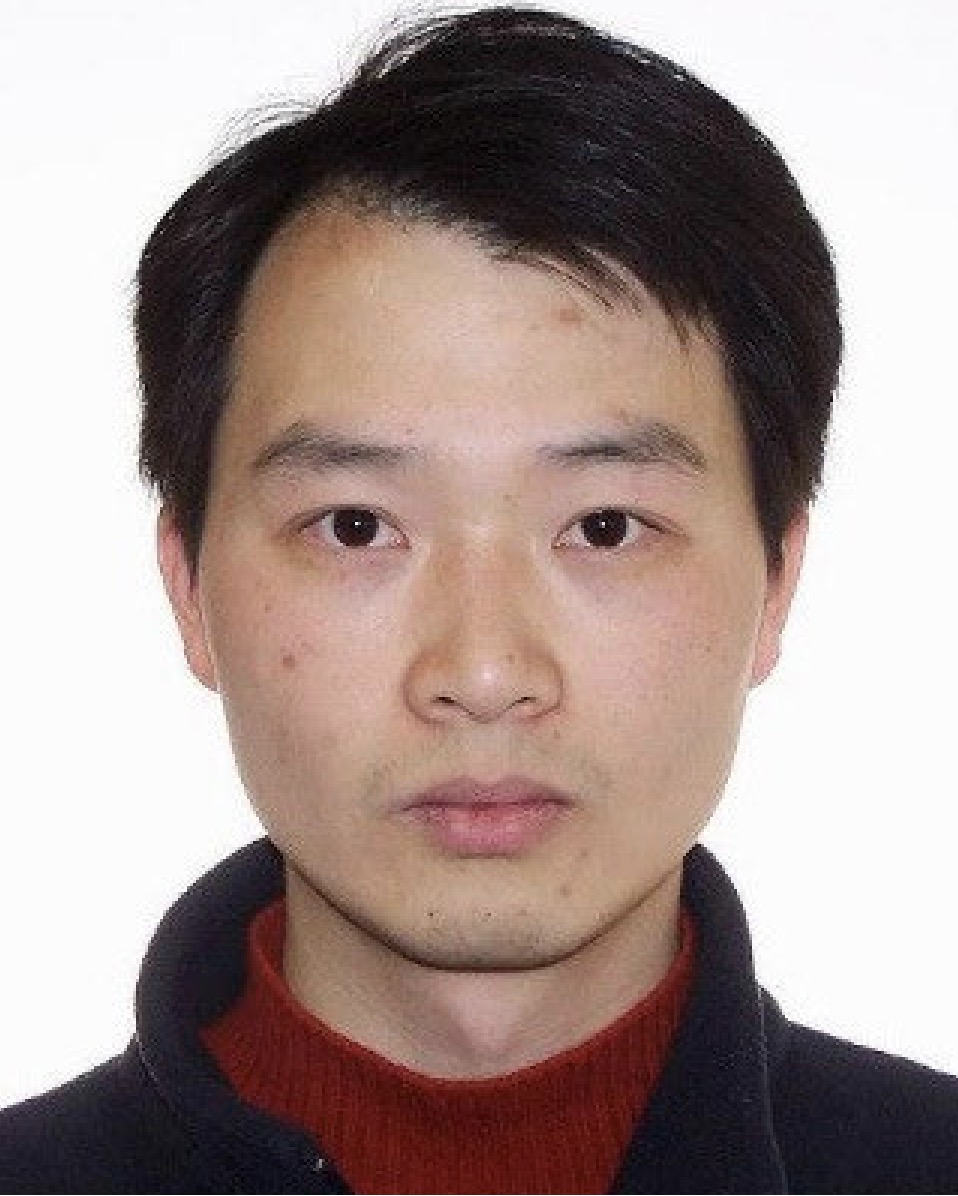}}]{Wei Ni}(Fellow, IEEE) received the B.E. and Ph.D. degrees in communication science and engineering from Fudan University, Shanghai, China, in 2000 and 2005, respectively. He was a Post-Doctoral Research Fellow at Shanghai Jiao Tong University, Shanghai, from 2005 to 2008; the Deputy Project Manager of the Bell Laboratories, Alcatel/Alcatel-Lucent, Shanghai, from 2005 to 2008; and a Senior Researcher with Devices Research and Development, Nokia, from 2008 to 2009. He is currently the Principal Research Scientist of the Commonwealth Scientific and Industrial Research Organisation (CSIRO), Sydney, NSW, Australia; a Conjoint Professor with the University of New South Wales, Sydney; an Adjunct Professor with the University of Technology Sydney, Sydney; and an Honorary Professor with Macquarie University, Sydney. He has authored seven book chapters, more than 300 journal articles, 100 conference papers, 26 patents, and ten standard proposals accepted by IEEE. His research interests include machine learning, online learning, stochastic optimization, and their applications to system efficiency and integrity. Dr. Ni has served first as the Secretary and then the Vice-Chair for IEEE New South Wales (NSW) Vehicular Technology Society (VTS) Chapter from 2015 to 2019, the Track Chair for VTC-Spring 2017, the Track Co-Chair for IEEE VTC-Spring 2016, the Publication Chair for BodyNet 2015, and the Student Travel Grant Chair for WPMC 2014. He has been the Chair of the IEEE VTS NSW Chapter since 2020, an Editor of IEEE TRANSACTIONS ON WIRELESS COMMUNICATIONS since 2018, and an Editor of IEEE TRANSACTIONS ON VEHICULAR TECHNOLOGY.
\end{IEEEbiography}

\begin{IEEEbiography}[{\includegraphics[width=1in,height=1.25in,clip,keepaspectratio]{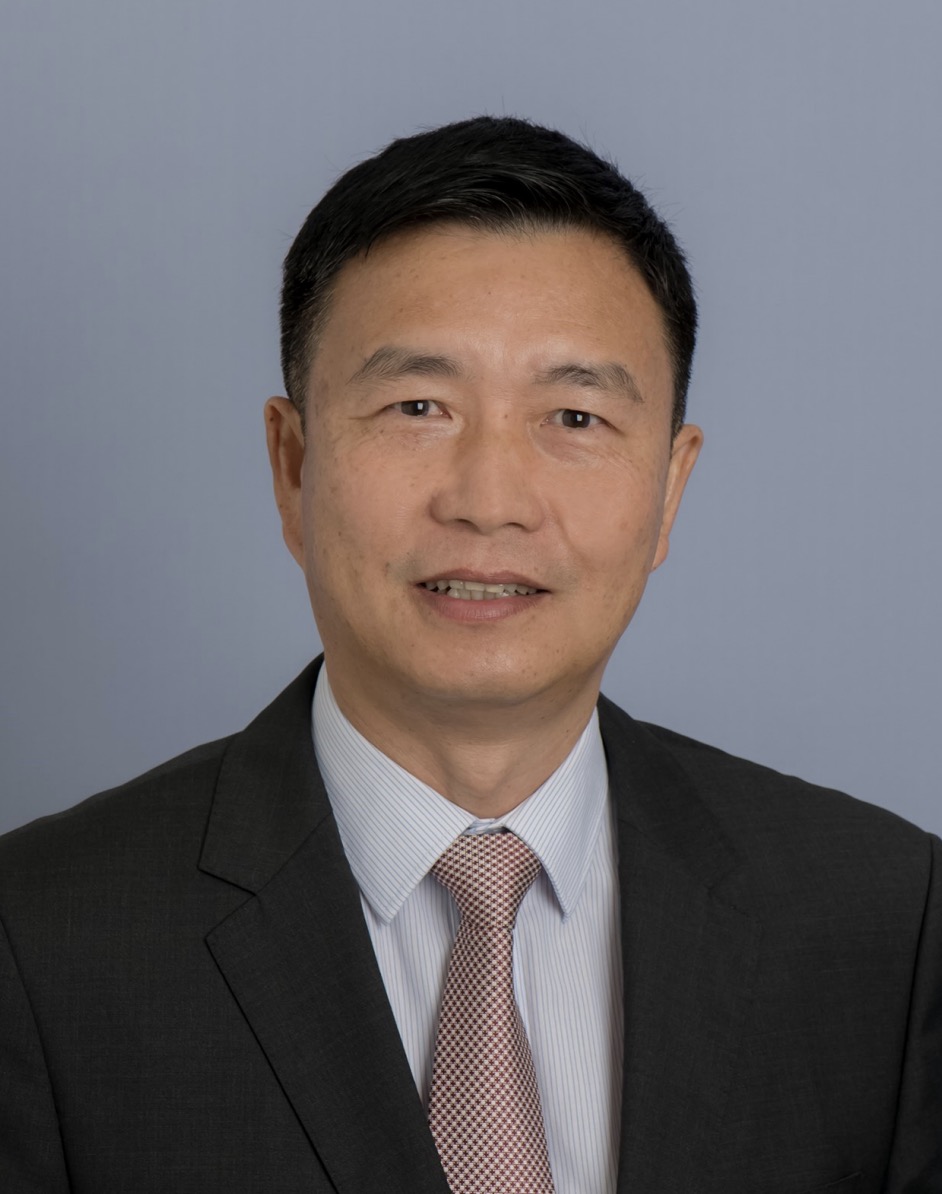}}]{Yong Xiang }(Senior Member, IEEE) received the Ph.D. degree in Electrical and Electronic Engineering from The University of Melbourne, Australia. He is a Professor at the School of Information Technology, Deakin University, Australia. His research interests include distributed computing, cybersecurity and privacy, machine learning and AI, and communications technologies. He has published 7 authored books, over 230 refereed journal articles, and over 100 conference papers in these areas. Professor Xiang is the Senior Area Editor of IEEE Signal Processing Letters, the Associate Editor of IEEE Communications Surveys and Tutorials, and the Associate Editor of Computer Standards and Interfaces. He was the Associate Editor of IEEE Signal Processing Letters and IEEE Access, and the Guest Editor of IEEE Transactions on Industrial Informatics, IEEE Multimedia, etc. He has served as Honorary Chair, General Chair, Program Chair, TPC Chair, Symposium Chair and Track Chair for many conferences, and was invited to give keynotes at numerous international conferences.
\end{IEEEbiography}

\begin{IEEEbiography}[{\includegraphics[width=1in,height=1.25in,clip,keepaspectratio]{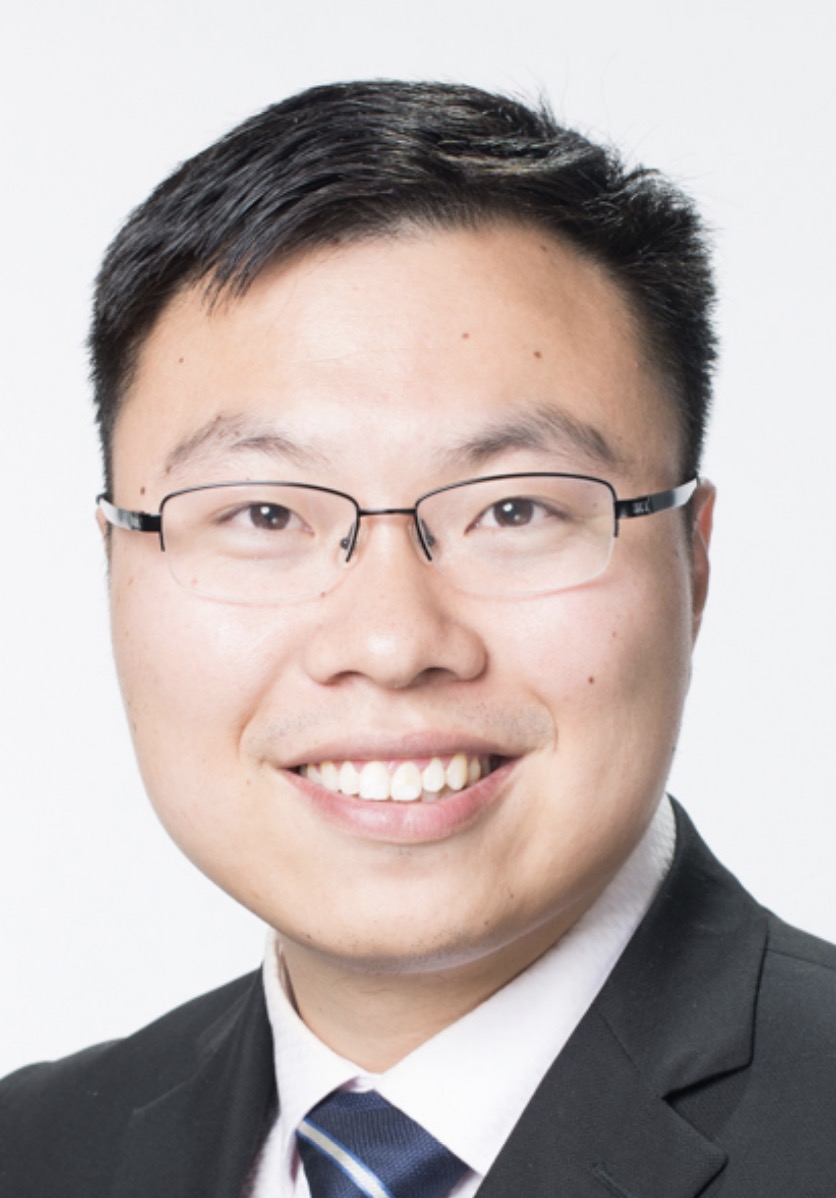}}]{Longxiang Gao }(Senior Member, IEEE) received his PhD in Computer Science from Deakin University, Australia. He is currently a Professor at Qilu University of Technology (Shandong Academy of Sciences) and Shandong Computer Science Center(National Supercomputer Center in Jinan). He was a Senior Lecturer at School of Information Technology, Deakin University and a post-doctoral research fellow at IBM Research \& Development, Australia. His research interests include Fog/Edge computing, Blockchain, data analysis and privacy protection. Dr. Gao has over 100 publications, including patent, monograph, book chapter, journal and conference papers. Some of his publications have been published in the top venue, such as IEEE TMC, IEEE TPDS, IEEE IoTJ, IEEE TDSC, IEEE TVT, IEEE TCSS, IEEE TII and IEEE TNSE. He has been Chief Investigator (CI) for more than 20 research projects (the total awarded amount is over \$5 million), from pure research project to contracted industry research. He is a Senior Member of IEEE.
\end{IEEEbiography}

\begin{IEEEbiography}[{\includegraphics[width=1in,height=1.25in,clip,keepaspectratio]{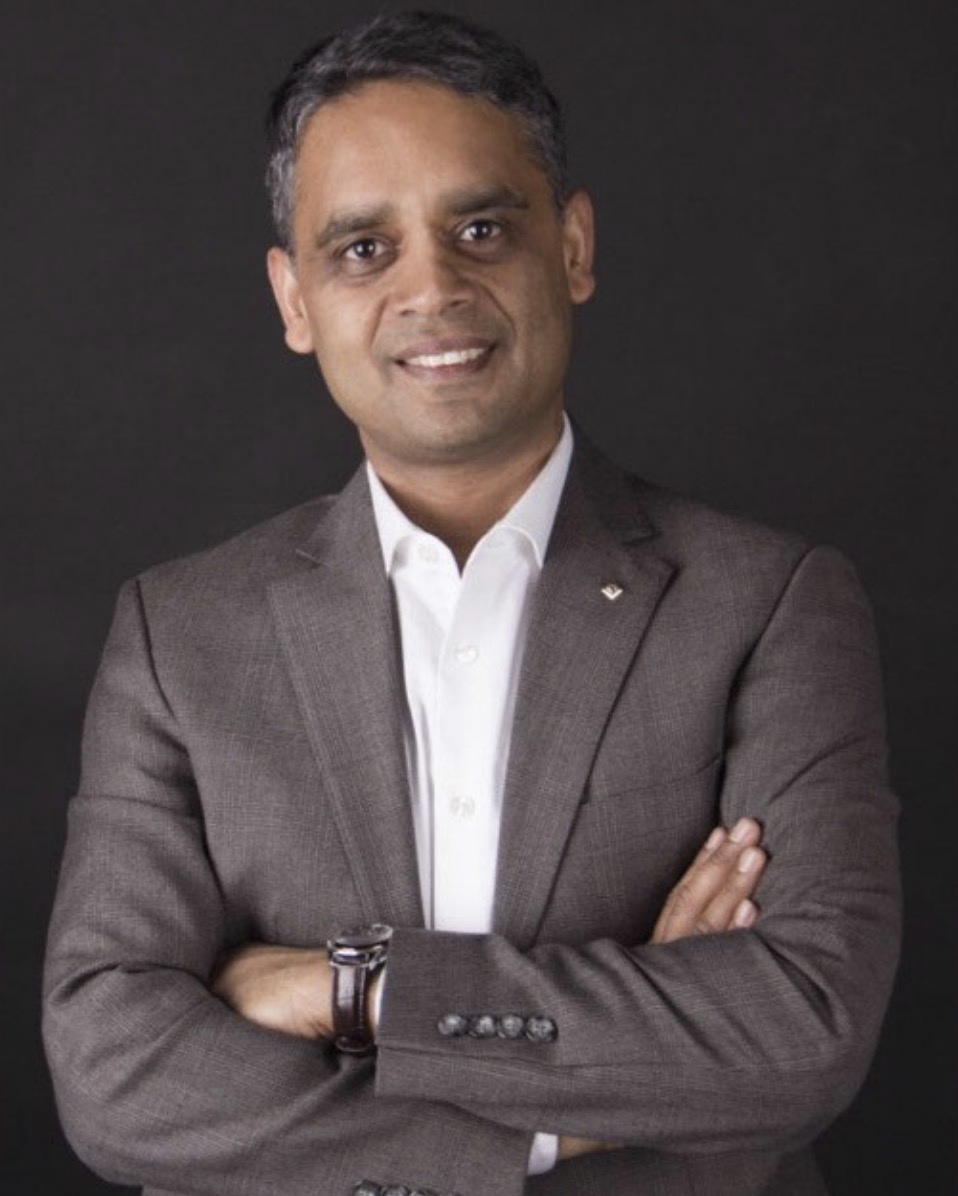}}]{Ekram Hossain}(Fellow, IEEE) is a Professor and the Associate Head (Graduate Studies) of the Department of Electrical and Computer Engineering, University of Manitoba, Canada. He is a member(Class of 2016) of the College of the Royal Society of Canada. He is also a fellow of the Canadian Academy of Engineering and the Engineering Institute of Canada. His current research interests include design, analysis, and optimization beyond 5G/6G cellular wireless networks. He was elevated to an IEEE fellow, for contributions to spectrum management and resource allocation in cognitive and cellular radio networks. He was an Elected Member of the Board of Governors of the IEEE Communications Society for the term 2018–2020. He received the 2017 IEEE ComSoc TCGCC (Technical Committee on Green Communications and Computing) Distinguished Technical Achievement Recognition Award, for outstanding technical leadership and achievement in green wireless communications and networking. He has won several research awards, including the 2017 IEEE Communications Society Best Survey Paper Award and the 2011 IEEE Communications Society Fred Ellersick Prize Paper Award. He was listed as a Clarivate Analytics Highly Cited Researcher in Computer Science in 2017, 2018, 2019, 2020, 2021, and 2022. Currently, he serves as an Editor for the IEEE TRANSACTIONS ON MOBILE COMPUTING and the IEEE ComSoc Director of Online Content. Previously, he served as the Editor-inChief (EiC) for the IEEE Press (2018–2021) and the IEEE Communications Surveys and Tutorials (2012–2016). He was a Distinguished Lecturer of the IEEE Communications Society and the IEEE Vehicular Technology Society. He served as the Director of Magazines for the IEEE Communications Society(2020–2021).
\end{IEEEbiography}

\vspace{11pt}

\end{document}